\documentclass[fleqn,usenatbib]{mnras}

\usepackage{newtxtext,newtxmath}
\usepackage[T1]{fontenc}

\DeclareRobustCommand{\VAN}[3]{#2}
\let\VANthebibliography\thebibliography
\def\thebibliography{\DeclareRobustCommand{\VAN}[3]{##3}\VANthebibliography}

\usepackage{graphicx}	% Including figure files
\usepackage{amsmath}	% Advanced maths commands
\usepackage[flushleft]{threeparttable}
\usepackage{longtable}

\newcommand{\quotes}[1]{``#1''} % double quotes

\title[On the thermal structure of proto-SSC\,13]{On the thermal structure of the proto-Super Star Cluster 13 in NGC\,253}

\author[F. Rico-Villas et al.]{
F. Rico-Villas$^{1}$\thanks{E-mail: frico@cab.inta-csic.es},
E. González-Alfonso$^{2}$,
J. Martín-Pintado$^{1}$,
V. M. Rivilla$^{1}$,
and S. Mart\'in$^{4, 5}$
\\
$^{1}$Centro de Astrobiolog\'ia (CSIC-INTA), Ctra de Ajalvir, km 4, Torrej\'on de Ardoz, 28850, Madrid, Spain\\
$^{2}$Universidad de Alcalá, Departamento de Física y Matemáticas, Campus Universitario, Alcalá de Henares, 28871, Madrid, Spain\\
$^{3}$European Southern Observatory, Alonso de C\'ordova, 3107, Vitacura, Santiago 763-0355, Chile\\
$^{4}$Joint ALMA Observatory, Alonso de C\'ordova, 3107, Vitacura, Santiago 763-0355, Chile
}

\date{Accepted 2022 July 18. Received YYY; in original form ZZZ}

\pubyear{2022}

\begin{document}
\label{firstpage}
\pagerange{\pageref{firstpage}--\pageref{lastpage}}
\maketitle

\begin{abstract}
Using high angular resolution ALMA observations ($0.02\arcsec\approx0.34$\,pc), we study the thermal structure and kinematics of the proto super star cluster $13$ in the central region of NGC\,253 through their continuum and vibrationally excited HC$_3$N emission  from  $J=24-23$ and $J=26-25$ lines arising from vibrational states up to $v_4=1$.
We have carried 2D-LTE and non-local radiative transfer modelling of the radial profile of the HC$_3$N and continuum emission in concentric rings of $0.1$\,pc width. From the 2D-LTE analysis, we found a Super Hot Core (SHC) of $1.5$\,pc with very high vibrational temperatures ($>500$\,K), and a jump in the radial velocity ($21$\,km\,s$^{-1}$) in the SE-NW direction. From the non-local models, we derive the HC$_3$N column density, H$_2$ density and dust temperature ($T_\text{dust}$)  profiles. Our results show that the thermal structure of the SHC is dominated by the greenhouse effect due to the high dust opacity in the IR, leading to an overestimation of the LTE $T_\text{dust}$ and its derived luminosity. The kinematics and $T_\text{dust}$ profile of the SHC suggest that star formation was likely triggered by a cloud-cloud collision. We compare proto-SSC\,$13$ to other deeply embedded star-forming regions, and discuss the origin of the $L_\text{IR}/M_{\text{H}_2}$ excess above $\sim100$\,L$_\odot$\,M$_\odot^{-1}$ observed in (U)LIRGs.

\end{abstract}

\begin{keywords}
galaxies: individual: NGC 253 -- galaxies: ISM -- galaxies: nuclei -- galaxies: star clusters: general -- galaxies: star formation
\end{keywords}

%%%%%%%%%%%%%%%%%%%%%%%%%%%%%%%%%%%%%%%%%%%%%%%%%%

%%%%%%%%%%%%%%%%% BODY OF PAPER %%%%%%%%%%%%%%%%%%

\section{Introduction}

Using high angular resolution ALMA observations ($0.02\arcsec\approx0.34$\,pc), we study the thermal structure and kinematics of the proto super star cluster $13$ in the central region of NGC\,253 through their continuum and vibrationally excited HC$_3$N emission  from  $J=24-23$ and $J=26-25$ lines arising from vibrational states up to $v_4=1$.
We have carried 2D-LTE and non-local radiative transfer modelling of the radial profile of the HC$_3$N and continuum emissions. From the 2D-LTE analysis, we found a Super Hot Core (SHC) with high vibrational temperatures ($>500$\,K), and a jump in the radial velocity ($21$\,km\,s$^{-1}$) in the SE-NW direction. From the non-local models, we derive the HC$_3$N column density, H$_2$ density and dust temperature ($T_\text{dust}$) profiles. The thermal structure of the SHC is dominated by the greenhouse effect leading to an overestimation of the LTE $T_\text{dust}$ and the corresponding luminosities. Star formation seems to be centrally-peaked, supporting the competitive accretion scenario. The formation of the super star cluster was most likely triggered by a cloud-cloud collision as suggested by the kinematics of the SHC. Finally, we compare proto-SSC\,$13$ to other deeply embedded star-forming regions, and discuss the origin of the $L_\text{IR}/M_{\text{H}_2}$ excess above $\sim100$\,L$_\odot$\,M$_\odot^{-1}$ observed in (U)LIRGs.

Even though a large number of SSCs have been observed in different starbursting galaxies \citep[e.g.][]{Melnick1985SSC, Whitmore1993SSC, Oconnell1994SSC, Meurer1995SSC, Whitmore1995, HoFilippenko1996, HoFilippenko1996bSSC, Watson1996, Whitmore1999SSC, Gelatt2001SSC, Tremonti2001SSC, Vanzi2003SSC, Turner2004SSC, McCrady2005SSC, Melo2005SSC, Mengel2008SSC}, most of these studies have been carried out in the UV, optical or near-IR wavelengths \citep[see ][for a review]{Portegies2010}, detecting relatively evolved SSCs. This is because SSCs remain entirely enshrouded by the large column densities of their natal giant molecular cloud during their formation and earliest phases of evolution, preventing the direct observations of these stages at the aforementioned wavelengths.   Therefore, little is known about SSCs during their early evolutionary phases apart from the fact that they need to form most of the stars in a very short timescale before the feedback from massive stars starts to disrupt and disperse the natal cloud \citep[supernovae from the most massive stars take place after $\sim10^5$\,yr of stellar evolution;][]{Bressert2012, Krumholz2014}, requiring relatively high star formation rates (SFRs) \citep[][]{Beck2015}. 

Thanks to ALMA and its high angular resolution and sensitivity, we can now observe high energy molecular lines in the (sub)millimeter range, sampling the hot molecular environments within the obscuring dust regions. This allow us to study the embedded phase of proto-SSCs during their super hot core phase (SHC), a scaled-up version of galactic molecular hot cores \citep[see][]{RicoVillas2020}. 
The rotational transitions from vibrationally excited states of HC$_3$N and HCN, usually excited by mid-IR  \citep[e.g.][]{GA19}, are excellent tracers of the SHC phase.
Since HC$_3$N has a much lower rotational constant ($B_\nu\sim4.5$\,GHz) than HCN does ($B_\nu\sim44.4$\,GHz) and allows up to seven normal vibrational modes (four stretching modes: $v_1$, $v_2$, $v_3$, $v_4$; and three bending modes: $v_5$, $v_6$ $v_7$), it has more rotational transitions in a given wavelength range ($1$ every $\sim 9$\,GHz), covering a wider energy range. With such many rotational lines from different vibrationally excited states, we are able to study the thermal structure and the kinematics of proto-SSCs with unprecedented detail. 
This provides the only way to estimate the stellar content in the earliest evolutionary phases of SSCs (proto-SSCs) through the determination of the luminosity of the SHC phase. However, as discussed by \citet{RicoVillas2020,RicoVillas2021NGC1068}, the determination of the luminosity from vibrationally excited HC$_3$N (hereafter HC$_3$N*) emission is not straightforward due to the back-warming of radiation \citep{Donnison1976, RowanRobinson1982, Ivezic1997}. This greenhouse effect produced by the back-warming \citep{GA19} is caused by the high column densities ($\gtrsim 10^{25}$\,cm$^{-2}$) and thus high dust opacities, which increase the inner dust temperatures well above the expected dust temperature in the optically thin case. This effect, by increasing the dust temperature and the mid-IR (wavelength responsible for the vibrational excitation of HC$_3$N) radiation field  in the innermost regions, leads to an overestimation of the luminosity of the SHCs. To fully account for this effect and to derive reliable luminosities, the inner thermal structure needs to be derived from spatially resolved observations of the HC$_3$N* emission.

Such study was done in \citet{RicoVillas2020} at a resolution of $0.29\arcsec\times0.19\arcsec$ ($4.9\,\text{pc}\times3.2\,\text{pc}$) for the $14$ forming SSCs previously analyzed by \citet{Leroy2018} in the nuclear region of the galaxy NGC\,253. This galaxy, located at $3.5$\,Mpc\footnote{We have adopted the most common distance assumed in the literature for NGC\,253 of $D=3.5$\,Mpc \citep{Rekola2005}. However, recently \citet{ngc253_dist_anand} and \citet{ngc253_dist_kara} measured $D=3.70\pm0.12$.}, is one of the closest galaxies hosting a nuclear starburst \citep[for details see][]{Mills2021}.  Through the detection of HC$_3$N* in the $v_7=1$, $v_7=2$ and $v_6=1$ states, \citet{RicoVillas2020} estimated the luminosity and age of the SSCs in the nuclear region. The high luminosities and young ages proved that they were forming stars at high rates with high star formation efficiencies (SFEs). 

In this work, thanks to new ALMA observations at an order of magnitude higher angular resolution of $0.022\arcsec\times0.020\arcsec$ ($0.37\,\text{pc}\times0.34\,\text{pc}$), we are able to resolve most of the forming SSCs studied in \citet{RicoVillas2020} and study in detail the inner thermal structure and kinematics from HC$_3$N* emission in the proto-SSC\,$13$. We selected to study this source in detail because it has one of the brightest continuum emission, was among the youngest sources in \citet{RicoVillas2020} and has no clear outflow signatures  \citep{Levy2021}. We combine the continuum data (at $219$\,GHz and $345$\,GHz) and the HC$_3$N* line emission with non-local radiative transfer models (which include all the high vibrationally excited states detected), to obtain the temperature and density profiles of a proto-SSC, and confirm the presence of the back-warming effect as expected from the high column densities. We also establish that the IR luminosities derived from the HC$_3$N* emission are overestimated by about an order of magnitude if the greenhouse effect is ignored.

\section{Observations}

We observed the nuclear region of NGC\,253 using the ALMA $12$\,m Array at very high angular resolution. We performed Band 6 ($211-275$\,GHz coverage) observations in ALMA configuration C43-9 (ALMA project 2018.1.01395.S, PI: Rico-Villas, Fernando), which provides the required high angular resolution to resolve most of the proto-SSCs studied in \citet{RicoVillas2020}. In order to observe the maximum number of HC$_3$N* lines as possible, we optimized the correlator set up by tuning the Local Oscillator at $227.889$\,GHz, allowing us to record in the lower and upper side bands the  frequencies of the $J=24-23$ ($\sim220$\,GHz) and $J=26-25$ ($\sim237$\,GHz) HC$_3$N* rotational transitions. The four spectral windows, with a bandwidth of  $1.875$\,GHz and a channel resolution of $1.3$\,km\,s$^{-1}$, were tuned to cover the frequencies  $218.16-220.03$\,GHz and $220.01-221.88$\,GHz for the lower side band, and $234.32-236.19$\,GHz and $236.17-238.04$\,GHz for the upper side band. 

The observations were carried out in six different execution blocks between 2019 June and 2019 July, with an average exposure time on source  per execution block of $44.45$\,minutes, and a total integration time of $4.45$\,hr. During the observations, short scans of J$0006-0623$ were used for bandpass and flux calibration; J$0038-2459$ was used for phase calibration.  A summary of the observation details can be found in Table~\ref{tab:HRNGC253_obslog}. The table shows the shortest baselines for each execution block, which set the maximum recoverable scale (MRS\footnote{The maximum recoverable scale is the largest angular scale structure that can be recovered by an interferometer and varies with the smallest available baseline ($D_\text{min}$) and observed wavelength  as $\theta_\text{MRS}\approx0.6\lambda/D_\text{min}$ \citep{ALMAbook}.})   between $0.35\arcsec$ and $0.38\arcsec$ ($5.9$\,pc and $6.4$\,pc). This MRS should be enough to assume that the entire flux all the compact SHCs \citep[with sizes $<2\,\text{pc}$][]{RicoVillas2020} is recovered. 
 To obtain further information about the dust emission, we used also Band 7 observations from the ALMA project ID: 2017.1.00433.S (PI: Bolatto, Alberto) at $\sim345$\,GHz. For further details on this observations see \citet{Levy2021}.

\begin{table}
  \caption[NGC\,253 ALMA Cycle 6 observations log]{Log of ALMA Cycle 6 observations  NGC\,253.}
  \centering
	\centering
	\label{tab:HRNGC253_obslog}
	\begin{tabular}{lcccc} 
		\hline
		Date  & N. Ant. & Short. Base. & Long. Base. & Int. Time \\
         (UT) &   & (m) & (m) & (minutes) \\
		\hline
		2019-06-06  	& 43 & 210 & 15238    &  44.30 \\
        2019-06-08		& 42 &  83 & 16196    &  44.43 \\
        2019-06-09  	& 43 &  83 & 16196    &  44.52 \\
        2019-06-10  	& 47 &  83 & 16196    &  44.48 \\
        2019-06-12  	& 41 &  83 & 16196    &  44.62 \\
        2019-07-10  	& 41 & 138 & 13894    &  44.38 \\
		\hline
	\end{tabular}
\end{table}

\subsection{Imaging}
\label{subsec:imaging}

The visibilities were calibrated using the 
 \texttt{CASA}\footnote{\url{https://casa.nrao.edu/}} 5.6.1-8 package \citep[Common Astronomy Software Applications,][]{McMullin2007} pipeline version $42866$. For the imaging we then combined the visibilities of the six execution blocks using the \texttt{tclean} task from the same \texttt{CASA} 5.6.1-8 version. For image cleaning we applied a Briggs weighting scheme, setting the \texttt{robust} parameter to $0.5$. For masking we used the option \texttt{auto-multithresh} to automatically mask regions during the deconvolution \citep{Kepley2020automask}. The channel width was set to  the original spectral resolution of $1$\,MHz (i.e. $\sim1.3$\,km\,s$^{-1}$). Considering the resulting angular resolution of $\sim0.025\arcsec$, we set the pixel size to $0.007\arcsec$ ($0.12$\,pc) to have at least $9$ pixels per beam solid angle (i.e. 1.5 times Nyquist sampling). After the imaging, we applied the primary beam correction.  Due to the large number of emission and absorption lines expected towards the proto-SSCs, the spectral cubes were imaged from the visibilities without  applying a $uv$-plane continuum subtraction. The resulting synthesized beams, channel resolution, covered frequencies and rms for each spectral cube are listed in  Table~\ref{tab:HRNGC253_obsima}. For the spectral cubes, the continuum was subtracted by fitting order 0  baselines. A continuum map (Figure~\ref{fig:NGC253_HR_219GHz_cont}), was generated from apparently line-free channels in the $uv$-plane with a synthesized beam of $0.020\arcsec\times0.019\arcsec$ ($0.34\,\text{pc}\times0.32\,\text{pc}$)  and a rms noise of $13$\,$\upmu$Jy\,beam$^{-1}$. However, due to the significant spectral feature contribution mentioned above, the continuum emission in these regions has to be treated with caution. 

  \begin{table}
  \caption[NGC\,253 ALMA ID 2018.1.01395.S project spectral windows]{ALMA ID 2018.1.01395.S project NGC\,253 reduced spectral cubes properties. The table lists the covered rest frequencies, the synthesized beam, position angle, channel width and channel rms for each spectral cube.}
  \centering
	\centering
	\label{tab:HRNGC253_obsima}
	\begin{tabular}{ccccc} 
		\hline
		Rest Freq.  & Syn. Beam & PA & Ch. width & rms   \\
         (GHz) &  (arcsec$^2$) & ($\deg$) & (km\,s$^{-1}$) & (mJy\,beam$^{-1}$) \\
		\hline
		 $218.16-220.03$	& $0.026\times0.022$  & $42$ & $1.34$    &  $0.38$ \\
         $220.01-221.88$    & $0.023\times0.022$  & $64$ & $1.33$    &  $0.35$ \\
         $234.32-236.19$  	& $0.022\times0.020$  & $53$ & $1.24$    &  $0.37$ \\
         $236.17-238.04$  	& $0.022\times0.020$  & $47$ & $1.23$    &  $0.39$ \\
		\hline
	\end{tabular}
\end{table}
 
 For the $\sim345$\,GHz data reduction and calibration of the visibilities we used the \texttt{CASA} 5.1.1-5 pipeline version $40896$. For the imaging of the $uv$-plane continuum, we selected line emission free channels and used the \texttt{tclean} task with the \texttt{robust} parameter set to $0.5$ and left the default pixel from the pipeline (i.e. $0.005\arcsec=0.085$\,pc). The resulting continuum map at $345$\,GHz has a resolution of $0.028\arcsec\times0.034\arcsec$ ($0.47\,\text{pc}\times0.58\,\text{pc}$)  and a rms noise of $37$\,$\upmu$Jy\,beam$^{-1}$.
 A summary of the image details of the continuum maps used in this work is shown in Table~\ref{tab:HRNGC253_continuums}.

 \begin{table}
  \caption[NGC\,253 ALMA continuums]{NGC\,253 continuum map images parameters.}
  \centering
	\centering
	\label{tab:HRNGC253_continuums}
	\begin{tabular}{ccccc} 
		\hline
		Project ID & Rest Freq.  & Syn. Beam & PA  & rms   \\
                   & (GHz) &  (arcsec$^2$) & ($\deg$)  & ($\upmu$Jy\,beam$^{-1}$) \\
		\hline
		 2018.1.01395.S	& $219$ &   $0.020\times0.019$  & $70$     &  $13$ \\
		 2017.1.00433.S	& $345$ &   $0.028\times0.034$  & $79$     &  $37$ \\
		\hline
	\end{tabular}
\end{table}

\section{The spatial structure of SSCs}
\label{sec:spatial_structure}

\begin{figure*}
\centering
    \includegraphics[width=0.8\linewidth]{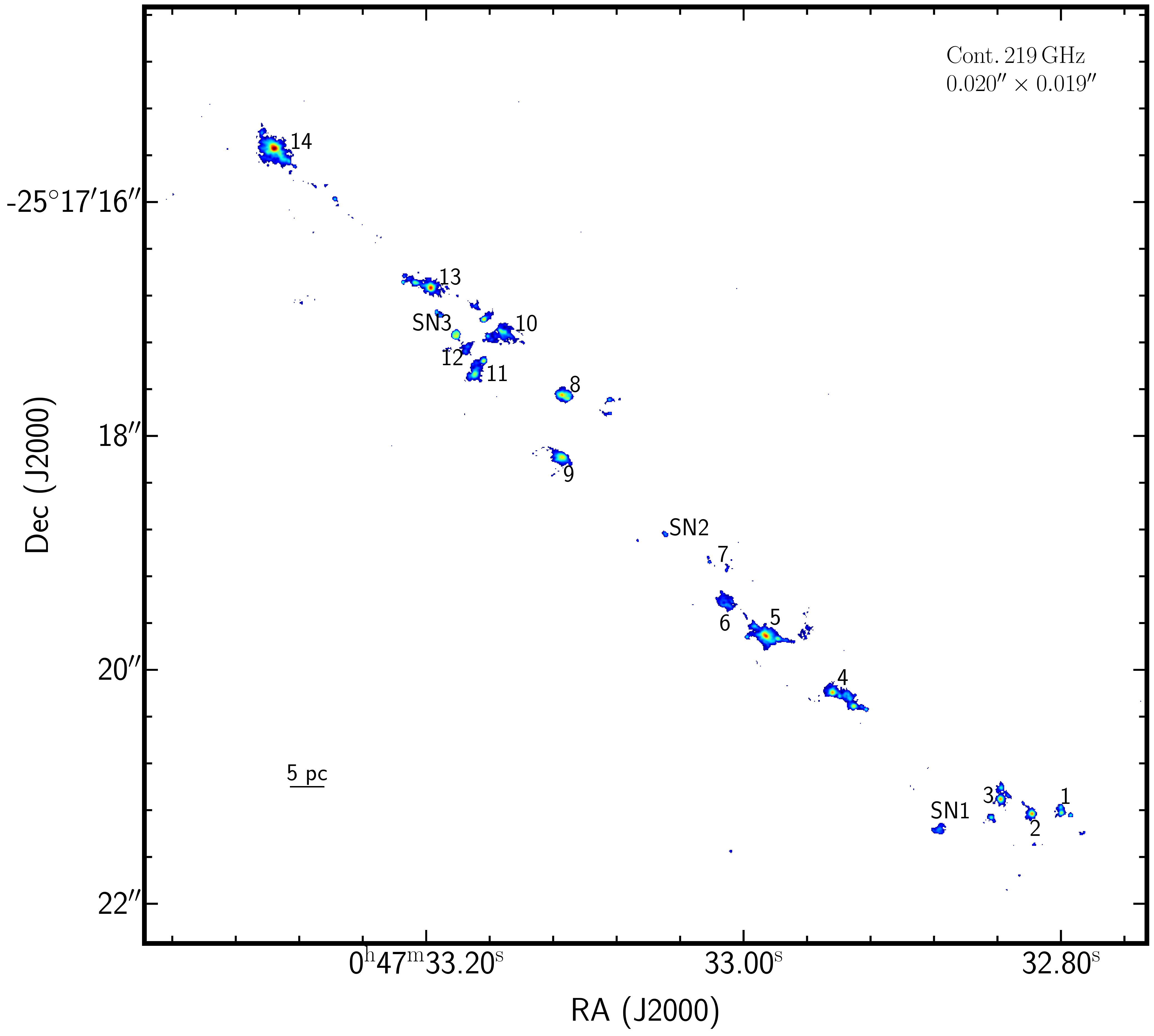}
  \caption[NGC\,253 high-resolution continuum map at $219$\,GHz]{Continuum map at $219$\,GHz of the NGC\,253 nuclear region with a resolution of $0.020\arcsec \times 0.019\arcsec$ above $3\sigma_{219\text{GHz}}$ ($\sigma_{219\text{GHz}}=13$\,$\upmu$Jy). Labelled are the proto-SSCs studied in \citet{RicoVillas2020}. 
  }
  \label{fig:NGC253_HR_219GHz_cont}
\end{figure*}

\begin{figure*}
\centering
    \includegraphics[width=\linewidth]{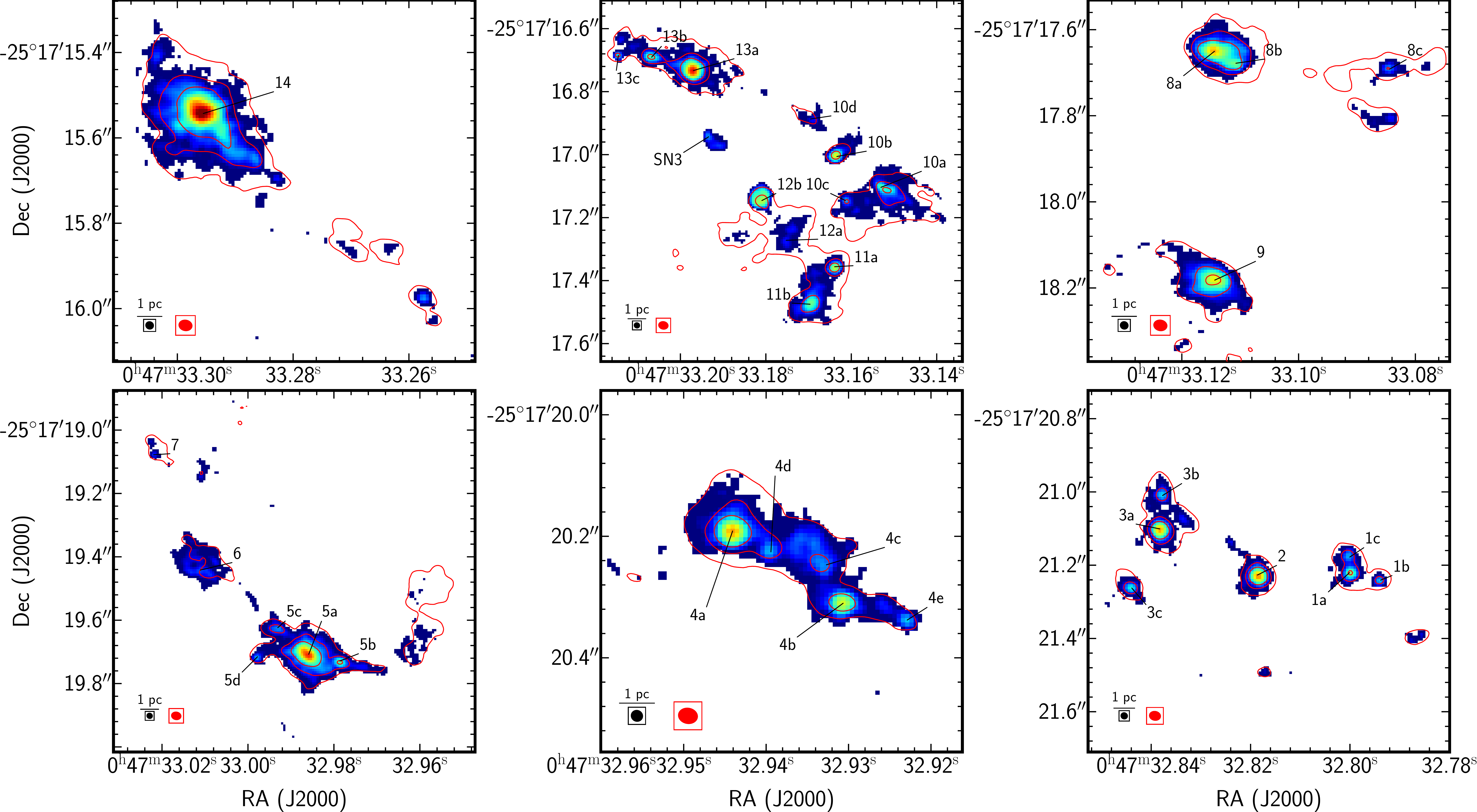}
  \caption[NGC\,253 high-resolution zoom of the $219$\,GHz and $345$\,GHz continuum emission with proto-SSCs indicated]{Zoom of the continuum map at $219$\,GHz above $5\sigma_{219\text{GHz}}$ ($\sigma_{219\text{GHz}}=13$\,$\upmu$Jy\,beam$^{-1}$)  for the different regions containing the proto-SSCs studied in \citet{RicoVillas2020}. Overlaid, the continuum emission at $345$\,GHz is shown in red with contour levels at $5\sigma_{345\text{GHz}}$, $10\sigma_{345\text{GHz}}$, $25\sigma_{345\text{GHz}}$ and $50\sigma_{345\text{GHz}}$ ($\sigma_{345\text{GHz}}=37$\,$\upmu$Jy\,beam$^{-1}$). The different components of each SSC (in which a given SSC is fragmented) are labelled with letters. The $219$\,GHz beam size ($0.020\arcsec \times 0.019\arcsec$; in black) and the $345$\,GHz beam size ($0.028\arcsec \times 0.0234\arcsec$; in red)  are plotted, along with a $1$\,pc scale, on the bottom left-hand corner for reference.  
  }
  \label{fig:NGC253_HR_219GHz_cont_zoom}
\end{figure*}

The very high resolution ALMA data ($\sim0.40\,\text{pc}\times0.35\,\text{pc}$) at $\sim220-235$\,GHz allows us to study in detail the structure and kinematics of the proto-SSCs analysed in \citet{Leroy2018} and \citet{RicoVillas2020}. The study carried out in \citet{RicoVillas2020} was done at a resolution of $0.29\arcsec \times 0.19\arcsec$ ($4.9\,\text{pc}\times3.2\,\text{pc}$) at $\sim219$\,GHz.
At similar frequencies ($218-238$\,GHz), this new data set provides a resolution up to one order of magnitude higher ($0.026\arcsec \times 0.022\arcsec=0.44\,\text{pc}\times0.38\,\text{pc}$, see Table~\ref{tab:HRNGC253_obsima}) than our previous analysis. Some of the $14$ proto-SSCs studied in \citet{RicoVillas2020} show clear inner structure with several components (SSCs $14$, $13$, $12$, $11$, $10$, $8$, $5$, $4$, $3$, $1$; see Figure~\ref{fig:NGC253_HR_219GHz_cont}).
This can be seen in the continuum map at $219$\,GHz  in Figure~\ref{fig:NGC253_HR_219GHz_cont}, encompassing all the proto-SSCs. The structure of the fragmented proto-SSCs is better illustrated in Figure~\ref{fig:NGC253_HR_219GHz_cont_zoom}, which shows a zoom-in of the different proto-SSCs.
In those cases where the marginally resolved proto-SSCs seen in \citet{RicoVillas2020} fragment at higher resolution, we label with letters the different components, leaving the label \quotes{a} for the region coincident with the position given by \citet{Leroy2018}.
The coordinates for these regions, based on their peak emission location at $219$\,GHz, are listed in Table~\ref{tab:219HR_positions}.
In Figure~\ref{fig:NGC253_HR_219GHz_cont} and Table~\ref{tab:219HR_positions}, we also show three positions with $219$\,GHz continuum emission but a weak counterpart $345$\,GHz continuum emission labelled SN\,1, SN\,2 and SN\,3. The lower $345$\,GHz continuum emission indicates a negative spectral index, not associated to dust clumps but rather synchrotron dominated supernovae remnants (SNRs), which seem to be associated to the SNRs in \citet{Ulvestad1997}.

\begin{table}
\begin{center}
\caption[NGC\,253 high-resolution $219$\,GHz continuum emission peak positions coordinates]{Coordinates of the peak positions at $219$\,GHz.}
\label{tab:219HR_positions}
\begin{tabular}{lcc}
\hline \noalign {\smallskip}
 Position & RA (J2000) & Dec (J2000) \\
          & $00^\text{h}47^\text{m}$ & $-25^{\circ}17^{\prime}$ \\
\hline \noalign {\smallskip}
1a  & $ 32^\text{s}.8001$ & $21\arcsec.220$ \\ 
1b  & $ 32^\text{s}.7939$ & $21\arcsec.241$ \\ 
1c  & $ 32^\text{s}.8001$ & $21\arcsec.178$ \\ 
2   & $ 32^\text{s}.8186$ & $21\arcsec.227$ \\ 
3a  & $ 32^\text{s}.8383$ & $21\arcsec.101$ \\ 
3b  & $ 32^\text{s}.8377$ & $21\arcsec.010$ \\ 
3c  & $ 32^\text{s}.8439$ & $21\arcsec.262$ \\ 
SN1 & $ 32^\text{s}.8770$ & $21\arcsec.360$ \\ 
4a  & $ 32^\text{s}.9441$ & $20\arcsec.191$ \\ 
4b  & $ 32^\text{s}.9306$ & $20\arcsec.310$ \\ 
4c  & $ 32^\text{s}.9327$ & $20\arcsec.247$ \\ 
4d  & $ 32^\text{s}.9394$ & $20\arcsec.226$ \\ 
4e  & $ 32^\text{s}.9229$ & $20\arcsec.338$ \\ 
5a  & $ 32^\text{s}.9859$ & $19\arcsec.708$ \\ 
5b  & $ 32^\text{s}.9787$ & $19\arcsec.729$ \\ 
5c  & $ 32^\text{s}.9931$ & $19\arcsec.624$ \\ 
5d  & $ 32^\text{s}.9977$ & $19\arcsec.715$ \\ 
6   & $ 33^\text{s}.0112$ & $19\arcsec.442$ \\ 
SN2 & $ 33^\text{s}.0493$ & $18\arcsec.833$ \\ 
7   & $ 33^\text{s}.0220$ & $19\arcsec.078$ \\ 
8a  & $ 33^\text{s}.1144$ & $17\arcsec.650$ \\ 
8b  & $ 33^\text{s}.1108$ & $17\arcsec.678$ \\ 
8c  & $ 33^\text{s}.0844$ & $17\arcsec.692$ \\ 
9   & $ 33^\text{s}.1144$ & $18\arcsec.182$ \\ 
10a & $ 33^\text{s}.1531$ & $17\arcsec.104$ \\ 
10b & $ 33^\text{s}.1634$ & $17\arcsec.006$ \\ 
10c & $ 33^\text{s}.1614$ & $17\arcsec.146$ \\ 
10d & $ 33^\text{s}.1696$ & $16\arcsec.887$ \\ 
11a & $ 33^\text{s}.1639$ & $17\arcsec.356$ \\ 
11b & $ 33^\text{s}.1696$ & $17\arcsec.475$ \\ 
12a & $ 33^\text{s}.1753$ & $17\arcsec.272$ \\ 
12b & $ 33^\text{s}.1810$ & $17\arcsec.146$ \\ 
SN3 & $ 33^\text{s}.1934$ & $16\arcsec.943$ \\ 
13a & $ 33^\text{s}.1970$ & $16\arcsec.733$ \\ 
13b & $ 33^\text{s}.2068$ & $16\arcsec.691$ \\ 
13c & $ 33^\text{s}.2145$ & $16\arcsec.684$ \\ 
14  & $ 33^\text{s}.2955$ & $15\arcsec.543$ \\  
\hline \noalign {\smallskip}
\end{tabular}
\end{center}
\end{table}

\section{Proto-SSC 13: A case study}

Our high spatial resolution data clearly shows a large fragmentation on the formation of massive star clusters and SSCs. To better understand the formation of SSCs and their evolution, we need to study the SSCs in their SHC phase (i.e. proto-SSC) in great detail to determine the properties of the different components detected in our maps.

In this work we carry out a detailed study of proto-SSC\,$13$a (i.e. SHC\,13). 
The analysis of the other SHCs in forthcoming works will follow the procedure implemented for proto-SSC\,$13$a.
Between the two brightest sources at $219$\,GHz and $345$\,GHz continuum emission, proto-SSC\,13a was selected instead of proto-SSC\,14 because the latter, despite showing brighter continuum emission at both frequencies, presents spatially asymmetric emission and complex spectral absorption features in its central region at the HC$_3$N$^*$ spectral lines. In fact, \citet{Levy2021} recently observed P-Cygni line profiles in the CS $7-6$ and H$^{13}$CN $4-3$ lines in the central region of SSC\,14 (also in SSC\,4a and SSC\,5a), indicative of outflowing gas suggesting a more evolved state of evolution. Since we are interested in understanding the triggering mechanism for the formation of SSCs in galaxies and since there is no evidence for mass ejection in proto-SSC\,$13$a, it is expected to provide a better representation of the very early stages of SSC formation.

In particular, we will address one of the main uncertainties in the previous analysis, the actual luminosities of the SHCs strongly affected by the back-warming (i.e. the greenhouse effect) and the inner kinematics unresolved by previous lower spatial resolution data. In the following sections we present the first study of internal heating of the SHCs combining a multitransition analysis of the spatially resolved HC$_3$N* and dust emissions with the modelling of the backwarming effect for different heating scenarios.

\subsection{\texorpdfstring{HC$_3$N*}{HC3N*} emission: 2D LTE analysis of proto-SSC\,13}
\label{subsec:2d_analysis}

Following \citet{RicoVillas2020}, a LTE analysis of the HC$_3$N* emission from proto-SSC\,$13$ was carried out with the Spectral Line Identification and Modelling (\texttt{SLIM}) module \citep{Madcuba2019} within \textsc{Madcuba}.
\texttt{SLIM} is a very convenient tool for high spatial resolution spectral analysis since it allows for line identification of spectral features and its LTE analysis (including optical depth effects) directly onto spectral data.
We have developed scripts to deal with the 2D  determination of the LTE parameters with the \texttt{SLIM} multi line analysis including upper limits in both HC$_3$N column densities and $T_\text{vib}$. %These scripts, will be coded in \texttt{SLIM} cubes for automatic interactive analysis directly on spectral cubes.

\begin{figure*}
\centering
    \includegraphics[width=0.9\linewidth]{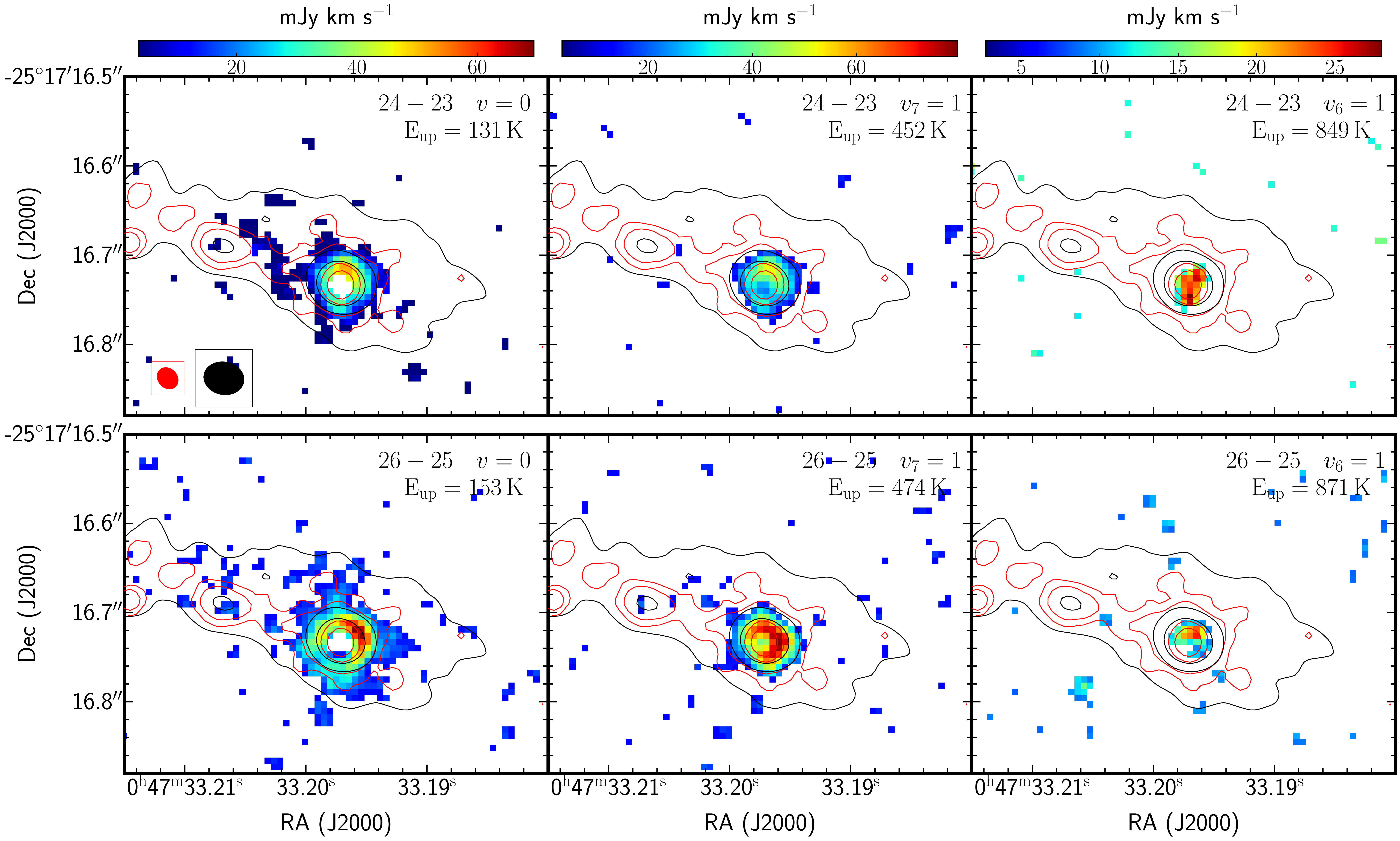}
  \caption[NGC\,253 proto-SSC\,$13$ HC$_3$N $v=0$, $v_7=1$ and $v_6=1$ moment 0 maps]{SSC\,13 HC$_3$N* moment 0 maps. The left panels show the $J=24-23$ (top) and the $J=26-25$ (bottom) rotational transitions from the $v=0$ state. The middle panels show the  $J=24-23$ $l_\text{up}=1$ (top) and the $J=26-25$ $l_\text{up}=1$ (bottom) rotational transitions from the $v_7=1$ vibrational state.  The right panels show the $J=24-23$ $l_\text{up}=-1$ (top) and the $J=26-25$ $l_\text{up}=-1$ (bottom) rotational transitions from the $v_6=1$ vibrational state. Overlaid with red and black contours are the $5\sigma$, $10\sigma$, $50\sigma$ and $100\sigma$ levels from the $219$\,GHz and $345$\,GHz continuum emission, respectively. Top left panel shows the beam sizes of the $219$\,GHz (in red) and the $345$\,GHz (in black) continuum emission.
  }
  \label{fig:NGC253_HR_mom0_SHC13}
\end{figure*}

Figure~\ref{fig:NGC253_HR_mom0_SHC13} shows the integrated intensity (i.e. moment 0) maps of SSC\,13 from the HC$_3$N $J=24-23$ and $J=26-25$ rotational transitions in the $v=0$ ground state, the $v_7=1$ and $v_6=1$ vibrationally excited states. Among the different emitting regions that conform the proto-SSC\,13 (i.e. $13$a, $13$b and $13$c), the only region that shows HC$_3$N* emission is $13$a, where we will focus our analysis. From the extent of the continuum emission at $219$\,GHz and $345$\,GHz, and the emission of the $v=0$ transitions, we estimate the radii of the   proto-SSC\,$13$a to be $r\approx1.5$\,pc.
In the moment 0 maps, in the region where the $219$\,GHz continuum emission peaks (the central region of SSC\,$13$a) we observe a hole in the emission of the $v=0$ lines, and also some decrease of the flux in the $v_7=1$ lines. However, the moment 0 maps from the $v_6=1$ lines do not show such feature. Since the $v_6=1$ transitions arise from higher energies ($E_\text{up}>849$\,K) than the $v=0$ and $v_7=1$ lines ($E_\text{up}<475$\,K), we consider the decrease in the $v=0$ and $v_7=1$ to be due to  absorption of the background continuum and self-absorption in the low-energy transitions instead of a true cavity at the center of proto-SSC\,$13$a. Also, we remind that like in galactic HCs, some uncertainty in estimating the continuum emission is present in the inner regions, where a high spectral density of absorption and emission lines typical in SHCs hamper its estimation.  

Besides the rotational transitions arising from the $v=0$, $v_7=1$ and $v_6=1$ states, other HC$_3$N* $J=24-23$ and $J=26-25$ rotational  transitions from even higher vibrationally excited states also lie in the analysed spectral range. These vibrational excited states are $v_7=2$, $v_5=1/v_7=3$, $v_6=v_7=1$, $v_7=4/v_5=v_7=1$, $v_6=2$, $v_4=v_7=1$ and $v_4=1,v_7=2/v_5=2^0$ (see Table~\ref{tab:HC3N_lines_info}). 
However, as shown in \citet{RicoVillas2020}, the proto-SSCs like proto-SSC\,$13$a have an extremely rich chemistry and a very prominent molecular line emission from many other species, making several HC$_3$N* transitions to be blended with transitions from other molecules. This makes that some of the weaker high-energy\footnote{We will refer to low-energy HC$_3$N* lines to the rotational transitions arising from the ground state $v=0$ and the $v_7=1$ vibrationally excited state. Similarly, high-energy transitions will be those with $E_\text{lo}>1000$\,K, i.e. those from the  $v_5=1/v_7=3$, $v_6=v_7=1$, $v_4=1$ and $v_6=2$ vibrational states.} HC$_3$N* emission lines to be blended with other species.
We have been able to confirm the detection of clean or only slightly-blended HC$_3$N* emission in the $v=0$ vibrational ground state and in the $v_7=1$, $v_7=2$, $v_6=1$, $v_5=1/v_7=3$, $v_6=v_7=1$ and $v_4=1$ vibrationally excited states (see Figure~\ref{ap:fig:NGC253_HR_SLIM_SHC13_ring0p15}). Transitions from the $v_7=4/v_5=v_7=1$ vibrational state are strongly blended with other lines. For higher vibrational states ($v_6=2$ and $v_4=v_7=1$), the expected line strength \citep[from the CDMS\footnote{\url{http://www.astro.uni-koeln.de/cgi-bin/cdmssearch}} catalogue;][]{CDMS1,CDMS2} decreases by more than a $45\%$ with respect to that of the $v_4=1$\:$J=26-25$ transition (detected only in the most central region), and therefore are not detected. A summary of the HC$_3$N* transitions in the observed spectral range, with their quantum numbers, energy of the upper state of the transition, frequencies, Einstein-$A_{ul}$ coefficients, line strengths and if they are detected and contaminated is shown in Table~\ref{tab:HC3N_lines_info}.

%\begin{figure*}
%\centering
%    \includegraphics[width=0.95\linewidth]{images/appendix/SHC_13_plot_d0p15pc_mJy_fill_restfreqs_sty3_finallim.pdf}
%  \caption[NGC\,253 proto-SSC\,$13$a HC$_3$N* averaged spectra]{Proto-SSC\,$13$a HC$_3$N* averaged spectra (black histogram) from the ring enclosing the pixels with distances to the center between $0.1$\,pc and $0.2$\,pc. The fitted HC$_3$N* LTE profiles are shown with solid green lines for the $J=24-23$ transitions and with solid blue lines for the $J=26-25$ transitions. The sum of the \texttt{SLIM} fitted LTE profiles for all species is shown with dashed red lines. Labels for HC$_3$N transitions are shown in black, indicating the vibrational state and the upper level quantum numbers in parenthesis (i.e. ($l_\text{up}$) or ($l_\text{up}$, $k_\text{up}$)). Labels for transitions from other species are indicated in grey. 
%  }
%  \label{fig:NGC253_HR_SLIM_SHC13_ring0p15}
%\end{figure*}

To improve the signal-to-noise ratio for the LTE modelling, we have smoothed the spectral resolution to half the original value, down to $\sim2.5$\,km\,s$^{-1}$.
Figures inside Appendix~\ref{ap:HR_spec} show the averaged spectra within rings centered on proto-SSC\,$13$, (see Section~\ref{subsec:rings} for the definition of these rings) with all the HC$_3$N* transitions and other identified species also indicated for completeness.
%Figure~\ref{ap:NGC253_HR_SLIM_SHC13_ring0p15} shows the averaged spectra  within a ring with distances between $0.1$\,pc and $0.2$\,pc to the center (see Section~\ref{subsec:rings} for the definition of these rings and Appendix~\ref{ap:HR_spec} for the averaged spectra of all rings, with distances up to $1.5$\,pc, available in the online version) with all the HC$_3$N* transitions and other identified species also indicated for completeness.
Those HC$_3$N* lines that are blended with a line from another species but with only one transition observed for this species, are not considered for the \texttt{SLIM} LTE modelling since we cannot properly estimate the degree of blending. In case it is partially blended with another species with more than one observed transition, we  also model these species and deblend its emission from that of the HC$_3$N* emission with \texttt{SLIM} assuming the emission from the other species can be properly fitted under LTE.
Also, for the central pixels within a region of $\approx0.45$\,pc, we ignored the HC$_3$N rotational lines from the lowest vibrational states $v=0$ and $v_7=1$ because they are affected by strong absorption of the continuum and self-absorption (see Figure~\ref{fig:NGC253_HR_mom0_SHC13}). Therefore, for the \texttt{SLIM} LTE modelling of the central region, we only considered transitions from vibrationally excited states with the highest energies, which are expected to be less affected by absorption. 

\begin{figure*}
\centering
    \includegraphics[width=0.65\linewidth]{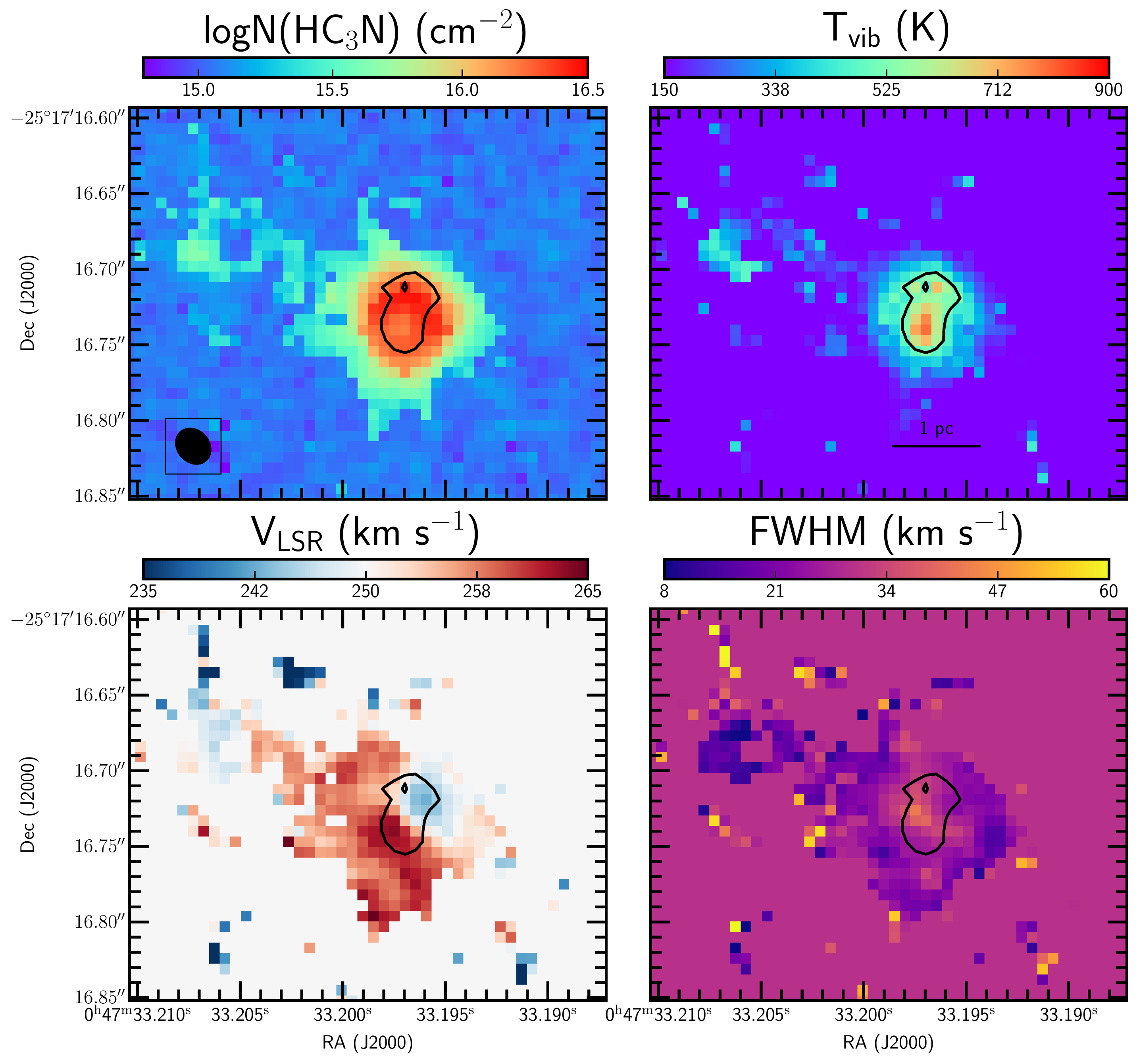}
  \caption[NGC\,253 proto-SSC\,$13$ LTE fitted values with SLIM]{Proto-SSC\,$13$ LTE fitted values with SLIM. Top left panel shows the HC$_3$N column density, top right panel the vibrational temperature, bottom left the $V_\text{LSR}$ and bottom right the FWHM. The black contour indicates the region with $T_\text{vib}>500$\,K. To estimate upper limits to the column density we have fixed $T_\text{vib}$ to $150$\,K, the $V_\text{LSR}$ to $250$\,km\,s$^{-1}$ and the FWHM to $30$\,km\,s$^{-1}$, and therefore these values appear as background pixels.
  }
  \label{fig:NGC253_HR_SLIM_SHC13}
\end{figure*}

Following the above procedure, we have developed \textsc{Madcuba} scripts to carry the HC$_3$N* LTE modelling of the HC$_3$N* emission with \texttt{SLIM} for all pixels of the proto-SSC\,$13$a emitting region. In order to obtain the vibrational temperature ($T_\text{vib}$) and the total HC$_3$N column density ($N$(HC$_3$N)), we have modelled the HC$_3$N* $J=24-23$ and $J=26-25$ rotational transitions from all (excluding the $v=0$ and $v_7=1$ transitions in the central pixels, see above) the vibrationally excited states and the ground state simultaneously with a single temperature (i.e. the vibrational temperature $T_\text{vib}$). Unfortunately, the small energy difference between the $J=24-23$ and the $26-25$ lines and the high $T_\text{vib}$ did not allow us to estimate the rotational temperature $T_\text{rot}$. However, as seen in \citet{RicoVillas2020, RicoVillas2021NGC1068, Costagliola2010}, we expect $T_\text{rot}$ to be lower than $T_\text{vib}$. For those pixels where HC$_3$N from the $v=0$ is detected but HC$_3$N* (from the $v_7=1$) is not, we estimated an upper limit for $T_\text{vib}$ (\texttt{SLIM} calculates these upper limits by using the $3\sigma$ upper limit to the $v_7=1$ column density and the fitted column density of the $v=0$ transition). The pixels where no HC$_3$N emission is detected (i.e. no HC$_3$N from the ground and vibrationally states) were masked with the estimated upper limits  to the HC$_3$N column density with \texttt{SLIM} \citep[based on the rms noise of the integrated line intensity; for details, see][]{Madcuba2019} assuming a $T_\text{vib}\sim150$\,K  (close to the median  upper limit $T_\text{vib}\sim161$\,K), and fixing the full width at half maximum (FWHM) to $30$\,km\,s$^{-1}$ (close to the values where HC$_3$N is detected) and  the velocity of the local standard of rest ($V_\text{LSR}$) to $250$\,km\,s$^{-1}$ (the mean $V_\text{LSR}$ for SSC\,13). We note that these N(HC$_3$N) upper limits are conservative and probably overestimate the column density since they are obtained assuming a high excitation temperature.
Figure~\ref{fig:NGC253_HR_SLIM_SHC13} shows the HC$_3$N column density, vibrational temperature, $V_\text{LSR}$ and FWHM map values derived with \texttt{SLIM}.

%\subsubsection[Results of the 2D analysis]{\Large{Results of the 2D analysis}}
%\label{subsec:3DLTEresults}

From Figure~\ref{fig:NGC253_HR_SLIM_SHC13} we can see that the region with highest HC$_3$N column densities ($\gtrsim 10^{16}$\,cm$^{-2}$) is the region where we also find the highest temperatures ($T_\text{vib}\gtrsim400$\,K). There is a decrease in the column density towards the center of proto-SSC\,$13$a coincident with the highest temperatures. This is a spurious effect probably caused by the high opacity towards the center. Also, while the temperature map appears symmetric, the column density map seems to be slightly elongated in the NE-SW direction along the galaxy disk, where we find a bridge of emission joining $13$a with $13$b, see Fig.~\ref{fig:NGC253_HR_219GHz_cont_zoom}.
Lower left and lower right panels of Figure~\ref{fig:NGC253_HR_SLIM_SHC13} show the velocity and the FWHM maps derived from HC$_3$N* \texttt{SLIM} modelling. 
As derived in \citet{RicoVillas2020}, proto-SSC\,$13$a has a systemic velocity of $250$\,km\,s$^{-1}$.
Figure~\ref{fig:NGC253_HR_SLIM_SHC13} clearly shows a sharp jump in the velocity along the SE-NW direction \citep[i.e. perpendicular to NGC\,253 rotation disk, with PA$=55^\circ$;][]{Krieger2019}, dividing the region with $T_\text{vib} \gtrsim 500$\,K. 
Perpendicular to this local velocity change, i.e. the region with velocities $\sim250$\,km\,s$^{-1}$ along NGC\,253 disk, we can see that the HC$_3$N* FWHMs are also larger ($\sim35$\,km\,s$^{-1}$). 
The possible nature and implications of the $\sim21$\,km\,s$^{-1}$ jump between the minimum ($242$\,km\,s$^{-1}$)  and maximum ($263$\,km\,s$^{-1}$) velocities is discussed in Section~\ref{subsec:vel_grad}.
% The separation of $~21$\,km\,s$^{-1}$ between the minimum ($242$\,km\,s$^{-1}$)  and maximum ($263$\,km\,s$^{-1}$) velocities in this velocity gradient could be attributed to rotational motions or to a cloud-cloud collision (which would have triggered the star formation inside proto-SSC\,$13$a). Also, infalling or outflowing gas cannot be completely ruled out as indicated by \citet{Levy2021}, since the large number of vibrationally excited emission lines could be concealing the P-Cygni profiles they observe in the CS $7-6$ and H$^{13}$CN $4-3$ transitions. However we will leave an in-depth analysis of the proto-SSCs analysis for a future work.

%Proto-SSC\,13 velocity-integrated line intensities in Jy\;km\,s$^{-1}$\;beam$^{-1}$ as a function of the distance in pc to each averaged ring of the HC$_3$N* $v=0$, $v_7=1$, $v_7=2$ and $v_6=1$ transitions used for the radiative transfer modelling. The table also lists the quantum numbers and the frequency of the transition in GHz. Higher vibrational transitions are listed in Table~\ref{tab:HC3N_HR_fluxes2}

\begin{table*}
    \begin{center}
    \caption[Proto-SSC\,13 HC$_3$N* intensities]{Velocity integrated line intensities of ring averaged spectra in proto-SSC\,$13$.}
    \label{tab:HC3N_HR_fluxes}
    \begin{threeparttable}
    \setlength{\tabcolsep}{1.9pt}
    \begin{tabular}{ccr@{\hspace{0.7\tabcolsep}}lcr@{\hspace{0.7\tabcolsep}}lcr@{\hspace{0.7\tabcolsep}}lcr@{\hspace{0.7\tabcolsep}}lcr@{\hspace{0.7\tabcolsep}}lcr@{\hspace{0.7\tabcolsep}}lcr@{\hspace{0.8\tabcolsep}}lcr@{\hspace{0.7\tabcolsep}}lcr@{\hspace{0.7\tabcolsep}}lcr@{\hspace{0.7\tabcolsep}}lcr@{\hspace{0.7\tabcolsep}}l}
    \hline \noalign {\smallskip}
Dist.  & &
\multicolumn{2}{c}{$v=0$} & &
\multicolumn{2}{c}{$v_{7}=1$} &   & \multicolumn{2}{c}{$v_{7}=1$} &   & 
\multicolumn{2}{c}{$v_{7}=2$} &   & \multicolumn{2}{c}{$v_{7}=2$} &   & 
\multicolumn{2}{c}{$v_{6}=1$} &   & \multicolumn{2}{c}{$v_{6}=1$} &   &
\multicolumn{2}{c}{$v_{5}=1$/$v_{7}=3$} &   & \multicolumn{2}{c}{$v_{6}=v_{7}=1$} &   & 
\multicolumn{2}{c}{$v_{4}=1$} &   & \multicolumn{2}{c}{$v_{6}=2$}   	  \\ 

\hline  \hline \noalign {\smallskip}
%  & 
%\multicolumn{2}{c}{\small{26$-$25}} & &
%\multicolumn{2}{c}{\small{24,1$-$23,-1}} &   & \multicolumn{2}{c}{\small{26,1$-$25,-1}} &   & 
%\multicolumn{2}{c}{\small{24,0-23,0}} &   & \multicolumn{2}{c}{\small{26,0$-$25,0}} &   & 
%\multicolumn{2}{c}{\small{24,-1-23,1}}  &   & \multicolumn{2}{c}{\small{26,-1$-$25,1}} \\ 
 & \small{$J_\text{up}$-$J_\text{lo}$} & 
\multicolumn{2}{c}{\small{26$-$25}} & &
\multicolumn{2}{c}{\small{24$-$23}} &   & \multicolumn{2}{c}{\small{26$-$25}} &   & 
\multicolumn{2}{c}{\small{24$-$23}} &   & \multicolumn{2}{c}{\small{26$-$25}} &   & 
\multicolumn{2}{c}{\small{24$-$23}} &   & \multicolumn{2}{c}{\small{26$-$25}} &   & 
\multicolumn{2}{c}{\small{26$-$25}} &   & \multicolumn{2}{c}{\small{26$-$25}} &   & 
\multicolumn{2}{c}{\small{26$-$25}} &   & \multicolumn{2}{c}{\small{24$-$23}} \\ 

 & \small{$l_\text{up}$; $l_\text{lo}$} & 
\multicolumn{2}{c}{} & &
\multicolumn{2}{c}{\small{1;\;-1}}&   & \multicolumn{2}{c}{\small{1;\;-1}} &   & 
\multicolumn{2}{c}{\small{0;\;0}} &   & \multicolumn{2}{c}{\small{0;\;0}} &   & 
\multicolumn{2}{c}{\small{-1;\;1}}&   & \multicolumn{2}{c}{\small{-1;\;1}} &    &
\multicolumn{2}{c}{\small{1,0;\;-1,0}}&   & \multicolumn{2}{c}{\small{2,2;\;-2,2}} &   & 
\multicolumn{2}{c}{} &   & \multicolumn{2}{c}{\small{0;\;0}} \\ 

 (pc) & \footnotesize{$\nu$\,(GHz)} & 
\multicolumn{2}{c}{\small{236.51}} & &
\multicolumn{2}{c}{\small{219.17}} &   & \multicolumn{2}{c}{\small{237.43}} &   & 
\multicolumn{2}{c}{\small{219.68}} &   & \multicolumn{2}{c}{\small{237.97}} &   & 
\multicolumn{2}{c}{\small{218.68}} &   & \multicolumn{2}{c}{\small{236.90}} &   &
\multicolumn{2}{c}{\small{236.66}} &   & \multicolumn{2}{c}{\small{237.81}} &   & 
\multicolumn{2}{c}{\small{236.18}} &   & \multicolumn{2}{c}{\small{219.02}} \\

\hline  \hline \noalign {\smallskip}

$0.05$	& &	$\leqslant$	&	$9.3$	&	&	$27.2\,\pm$	&	$3.2$	&	&	$37.8\,\pm$	&	$1.9$	&	&	$28.6\,\pm$	&	$3.5$	&	&	$30.4\,\pm$	&	$2.2$	&	&	$27.9\,\pm$	&	$3.1$	&	&	$30.7\,\pm$	&	$2.2$		 & &	$19.1\:\pm$	&	$2.1$   &	&	$19.1\:\pm$	&	$1.6$	&	&	$12.9\:\pm$	&	$1.7$	&	&	$\leqslant$	&	$9.4$	\\
$0.15$	& &	$21.1\,\pm$	&	$2.3$	&	&	$31.1\,\pm$	&	$2.2$	&	&	$42.2\,\pm$	&	$1.4$	&	&	$28.1\,\pm$	&	$2.5$	&	&	$30.5\,\pm$	&	$1.6$	&	&	$26.7\,\pm$	&	$2.2$	&	&	$29.9\,\pm$	&	$1.6$		 & &	$17.8\:\pm$	&	$1.5$   &	&	$16.0\:\pm$	&	$1.1$	&	&	$10.1\:\pm$	&	$1.2$	&	&	$\leqslant$	&	$6.8$	\\
$0.25$	& &	$41.8\,\pm$	&	$1.5$	&	&	$37.0\,\pm$	&	$1.7$	&	&	$45.1\,\pm$	&	$1.0$	&	&	$24.4\,\pm$	&	$1.7$	&	&	$28.4\,\pm$	&	$1.0$	&	&	$23.6\,\pm$	&	$1.5$	&	&	$25.4\,\pm$	&	$1.1$		 & &	$15.8\:\pm$	&	$1.1$	&	&	$12.9\:\pm$	&	$0.7$	&	&	$5.8\:\pm$	&	$0.7$	&	&	$\leqslant$	&	$4.9$	\\
$0.35$	& &	$52.6\,\pm$	&	$1.3$	&	&	$40.9\,\pm$	&	$1.6$	&	&	$42.0\,\pm$	&	$0.8$	&	&	$20.0\,\pm$	&	$1.4$	&	&	$23.6\,\pm$	&	$0.8$	&	&	$18.7\,\pm$	&	$1.3$	&	&	$17.0\,\pm$	&	$0.9$		 & &	$10.2\:\pm$	&	$0.8$	&	&	$8.8\:\pm$	&	$0.6$	&	&	$3.5\:\pm$	&	$0.6$	&	&	$\leqslant$	&	$4.1$	\\
$0.45$	& &	$50.5\,\pm$	&	$1.3$	&	&	$33.7\,\pm$	&	$1.2$	&	&	$34.0\,\pm$	&	$0.8$	&	&	$17.1\,\pm$	&	$1.2$	&	&	$17.6\,\pm$	&	$0.8$	&	&	$11.9\,\pm$	&	$1.0$	&	&	$10.3\,\pm$	&	$0.9$		 & &	$5.4\:\pm$	&	$0.7$	&	&	$4.4\:\pm$	&	$0.5$	&	&	$1.7\:\pm$	&	$0.4$	&	&	$\leqslant$	&	$2.9$	\\
$0.55$	& &	$41.2\,\pm$	&	$1.1$	&	&	$22.0\,\pm$	&	$1.2$	&	&	$24.4\,\pm$	&	$0.7$	&	&	$10.9\,\pm$	&	$1.3$	&	&	$11.5\,\pm$	&	$0.7$	&	&	$6.0\,\pm$	&	$1.1$	&	&	$4.9\,\pm$	&	$0.8$		 & &	$\leqslant$	&	$1.9$	&	&	$\leqslant$	&	$1.4$	&	&	$1.0\:\pm$	&	$0.3$	&	&	$\leqslant$	&	$3.0$	\\
$0.65$	& &	$30.4\,\pm$	&	$1.0$	&	&	$14.4\,\pm$	&	$1.1$	&	&	$14.6\,\pm$	&	$0.6$	&	&	$6.4\,\pm$	&	$1.2$	&	&	$6.6\,\pm$	&	$0.7$	&	&	$\leqslant$	&	$2.9$	&	&	$3.9\,\pm$	&	$0.7$		 & &	$\leqslant$	&	$1.7$	&	&	$\leqslant$	&	$1.0$	&	&	$\leqslant$	&	$0.9$	&	&	$\leqslant$	&	$2.6$	\\
$0.75$	& &	$21.8\,\pm$	&	$0.8$	&	&	$7.0\,\pm$	&	$1.0$	&	&	$6.6\,\pm$	&	$0.5$	&	&	$\leqslant$	&	$3.3$	&	&	$3.9\,\pm$	&	$0.5$	&	&	$\leqslant$	&	$2.9$	&	&	$2.2\,\pm$	&	$0.6$		 & &	$\leqslant$	&	$1.4$	&	&	$\leqslant$	&	$0.9$	&	&	$\leqslant$	&	$0.8$	&	&	$\leqslant$	&	$2.4$	\\
$0.85$	& &	$18.3\,\pm$	&	$0.8$	&	&	$3.9\,\pm$	&	$0.9$	&	&	$2.4\,\pm$	&	$0.5$	&	&	$\leqslant$	&	$3.3$	&	&	$2.9\,\pm$	&	$0.6$	&	&	$\leqslant$	&	$3.0$	&	&	$\leqslant$	&	$1.9$		 & &	$\leqslant$	&	$1.5$	&	&	$\leqslant$	&	$0.9$	&	&	$\leqslant$	&	$0.8$	&	&	$\leqslant$	&	$2.7$	\\
$0.95$	& &	$14.7\,\pm$	&	$0.7$	&	&	$2.7\,\pm$	&	$0.9$	&	&	$1.7\,\pm$	&	$0.5$	&	&	$\leqslant$	&	$3.0$	&	&	$2.0\,\pm$	&	$0.5$	&	&	$\leqslant$	&	$2.7$	&	&	$\leqslant$	&	$1.7$		 & &	$\leqslant$	&	$1.4$	&	&	$\leqslant$	&	$0.8$	&	&	$\leqslant$	&	$0.8$	&	&	$\leqslant$	&	$2.8$	\\
$1.05$	& &	$10.6\,\pm$	&	$0.7$	&	&	$\leqslant$	&	$2.4$	&	&	$1.9\,\pm$	&	$0.5$	&	&		&		&	&	$\leqslant$	&	$1.6$	&	&		&		&	&	$\leqslant$	&	$1.3$	& &		&		&	&		&		&	&		&		&	&		&		\\
$1.15$	& &	$7.9\,\pm$	&	$0.7$	&	&	$\leqslant$	&	$2.0$	&	&	$1.6\,\pm$	&	$0.5$	&	&		&		&	&	$\leqslant$	&	$1.6$	&	&		&		&	&	$\leqslant$	&	$1.3$	& &		&		&	&		&		&	&		&		&	&		&		\\
$1.25$	& &	$7.3\,\pm$	&	$0.7$	&	&	$\leqslant$	&	$2.0$	&	&	$\leqslant$	&	$1.4$	&	&		&		&	&	$\leqslant$	&	$1.5$	&	&		&		&	&	$\leqslant$	&	$1.2$	& &		&		&	&		&		&	&		&		&	&		&		\\
$1.35$	& &	$7.5\,\pm$	&	$0.6$	&	&	$\leqslant$	&	$1.5$	&	&	$\leqslant$	&	$1.1$	&	&		&		&	&	$\leqslant$	&	$1.5$	&	&		&		&	&	$\leqslant$	&	$1.0$	& &		&		&	&		&		&	&		&		&	&		&		\\
$1.45$	& &	$5.8\,\pm$	&	$0.6$	&	&	$\leqslant$	&	$1.5$	&	&	$\leqslant$	&	$1.1$	&	&		&		&	&	$\leqslant$	&	$1.2$	&	&		&		&	&	$\leqslant$	&	$1.0$	& &		&		&	&		&		&	&		&		&	&		&		\\

        \hline \noalign {\smallskip}
    \end{tabular}
    \begin{tablenotes}
    \item \textbf{Note.} The table lists the distance in pc to each averaged ring (i.e. $(R_\text{int}+R_\text{out})/2$) with the line intensities in Jy\;km\,s$^{-1}$\;beam$^{-1}$. The table also lists the quantum numbers and the frequency of the transition in GHz. \\
    The $v_{5}=1$/$v_{7}=3$ and $v_{6}=v_{7}=1$ quantum numbers refer to $l_\text{up},\,k_\text{up}$; $l_\text{lo},\,k_\text{lo}$. \\
    The $v_6=2$\:$24,0-23,0$ transition is not detected in any ring, but its upper limits are included as a consistency test for the non-local radiative transfer models.
    \end{tablenotes}
    \end{threeparttable}
    \end{center}
    
\end{table*}

\subsection{Radial distribution of \texorpdfstring{proto-SSC\,13a}{proto-SSC13a}: LTE-\texttt{SLIM} analysis}
\label{subsec:rings}
\label{ring_subsec_SLIM_LTE}

%To make a robust LTE analysis, we have  averaged the spectra within rings of $0.1$\,pc thickness with radius rising in $0.1$\,pc up to a radius of $1.5$\,pc, increasing the signal-to-noise ratio specially of the higher energy lines. This can be done because proto-SSC\,$13$a appears to be rather symmetric and the rings are  centered on the pixel with the brightest continuum emission at $219$\,GHz (RA(J2000)$=00^\text{h}47^\text{m}33^\text{s}.1970$; Dec(J2000)$=-25^\circ17^\prime16\arcsec.733$). 
To increase the signal-to-noise ratio of the spectra, specially for the higher energy lines, and thus improve our analysis, we have  averaged the spectra within rings of $0.1$\,pc thickness with radius increasing in $0.1$\,pc up to $1.5$\,pc, as the noise decreases with the considered number of pixels for each ring like $ 1/\sqrt{N_\text{px}}$.
This is appropriate for proto-SSC\,$13$a since it appears to be rather symmetric (e.g. see Figure~\ref{fig:NGC253_HR_SLIM_SHC13}), with the rings   centered on the pixel with the brightest continuum emission at $219$\,GHz (RA(J2000)$=00^\text{h}47^\text{m}33^\text{s}.1970$; Dec(J2000)$=-25^\circ17^\prime16\arcsec.733$). 
Table~\ref{tab:HC3N_HR_fluxes} lists the integrated line intensities of the HC$_3$N* transitions within the rings. 

%\subsubsection{\texttt{SLIM} analysis}

With \texttt{SLIM}, we have modelled the emission of these higher signal-to-noise ratio averaged ring spectra following the same LTE procedure described above.  Figures in the Appendix~\ref{ap:HR_spec} (available in the online version) show the HC$_3$N* LTE modelled emission for the rings. We have observed that at short distances ($r<0.4$\,pc), HC$_3$N* transitions from the $v=0$ and $v_7=1$ states show strong absorption and therefore, as described above, were not taken into account for the fit. Although for these  self-absorbed, contaminated or at noise level line profiles are simulated with \texttt{SLIM} for the derived
parameters, they are not taken into account for the fitting. Therefore, some undetected lines from high vibrational states show modelled emission within noise level.
Obviously, the  \texttt{SLIM} models cannot reproduce all HC$_3$N* lines simultaneously due to non-local effects that are not included. In addition, since line absorption is more important for low-energy lines than for high-energy lines, \texttt{SLIM} considers their ratio as very high  $T_\text{vib}$. We can thus consider the \texttt{SLIM} $T_\text{vib}$ \citep[i.e. $T_\text{dust}$, see][]{RicoVillas2020} in the innermost regions as upper limits.

\begin{table*}
\begin{center}
\caption{Proto-SSC$\,13$a HC$_3$N* vibrational temperature and column density comparison between the average of fitted pixels and the fit to the average spectrum for each ring.}
\label{tab:NGC253_HR_SLIM_SHC13_TexLog_profiles}
\begin{threeparttable}
\begin{tabular}{cccclclc}
%\hline \noalign {\smallskip}
\cline{4-8} {\smallskip}
		&	& & \multicolumn{2}{c}{Pixels mean} & & \multicolumn{2}{c}{Averaged spectra}	 \\	
\cline{1-2} \cline{4-5} \cline{7-8}
{\smallskip}
	Dist.	&	\# Pixels	& &	\multicolumn{1}{c}{$T_\text{vib}$}	&	$\log{N}$(HC$_3$N)	& &	\multicolumn{1}{c}{$T_\text{vib}$}	&	$\log{N}$(HC$_3$N) \\		
	    	(pc) 	        &		        & &	\multicolumn{1}{c}{(K)}				    &	(cm$^{-2}$)		&	&	\multicolumn{1}{c}{(K)}		    &	(cm$^{-2}$)	       \\
\hline \noalign {\smallskip}
 $	0.05	$ & $	1	$& & $ \leqslant751	$ & $	16.21	\pm	0.02	$ & & $	\leqslant 751	$ & $	16.21	\pm	0.03	$	\\ 
 $	0.15	$ & $	8	$& & $	656	\pm	55	$ & $	16.28	\pm	0.04	$ & & $	\leqslant 611	$ & $	16.37	\pm	0.02	$	\\ 
 $	0.25	$ & $	12	$& & $	517	\pm	66	$ & $	16.33	\pm	0.06	$ & & $	\leqslant 555	$ & $	16.40	\pm	0.02	$	\\ 
 $	0.35	$ & $	16	$& & $	432	\pm	79	$ & $	16.30	\pm	0.11	$ & & $	505	\pm	26	    $ & $	16.39	\pm	0.02	$	\\ 
 $	0.45	$ & $	20	$& & $	404	\pm	111	$ & $	16.16	\pm	0.18	$ & & $	420	\pm	37	    $ & $	16.21	\pm	0.02	$	\\ 
 $	0.55	$ & $	24	$& & $	381	\pm	107	$ & $	15.93	\pm	0.20	$ & & $	384	\pm	35	    $ & $	16.04	\pm	0.03	$	\\ 
 $	0.65	$ & $	28	$& & $	306	\pm	96	$ & $	15.80	\pm	0.19	$ & & $	357	\pm	34	    $ & $	15.92	\pm	0.02	$	\\ 
 $	0.75	$ & $	36	$& & $	234	\pm	84	$ & $	15.66	\pm	0.15	$ & & $	280	\pm	30	    $ & $	15.71	\pm	0.02	$	\\ 
 $	0.85	$ & $	32	$& & $	202	\pm	52	$ & $	15.53	\pm	0.12	$ & & $	228	\pm	21	    $ & $	15.58	\pm	0.02	$	\\ 
 $	0.95	$ & $	44	$& & $	176	\pm	54	$ & $	15.49	\pm	0.13	$ & & $	214 \pm 22	    $ & $	15.46	\pm	0.02	$	\\ 
 $	1.05	$ & $	56	$& & $	179	\pm	47	$ & $	15.40	\pm	0.09	$ & & $	\leqslant	186	$ & $ \leqslant	15.23	$	\\  %	\pm	0.03	$	\\ 
 $	1.15	$ & $	48	$& & $	149	\pm	43	$ & $	15.38	\pm	0.08	$ & & $	\leqslant	186	$ & $ \leqslant	15.13 	$	\\ %	\pm	0.04	$	\\ 
 $	1.25	$ & $	48	$& & $	149	\pm	50	$ & $	15.35	\pm	0.12	$ & & $	\leqslant	186	$ & $ \leqslant 15.11 	$	\\ %	\pm	0.04	$	\\ 
 $	1.35	$ & $	64	$& & $	183	\pm	27	$ & $	15.28	\pm	0.12	$ & & $	\leqslant	186	$ & $ \leqslant	15.07 	$	\\ %	\pm	0.04	$	\\ 
 $	1.45	$ & $	60	$& & $	167	\pm	65	$ & $	15.30	\pm	0.15	$ & & $	\leqslant	186	$ & $ \leqslant	15.05 	$	\\  %	\pm	0.04	$	\\ 
\hline \noalign {\smallskip}
\end{tabular}
\begin{tablenotes}
      \small
      \item \textbf{Note.} (1) Ring distance. (2) Number of pixels in each ring. (3) Weighted mean of the LTE vibrational temperature fitted for each pixel within the ring. The errors, except for ring 1 with one single pixel, are the errors of the mean. (4) Weighted mean of the LTE column density fitted for each pixel within the ring. The errors, except for the first ring with one single pixel, are the errors of the mean. (5) Fitted LTE vibrational temperature to the ring averaged spectra (i.e. the spectra obtained from averaging all the spectra from the pixels within each ring). (6) Fitted LTE column density to the ring averaged spectra.
    \end{tablenotes}
\end{threeparttable}
\end{center}
\end{table*}

\begin{figure}
\centering
    \includegraphics[width=0.75\linewidth]{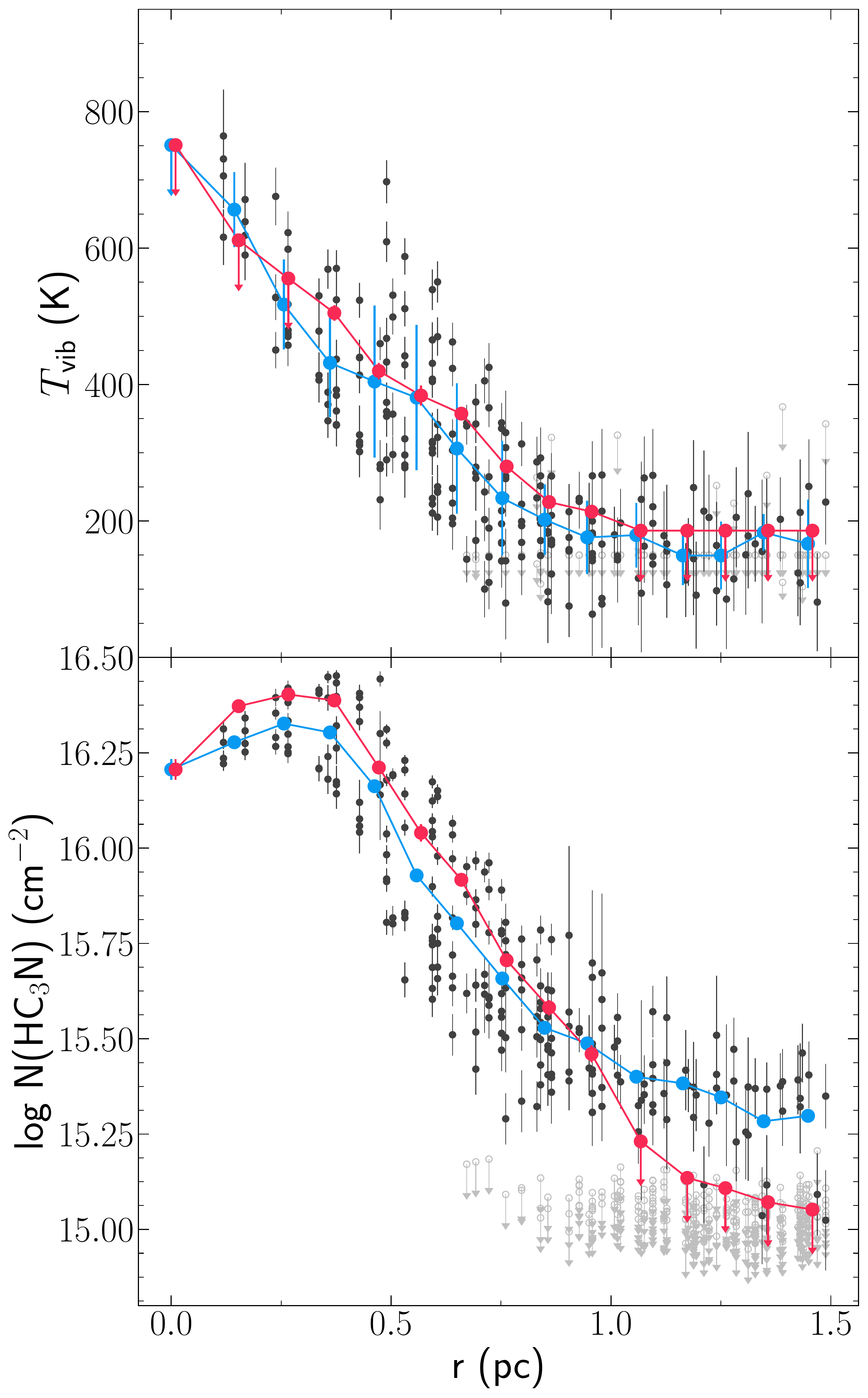}
  \caption[NGC\,253 proto-SSC HC$_3$N* LTE vibrational temperature and column density profiles]{Proto-SSC\,$13$a HC$_3$N* vibrational temperature ($T_\text{vib}$) and column density (log\,$N$(HC$_3$N)) profiles derived from the LTE model fitting with SLIM from Table~\ref{tab:NGC253_HR_SLIM_SHC13_TexLog_profiles}. Solid black dots represent the values fitted for each individual pixel, while grey markers with arrows represent upper limits for each individual pixel. Blue markers show the weighted mean of the fitted values for the pixels (not considering the upper limits) within each ring of $0.1$\,pc thickness. Red markers show the fitted values for the averaged spectra of each ring.
  }
  \label{fig:NGC253_HR_SLIM_SHC13_TexLog_profiles}
\end{figure}

Table~\ref{tab:NGC253_HR_SLIM_SHC13_TexLog_profiles} lists the distance to each ring, the number of pixels contained in the ring, and the weighted mean values (for the pixels within each ring that are not upper limits) of $T_\text{vib}$ and $N(\text{HC}_3\text{N})$  from the fit to the spectra of individual pixels in the image (for the pixels within each ring that are not upper limits), together with the corresponding values inferred from the averaged ring spectra. These values are plotted as a function of their distance to  proto-SSC\,$13$a center in Figure~\ref{fig:NGC253_HR_SLIM_SHC13_TexLog_profiles}. There is a good match between both profiles, but there is a clear improvement in the error of the estimated parameters in the averaged spectra, specially in the lower signal to noise regions.

Figure~\ref{fig:NGC253_HR_SLIM_SHC13_TexLog_profiles} show that $T_\text{vib}$ peaks at the central pixel ($T_\text{vib} \lesssim 751$\,K) and decreases with distance to $\sim200$\,K at $0.85$\,pc. At longer distances ($\gtrsim0.9$\,pc), most pixels have $T_\text{vib}$ upper limits (i.e. no HC$_3$N $v_7=1$ line was detected) with $T_\text{vib}\lesssim 186$\,K.
For the HC$_3$N* column density, its maximum value is obtained at $\sim0.25$\,pc. The lower column density values at smaller distances are probably an artifact produced by the strong absorption towards the central region, which still affects even the strength of the $v_6=1$ and $v_7=2$ lines. We note that the fitted averaged spectra values start to differ from the pixel mean values at distances $\gtrsim1.0$\,pc as a result of averaging all pixels (with and without HC$_3$N* column density upper limits in each individual pixel). 

\subsection{Radial distribution of \texorpdfstring{proto-SSC\,13a}{proto-SSC13a}: Non-local analysis}
\label{ring_subsec_nlocal_mod}

 As seen in \citet{RicoVillas2020}, LTE can be assumed for the excitation of the HC$_3$N* emission when there is a strong IR radiation field and a high dust optical depth in the mid-IR. This is because HC$_3$N molecules will be immersed in a local radiation field described by a blackbody at the dust temperature (i.e. $T_\text{vib}\sim T_\text{dust}$) and their ro-vibrational transitions (excited by $\sim 11-45$\,$\upmu$m radiation) will be thermalized independently from their Einstein-$A_{ul}$ coefficients and H$_2$ density.
 However, the large column densities that in principle validate LTE excitation, make non-local radiative transfer effects very important: the large amounts of gas and dust increase the probabilities for a photon emitted in some region at that local $T_\text{dust}$ to be absorbed in another region at $T^\prime_\text{dust}$, even more than once, before reaching the observer. 
 In addition, in the case of density and temperature gradients within the SHCs, the high line opacities will make every line to trace different excitation temperatures.
 We have then carried non-local radiative transfer models of the HC$_3$N* emission assuming LTE excitation for all states. We have assumed LTE excitation since, as already mentioned, all HC$_3$N* lines are expected to be thermalized at T$_\text{dust}$. Also, the $A_{ul}$ coefficients from many of the ro-vibrational bands are not available, preventing to carry a full non-LTE excitation modelling for all the observed lines.

The radiative transfer models are based on the radiative transfer code developed by \citet{GA97, GA99} and \citet{GA19}, with the inclusion of the $v_5=1/v_7=3$, $v_6=v_7=1$, $v_4=1$ and $v_6=2$ vibrationally excited states up to high $J$ values ($J=50$) and assuming thermalisation (i.e. local LTE excitation).
These models aim at reproducing the spatial profile of the emission from the detected, unblended and uncontaminated HC$_3$N* transitions listed in Table~\ref{tab:HC3N_HR_fluxes}. These transitions cover a wide range of energies, from $142$\,K to $1579$\,K.  The $v_6=2$\:$(24,0-23,0)$ transition is not detected and is included in the models as a control to the HC$_3$N column density and vibrational temperature. Emission lines are self-consistently fitted together with the radial profiles of the continuum emission at $235$\,GHz and $345$\,GHz obtained from the ring spectra (top left panel of Figure~\ref{fig:NGC253_HR_SLIM_SHC13_nLTE_conts}).

Regarding the thermal structure of the SHC, we have used different density and $T_\text{dust}$ radial profiles derived following \citet{GA19}, which take into account the greenhouse effect expected for the very large column densities ($N_{\text{H}_2}>10^{24}$) derived in \citet{RicoVillas2020}. The $T_\text{dust}$ profiles are obtained in a self-consistent way considering the heating and cooling of dust grains in a spherical cloud with a given density profile. Input model parameters are the luminosity surface density $\Sigma_\text{IR}=L_\text{IR}/(\pi R^2)$, the H$_2$ column density $N(\text{H}_2)$, the density profile (parameterized with $q$ as $n_{\text{H}_2}\propto r^{-q}$) and the absorption coefficient of dust as a function of wavelength $\kappa_\text{abs}(\lambda)$.
The derived temperature profiles also depend on the nature of the heating source (point source and centrally-peaked). For the case of SSCs, we use the centrally peaked starburst profiles from \citet{GA19}, which consider the deposition of energy due to star formation to be distributed across the cloud and proportional to the shell density and mass, in contrast with a single point source heating, expected for the AGN models \citep[][]{GA19}. The models are then described by the source radius $r_\text{out}$, a total $L_\text{IR}$, a total gas H$_2$ column density $N_{\text{H}_2}$ and the density profile power-law index $q$. These parameters define consistently the $T_\text{dust}$ profile for the type of heating (star formation or AGN) used in our radiative transfer predictions. 
Hence, the only free-parameters are the HC$_3$N abundance relative to H$_2$ ($X_{\text{HC}_3\text{N}}$), which we vary for each shell to fit the observed spatial profiles, and $X_\text{d}$ which reflects the amount of dust relative to gas (by mass).

We find that the ratio between the $235$\,GHz and $345$\,GHz continuum emission increases for  $r<0.85$\,pc (once corrected for the different beams), indicating the presence of free-free emission. 
To account for this free-free emission observed at the center we  use the $X_\text{d}$ parameter, which is boosted in the central region to mimic such free-free emission. For  $r<0.85$\,pc, the extra continuum added accounts for a total of $1.89$\,mJy at $235$\,GHz, very similar to the estimated non-dust contribution ($S_{219_\text{n-dust}}=1.99$\,mJy) obtained by assuming a spectral index expected for optically thin dust emission of $\alpha_{345-219}\sim3.5$. 
In the outer region (i.e. $r\gtrsim0.85$\,pc), $X_\text{d}$ is also adjusted to fit the observed $235$\,GHz emission, which translates into a higher column and mass of gas and dust relative to the power-law density profile.
This is because no single value of $q$ can describe the whole spatial profile of the observed continuum emission.
Accordingly, we consider the excess of dust added up to $0.85$\,pc as free-free emission and from $0.85$\,pc to $1.5$\,pc as an increase in mass that deviates from the $n_{\text{H}_2}\propto r^{-q}$ assumed profile.

%Regardless of the $q$ value used for the modelling of the source, we had to modify $X_\text{d}$ in both the innermost and outermost regions to reproduce the continuum emission at all radii.
%Since the modelled continuum emission does not include the contribution from free-free emission, $X_\text{d}$ is boosted in the central region to mimic such free-free emission. In the outer region (i.e. $r\gtrsim0.85$\,pc), $X_\text{d}$ is also adjusted to fit the observed $235$\,GHz emission, because no single value of $q$ can describe the whole spatial profile of the observed continuum emission. While we find that $q=1.5$ is the best fit (apart from the free-free emission) for $r<0.85$\,pc, the increased $X_\text{d}$ in the outer region translates into a higher column and mass of gas and dust relative to the power-law density profile.
%Taking into account that the ratio between the $235$\,GHz and $345$\,GHz continuum emission increases for  $r<0.85$\,pc (once corrected for the different beams), we can consider this as a clear evidence of free-free emission. 
%For  $r<0.85$\,pc, the extra dust added accounts for a total of $1.89$\,mJy at $235$\,GHz, very similar to the estimated non-dust contribution ($S_{219_\text{n-dust}}=1.99$\,mJy) obtained by assuming a spectral index expected for optically thin dust emission of $\alpha_{345-219}\sim3.5$. 
%Accordingly, we consider the excess of dust added up to $0.85$\,pc as free-free emission and from $0.85$\,pc to $1.5$\,pc as an increase in mass that deviates from the $n_{\text{H}_2}\propto r^{-q}$ assumed profile.

\begin{table*}
\begin{center}
\caption[Proto-SSC\,$13$a models parameters]{Parameters of the non-local radiative transfer models for the case of heating due to distributed star formation.}
\label{tab:nlocal_modpars}
\begin{threeparttable}
\begin{tabular}{cccccccccrc}
\hline \noalign {\smallskip}
 Model & $q$ & $L_\text{IR}$ & $\Sigma_\text{IR}$ & $M_{\text{H}_2}$ & $L_\text{IR}/M_{\text{H}_2}$ & $r_e$ & $N_{\text{H}_2}$ & $N_{\text{HC}_3\text{N}}$ & $\chi^2_\text{lines}$ & Type\\
       &(pc) &   (L$_\odot$) &(L$_\odot$\,pc$^{-2}$)& (M$_\odot$)    & (L$_\odot/$M$_\odot$)        & (pc)  &  (cm$^{-2}$)     & (cm$^{-2}$)   & & \\
\hline \noalign {\smallskip}
1  & $ 1.0 $ & $  9.4\times10^7 $ & $  1.3\times10^7 $ & $ 9.4\times10^5 $ & $ 100 $ & $ 0.8 $ & $ 1.0\times10^{25} $ & $ 1.1\times10^{18} $ & $5.1$ & Dist.\\
2  & $ 1.5 $ & $  9.4\times10^7 $ & $  1.3\times10^7 $ & $ 9.5\times10^5 $ & $ 99 $  & $ 0.4 $ & $ 1.2\times10^{25} $ & $ 3.1\times10^{17} $ & $4.5$ & Dist.\\
3  & $ 1.0 $ & $  3.9\times10^8 $ & $  5.5\times10^7 $ & $ 6.0\times10^5 $ & $ 656 $ & $ 0.8 $ & $ 9.2\times10^{24} $ & $ 1.3\times10^{17} $ & $16.6$ & Dist.\\
4  & $ 1.5 $ & $  3.9\times10^8 $ & $  5.5\times10^7 $ & $ 6.1\times10^5 $ & $ 638 $ & $ 0.4 $ & $ 1.0\times10^{25} $ & $ 9.7\times10^{16} $ & $11.0$ & Dist.\\
5  & $ 1.5 $ & $  9.8\times10^7 $ & $  1.4\times10^7 $ & $ 7.8\times10^5 $ & $ 126 $ & $ - $ & $ 1.1\times10^{25} $ & $ 1.3\times10^{17} $ & $7.2$ & Central\\
6  & $ 1.0 $ & $  9.8\times10^7 $ & $  1.4\times10^7 $ & $ 9.0\times10^5 $ & $ 109 $ & $ - $ & $ 1.0\times10^{25} $ & $ 1.3\times10^{17} $ & $10.9$ & Central\\
7  & $ 1.0 $ & $  9.8\times10^7 $ & $  1.4\times10^7 $ & $ 8.4\times10^5 $ & $ 117 $ & $ - $ & $ 6.9\times10^{24} $ & $ 1.9\times10^{17} $ & $6.0$ & Central\\
\hline \noalign {\smallskip}
\end{tabular}
\begin{tablenotes}
      \small
      \item \textbf{Notes:} All models have $r=1.5$\,pc. $q$ is the power-law index assumed for the density profile. $L_\text{IR}$ and $\Sigma_\text{IR}$ are the luminosity and the luminosity surface density. $M_{\text{H}_2}$ is the model total mass and $L_\text{IR}/M_{\text{H}_2}$ the  luminosity per unit of mass. $r_e$ is the radius containing half of the star formation for distributed star formation models, which is $\propto n_{\text{H}_2}M_{\text{H}_2}$ for each shell \citep[see][]{GA19}. For central star formation models, all the star formation is contained inside $r<0.05$\,pc.
      $N_{\text{H}_2}$ and $N_{\text{HC}_3\text{N}}$ are to the total H$_2$ and HC$_3$N column densities. 
      %$\tau_{100}$ is the radial optical depth at $100$\,$\upmu$m. 
      $\chi^2_\text{lines}$ is the average of the  $\chi^2_\text{line}=\sum\limits_{i} \frac{(o_i-m_i)^2}{n \delta_{o_i}^2}$ for each line, where $o_i$ and $m_i$ are the observed and modelled line intensities, $n$ the number of points taken into account, and $\delta_{o_i}$ is the observed error. Type refers to distributed or central star formation models.
\end{tablenotes}
\end{threeparttable}
\end{center}
\end{table*}

%Then, the predicted continuum and HC$_3$N emissions are  convolved with the beam of the observations ($0.022\arcsec \times 0.020\arcsec$ and $0.028\arcsec \times 0.034\arcsec$) and their intensities are averaged for the rings of $0.1$\,pc thickness previously defined.

From the continuum emission\footnote{Since the SSCs are being internally heated, the large column densities along with the greenhouse effect make the continuum emission to be much more intense at their center. Hence, the FWHMs do not reveal the total extension of the continuum emission.} at $219$\,GHz ($0.5$\,pc$\times0.5$\,pc) and at $345$\,GHz ($0.8$\,pc$\times0.8$\,pc) and the extent of the HC$_3$N $v=0$ emission (see Figures~\ref{fig:NGC253_HR_mom0_SHC13} and \ref{fig:NGC253_HR_SLIM_SHC13}), we modelled proto-SSC\,$13$a up to a radius of $1.5$\,pc considering several density and temperature profiles.

\subsubsection{Non-local modelling results}

Figures~\ref{fig:NGC253_HR_SLIM_SHC13_nLTE_conts} and ~\ref{fig:NGC253_HR_SLIM_SHC13_nLTE_lines} show the results for four representative models with star formation heating that best reproduce the continuum and the HC$_3$N* line emission.
%(individual figures for every model can be found in Appendix~\ref{ap:radtransf_HR}). 
The models have an H$_2$ column density of $\sim10^{25}$\,cm$^{-2}$ with two different density profiles ($q=1$ and $q=1.5$, modified as indicated above), and cover two different IR luminosities ($L_\text{IR}$) of $9.2\times10^7$\,L$_\odot$ and $3.9\times10^8$\,L$_\odot$.
%in order to reproduce and not overestimate the observed $235$\,GHz continuum emission, providing an upper limit to the luminosity of the source.
These luminosities are close to the protostar luminosity of $10^8$\,L$_\odot$ derived in \citet{RicoVillas2020} for proto-SSC\,$13$. A summary of the parameters of the most representative models, labelled $1$ to $4$, is shown in Table~\ref{tab:nlocal_modpars}. Since we have corrected the model masses to fit the $235$\,GHz continuum emission in the outer regions (i.e. $0.85-1.5$\,pc), models with the same luminosity but different $q$ index have similar final gas masses. 

Figure~\ref{fig:NGC253_HR_SLIM_SHC13_nLTE_conts} shows, in the upper left panel, the observed $235$\,GHz continuum emission at a $0.022\arcsec \times 0.020\arcsec$ resolution (filled black circles) and the $345$\,GHz continuum emission at a $0.028\arcsec \times 0.034\arcsec$ resolution (filled grey circles). 
The dashed lines represent the predicted continuum emission profiles of Model $2$ ($q=1.5$) at $235$\,GHz and $345$\,GHz for the corresponding $T_\text{dust}$ profiles from \citet{GA19}. The solid line show the continuum emission at $235$\,GHz derived from our modelling after varying $X_\text{d}$. Figure~\ref{fig:NGC253_HR_SLIM_SHC13_nLTE_conts} also compares the modelled $T_\text{dust}$ spatial profile considering the greenhouse effect for the different density profiles and luminosities. It also shows the HC$_3$N column density profiles (solid lines) with the values derived from \texttt{SLIM} in Section~\ref{ring_subsec_SLIM_LTE} (filled black circles) and the HC$_3$N abundance relative to H$_2$ ($X(\text{HC}_3\text{N})$).
Despite having similar H$_2$ column densities and/or  $L_\text{IR}$, the difference in the density power-law index $q$ ($q=1.0$ or $q=1.5$ for $n_{\text{H}_2}\propto r^{-q}$), translates into different H$_2$ densities and $T_\text{dust}$ within a given shell, resulting in the different $N_{\text{HC}_3\text{N}}$ and $X_{\text{HC}_3\text{N}}$ profiles plotted in Figure~\ref{fig:NGC253_HR_SLIM_SHC13_nLTE_conts}. 
All models reproduce well the observed $235$\,GHz continuum emission, which, as expected, clearly shows that the models constrained only by dust emission are degenerated. Further constraints to the models are provided by considering the emission from the multiple detected HC$_3$N* lines as a function of the radius.

\subsubsection{Breaking the model degeneracy using HC$_3$N* emission}

\begin{figure*}
\centering
    \includegraphics[width=0.75\linewidth]{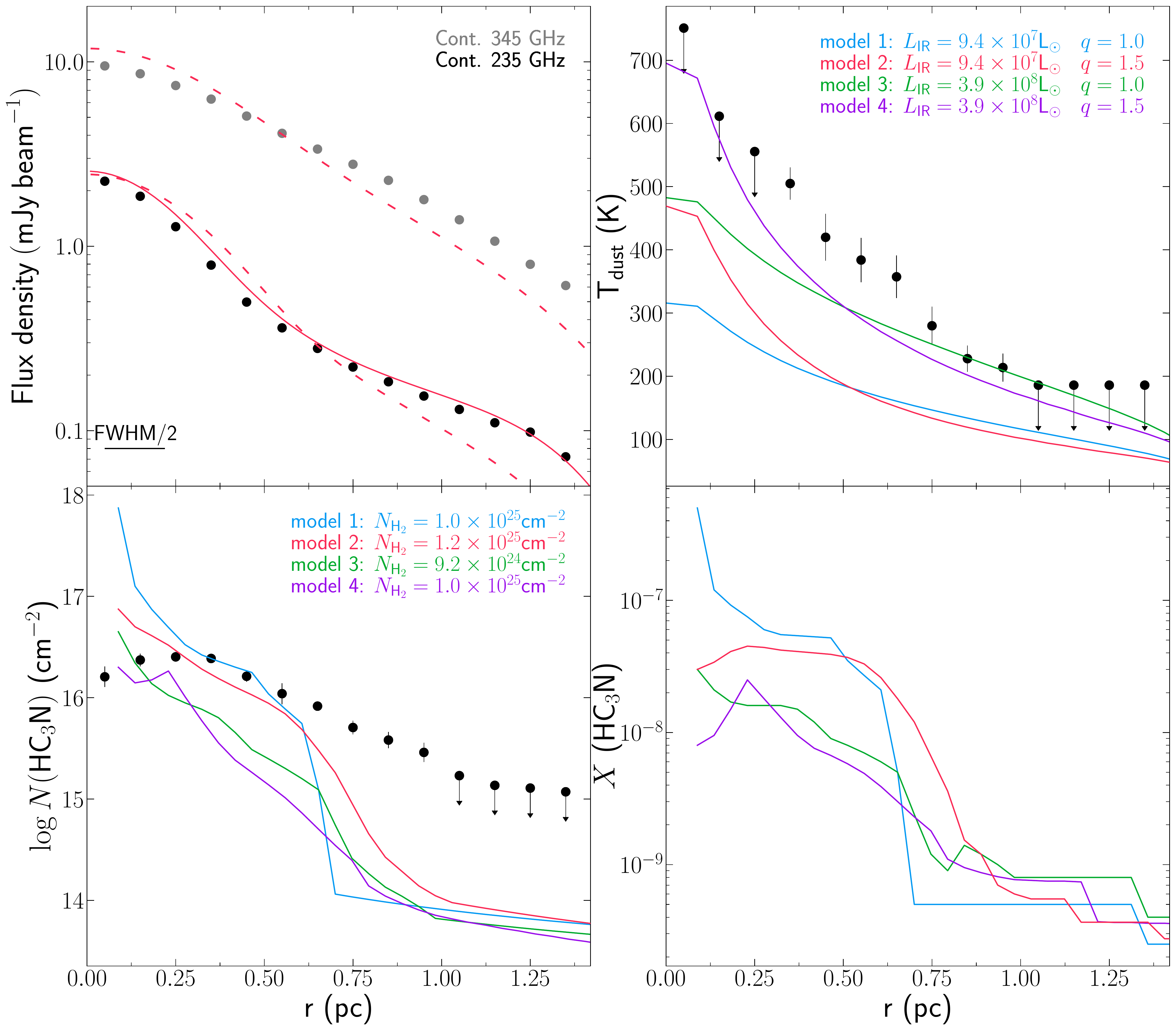}
  \caption[Proto-SSC\,$13$a observed and modelled continuum spatial emission]{Observed and modelled continuum emission and  parameters used each model (Model 1 in blue, Model 2 in red, Model 3 in green and Model 4 in purple) profiles. Upper left panel shows the rings continuum emission at $235$\,GHz with a resolution of $0.022\arcsec \times 0.020\arcsec$ (black filled circles) and at $345$\,GHz at a resolution of $0.028\arcsec \times 0.034\arcsec$ (grey filled circles). Dashed red lines show the continuum emission from Model 2 ($q=1.5$) resulting from the $T_\text{dust}$ profile for both frequencies, while the red solid line shows the continuum emission at $235$\,GHz resulting from the modelling after varying $X_\text{d}$ to account for both the free-free emission in the innermost region and the excess of continuum emission in the outermost region.
  Upper right and lower left panels show the dust temperature ($\sim T_\text{vib}$) and column density profiles obtained from the \texttt{SLIM} LTE model (black filled circles) for the averaged ring spectra and the solid colored lines indicates the profiles of each non-local model. Lower right panel solid colored lines show the HC$_3$N abundance profiles for each model. Half of the beam FWHM (i.e. $0.011\arcsec$)  is shown on the upper right panel.
  }
  \label{fig:NGC253_HR_SLIM_SHC13_nLTE_conts}
\end{figure*}

\begin{figure*}
\centering
    \includegraphics[width=\linewidth]{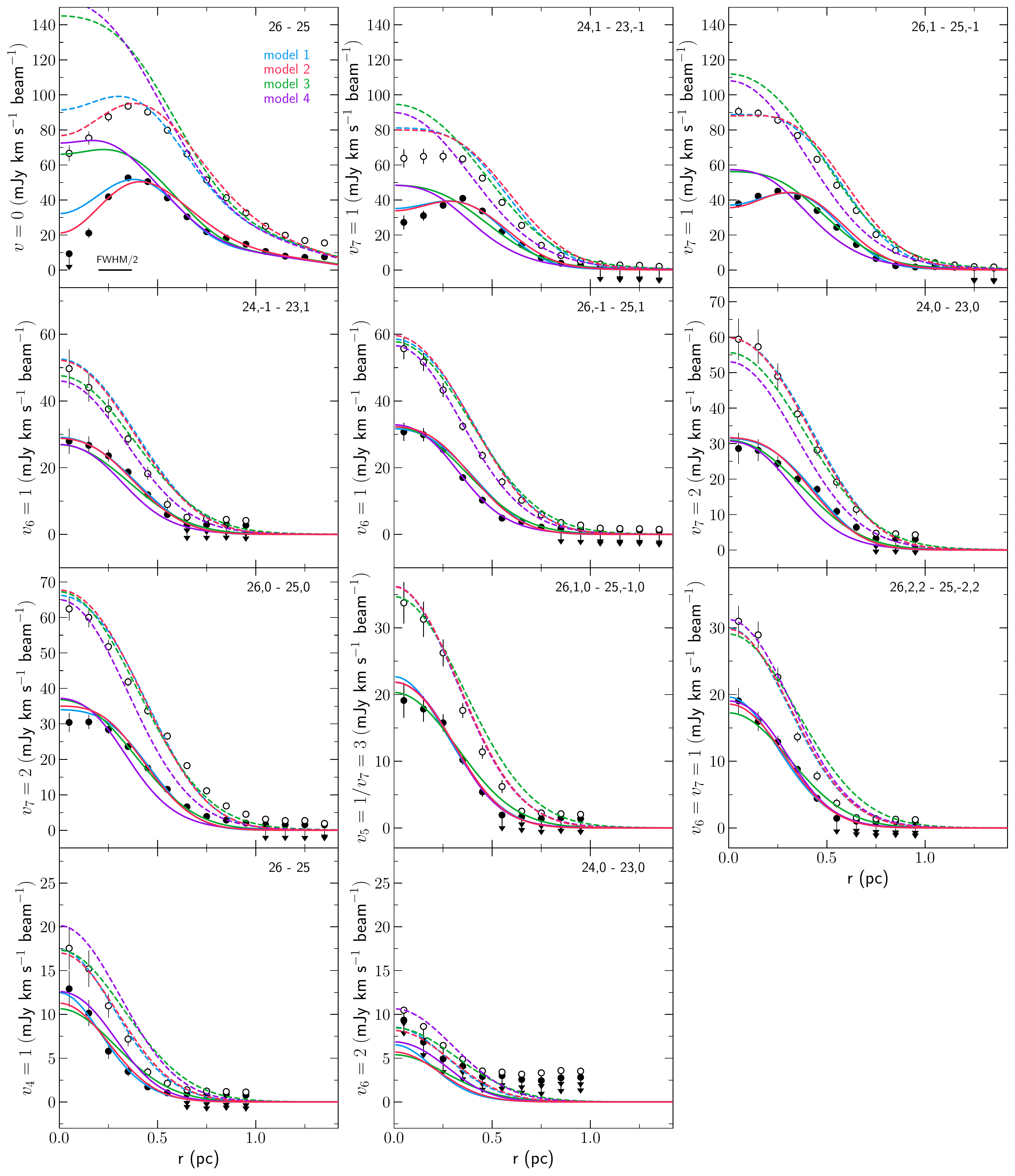}
  \caption[Proto-SSC\,$13$a observed and modelled HC$_3$N* line spatial emission]{Observed and modelled HC$_3$N* line emission spatial profiles of the transitions listed in Table~\ref{tab:HC3N_HR_fluxes}. Black filled and open circles show the integrated line intensity of the averaged ring spectra at $0.022\arcsec \times 0.020\arcsec$ and $0.028\arcsec \times 0.034\arcsec$ resolution, respectively, for each HC$_3$N* transition (quantum number of the transition are indicated on the upper right corner of each panel). Colored lines show the modelled emission for each model (Model 1 in blue, Model 2 in red, Model 3 in green and Model 4 in purple) at a resolution of $0.022\arcsec \times 0.020\arcsec$ (solid lines) and $0.028\arcsec \times 0.034\arcsec$ (dashed lines). 
  Half of the beam FWHM is shown on the first panel.
  }
  \label{fig:NGC253_HR_SLIM_SHC13_nLTE_lines}
\end{figure*}

Figure~\ref{fig:NGC253_HR_SLIM_SHC13_nLTE_lines} shows the observed HC$_3$N* emission radial profiles at a resolution of $0.022\arcsec \times 0.020\arcsec$ (filled black circles) and at a resolution of $0.028\arcsec \times 0.034\arcsec$ (open circles) for the HC$_3$N* transitions listed in Table~\ref{tab:HC3N_HR_fluxes}. The modelled HC$_3$N* radial profiles are also plotted at both resolutions with solid and dashed, respectively, colored lines for each model. Figure~\ref{fig:NGC253_HR_SLIM_SHC13_nLTE_lines} shows that most lines from high vibrational excited states (high-energy lines) are well reproduced by the four models.
Considering only these high-$v$ lines, there is still a degeneracy between the considered models. 
Lowering $T_\text{dust}$ (as in models $1$ and $2$, Fig.~\ref{fig:NGC253_HR_SLIM_SHC13_nLTE_conts}) decreases the emission of the high-energy transitions, which requires to significantly increase the HC$_3$N abundance in the hotter inner shells to reproduce the high-energy observed emission. Increasing $n_{\text{H}_2}$ rises the HC$_3$N column density  and therefore the HC$_3$N emission, which can be decreased by lowering its HC$_3$N abundance. However, we can favour one model over another by taking into account also the low-$v$ lines (including the $v=0$) and the physical constraints of the assumed model parameters.

\subsubsection[Model discrimination]{Model discrimination}
\label{subsec:model_disc}

As already mentioned, the models are somewhat degenerated if only the continuum emission and/or very high-energy HC$_3$N* lines are considered, but one can break the degeneration by taking into account also the HC$_3$N lines  arising from the $v=0$ and $v_7=1$ states.

\begin{figure}
\centering
    \includegraphics[width=0.8\linewidth]{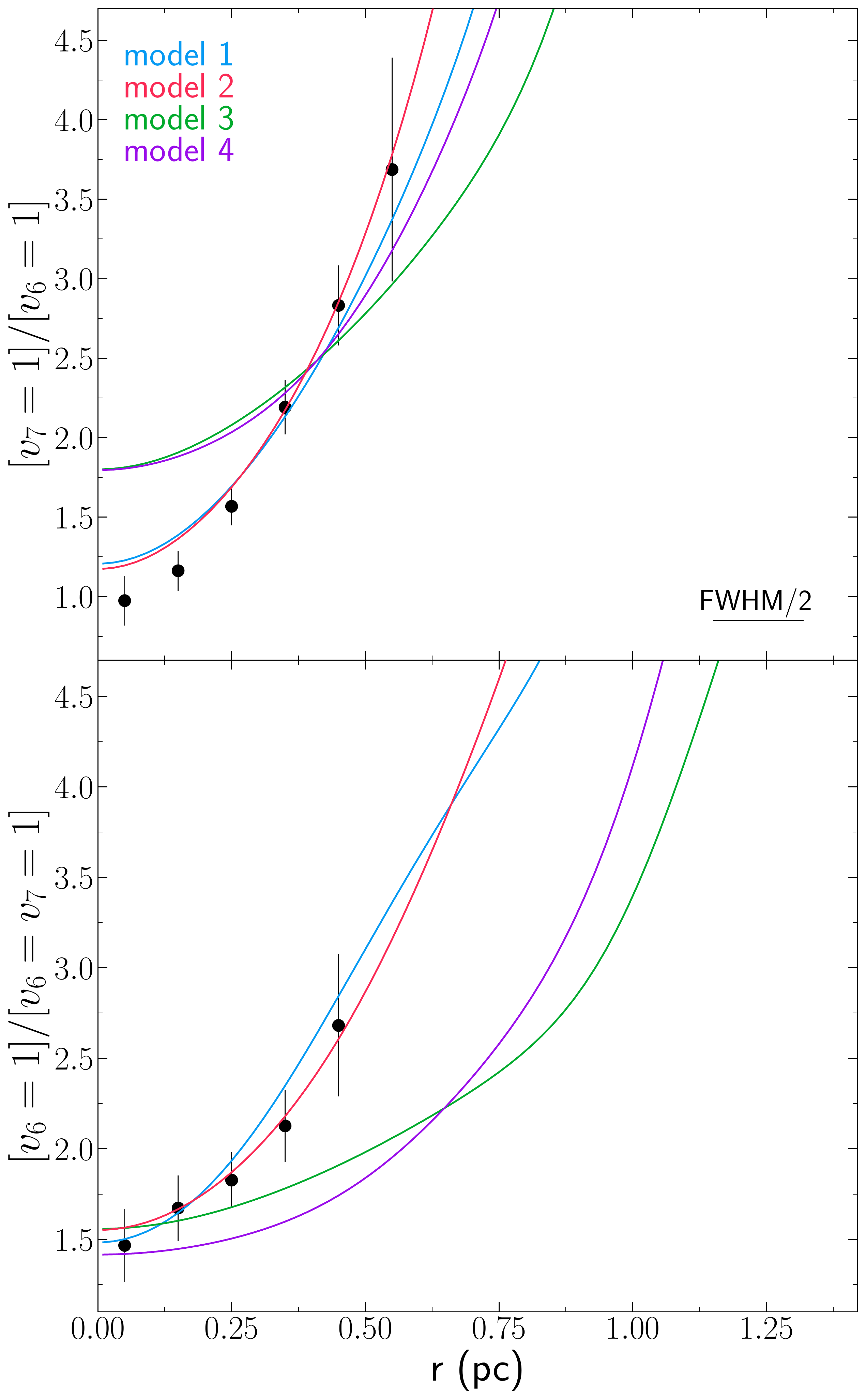}
  \caption[Proto-SSC\,$13$a observed and modelled HC$_3$N* line ratios profiles]{Observed and modelled HC$_3$N* line ratios radial profiles. Left panel shows the ratio between the $v_7=1$\:$(24,1-23,-1)$ and the $v_6=1$\:$(24,-1-23,1)$ transitions. Right right panel shows the ratio between the $v_6=1$\:$(24,-1-23,1)$ and the $v_6=v_7=1$\:$(26,2,2-25,-2,2$) transitions.
  }
  \label{fig:NGC253_HR_SLIM_SHC13_nLTE_ratios}
\end{figure}

We propose to consider simultaneously both the high-energy and the low-energy transitions ratios to increase the dynamic range in energy.
Such large energy range can be used thanks to the large number of HC$_3$N* transitions in the observed frequency range.
We show in Figure~\ref{fig:NGC253_HR_SLIM_SHC13_nLTE_ratios} the ratio between the $v_7=1$\,$(24,1-23,-1)$ and the  $v_6=1$\,$(24,\text{-1}-23,1)$ (i.e. $[v_7=1]/[v_6=1]$) as representative of the low-energy transitions and the ratio between the  $v_6=1$\,$(24,\text{-1}-23,1)$  and the $v_6=v_7=1$\,$(26,2,2-25,-2,2)$ (i.e. $[v_6=1]/[v_6=v_7=1]$)  as representative of the high-energy transitions.
We can see that Model 3 and Model 4 fail to reproduce both line ratio profiles simultaneously. In the innermost region the  $[v_6=1]/[v_6=v_7=1]$ is close to the observed ratio but the $[v_7=1]/[v_6=1]$ ratio is far from being reproduced; the opposite effect is found for the $[v_7=1]/[v_6=1]$ ratio. On the other hand, Model 1 and Model 2 reproduce both ratios at the same time for all radii, with Model 2 predicting values closer to the observed ratios. Clearly, Models 3 and 4 grossly overestimate the emission of the low-energy $v=0$ and $v_7=1$ from the innermost rings, while Models 1 and 2 agree much better with observations.

From the line integrated intensity radial profiles and from the $[v_7=1]/[v_6=1]$ and $[v_6=1]/[v_6=v_7=1]$ line ratios profiles (Figures~\ref{fig:NGC253_HR_SLIM_SHC13_nLTE_lines} and \ref{fig:NGC253_HR_SLIM_SHC13_nLTE_ratios}), Model 1 and Model 2 are the best in reproducing all spatial profiles (they have the lowest $\chi^2_\text{lines}$ values, see Table~\ref{tab:nlocal_modpars}). These include the absorption features in the inner rings from the low-energy ground state $v=0$ and $v_7=1$ transitions, and both line ratios. 
Between these two models, Model 2 is favored against Model 1 because Model 2 reproduces better the $[v_7=1]/[v_6=1]$ and $[v_6=1]/[v_6=v_7=1]$ line ratios. Also, for Model 1, the $v=0$ line is not reproduced. Even Model 2, while showing a stronger decrease in the $v=0$ flux towards the center, still cannot fully reproduce it, suggesting that the  HC$_3$N  column density towards the innermost rings is somewhat higher than in the model. Moreover, Model 1 requires a rather \quotes{strange} HC$_3$N abundance profile (with very high abundances towards the center, see Figure~\ref{fig:NGC253_HR_SLIM_SHC13_nLTE_conts}) in order to reproduce the distributions of the high-energy lines.

In summary,  assuming spherical symmetry with a temperature and density radial profiles obtained in a consistent way considering the heating from the greenhouse effect due to the large optical depths in the IR and distributed star formation heating, Model 2 is able to reproduce the radial distribution of the continuum and the HC$_3$N* emission.

\section{Discussion}

\subsection{Dust temperature profiles derived from local and non-local radiative transfer analysis}

If we compare the $T_\text{dust}$ and the $N(\text{HC}_3\text{N})$ profiles derived from Model 2, which take into account non-local radiative transfer effects, with those derived from the \texttt{SLIM} LTE analysis, we can see very significant differences in both profiles (see Fig.~\ref{fig:NGC253_HR_SLIM_SHC13_nLTE_conts}). 
The differences in the dust temperature (overall higher $T_\text{dust}$ values for the \texttt{SLIM} model) arise from the fact that the \texttt{SLIM} LTE model does not include non-local radiative transfer effects along the line of sight, which are crucial for the large column densities, H$_2$ density and dust temperature profiles expected in the very early phases of the SSCs formation. 
These effects give rise to the absorption features in the inner regions for the lines of the low $v=0$ and $v_7=1$ vibrational states, which decrease their intensities in relation to higher-$v$ transitions, indicating that lines from different vibrational states arise from different regions and thus sample different physical conditions along the line of sight. When these transitions are not taken into account to avoid this effect, as considered for the \texttt{SLIM} model inner regions, the covered energy range is reduced and the \texttt{SLIM} fit tends to rise $T_\text{dust}$ as the absorption also affects the $v_7=2$ and $v_6=1$, although to a lesser extent than for the $v=0$ and $v_7=1$.
Even at large radii, the LTE dust temperature is overestimated. This is due to the combination of different optical depths in the $v=0$ and the $v_7=1$ lines and the $T_\text{dust}$ radial gradient, making both lines to trace the region where they become optically thick and therefore tracing different $T_\text{dust}$ within the proto-SSC. Since the $v=0$ lines are optically thick at larger radii where the $v_7=1$ lines are optically thin, the $T_\text{dust}$ traced by the $v=0$ lines will be smaller than that from the $v_7=1$ lines. This effect will also appear for any pair of lines from different vibrationally excited levels. 
Also, since the rings width is smaller ($0.1$\,pc) than half of the beam FWHM ($0.17$\,pc), which is the true resolution power of the observations, each ring has also a contribution from the adjacent rings in the sky plane, however we expect it to affect to a lesser extent the differences in the $T_\text{dusts}$ profile than the contribution from other rings along the line of sight.
Regarding the HC$_3$N column density, its profile is similar at small radii to that of the non-local models, but at longer radii ($>0.7$\,pc) they start to differ significantly. This is partially due to subthermal excitation (difference between $T_\text{rot}$ and $T_\text{vib}$) found in \citet{Costagliola2010} and \citet{RicoVillas2021NGC1068}, which tends to overestimate the LTE HC$_3$N column densities when considering $T_\text{vib}$, instead of $T_\text{rot}$, for the calculation in \texttt{SLIM}.
Observations of less abundant isotopologues  will help to alleviate the optical depth problems in the LTE analysis. Unfortunately, the isotopologues lines are weaker and the potential lines to be detected  in extragalactic sources will be limited to the lines arising from the lowest vibrational levels.

\subsection{Central vs. distributed star formation}
\label{subsec:central_or_distr_SF}

\begin{figure*}
\centering
    \includegraphics[width=0.6\linewidth]{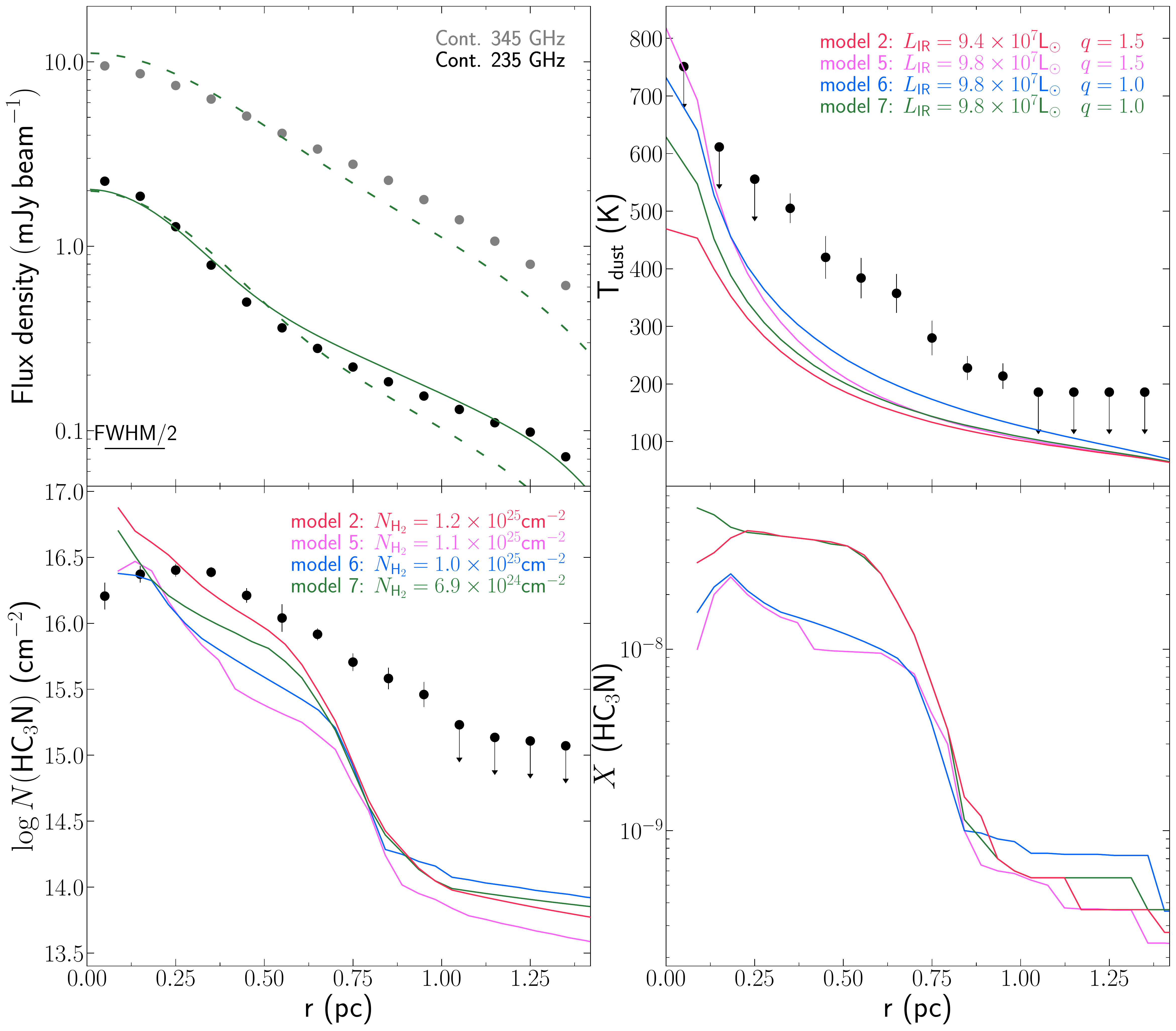}
  \caption[Proto-SSC\,$13$a modelled continuum spatial emission from central star formation]{Same as Figure~\ref{fig:NGC253_HR_SLIM_SHC13_nLTE_conts} but for the spatially distributed star formation Model 2 (in red), and the central star formation Model 5 (in magenta), Model 6 (in dark blue) and Model 7 (in dark green), see Section~\ref{subsec:central_or_distr_SF}. 
  }
  \label{fig:NGC253_HR_SLIM_SHC13_nLTE_conts_central}
\end{figure*}

The models described above assume that the star formation spatial profile is proportional to the gas density and mass, injecting energy from stars at different radii.
The energy deposition per unit of time and volume is proportional to the SFR per unit of volume \citep[see][for details]{GA19}.
This is reflected in the different $T_\text{dust}$ profiles for a given column density and power-law index $q$. Consequently, a $q=1.0$ index represents a more scattered star formation than a  $q=1.5$ index, as illustrated by $r_e$, the radius that contains half of the star formation (i.e. luminosity). Models with $q=1.0$ have a more extended $r_e$ of $\sim0.8$\,pc and models with $q=1.5$ a smaller $r_e\sim0.4$\,pc.  The selected model, i.e. Model 2, has a $q=1.5$ ($r_e\sim0.4$\,pc), indicating that a higher concentration of star formation (or the most massive stars) is located at the center of proto-SSC\,$13$a, although a fraction of the star formation also takes place at larger radii.

To test the assumption of distributed star formation in the proto-SSC\,$13$a, we have also modelled the HC$_3$N* and continuum emission assuming that the star formation only takes place in the very  central region ($\sim0.05$\,pc, i.e. central star formation). To do so, we have used the \quotes{AGN} model profiles for $T_\text{dust}$ from \citet{GA19}. 
Figure~\ref{fig:NGC253_HR_SLIM_SHC13_nLTE_conts_central} compares the distributed star formation Model 2 with three central star formation models (Models 5, 6 and 7). The three models have the same luminosity as Model 2 ($L_\text{IR}=9.2\times10^7$\,L$_\odot$), and Models 5 and 6 also the same column density ($N_{\text{H}_2}=10^{25}$\,cm$^{-2}$), but Model 7 has a lower column density ($N_{\text{H}_2}=5.6\times10^{24}$\,cm$^{-2}$).
For the same luminosity and column density as the models with distributed star formation from Section~\ref{ring_subsec_nlocal_mod}, the central star formation models have much steeper $T_\text{dust}$ profiles. 
Due to the small size of the region forming stars of $\sim0.05$\,pc located in the center, the greenhouse effect makes radiation to escape more difficult than if star formation were distributed all over the source for the same dust column density. In the outermost regions, however, the central and distributed star formation models have similar dust temperatures.

\begin{figure}
\centering
    \includegraphics[width=0.8\linewidth]{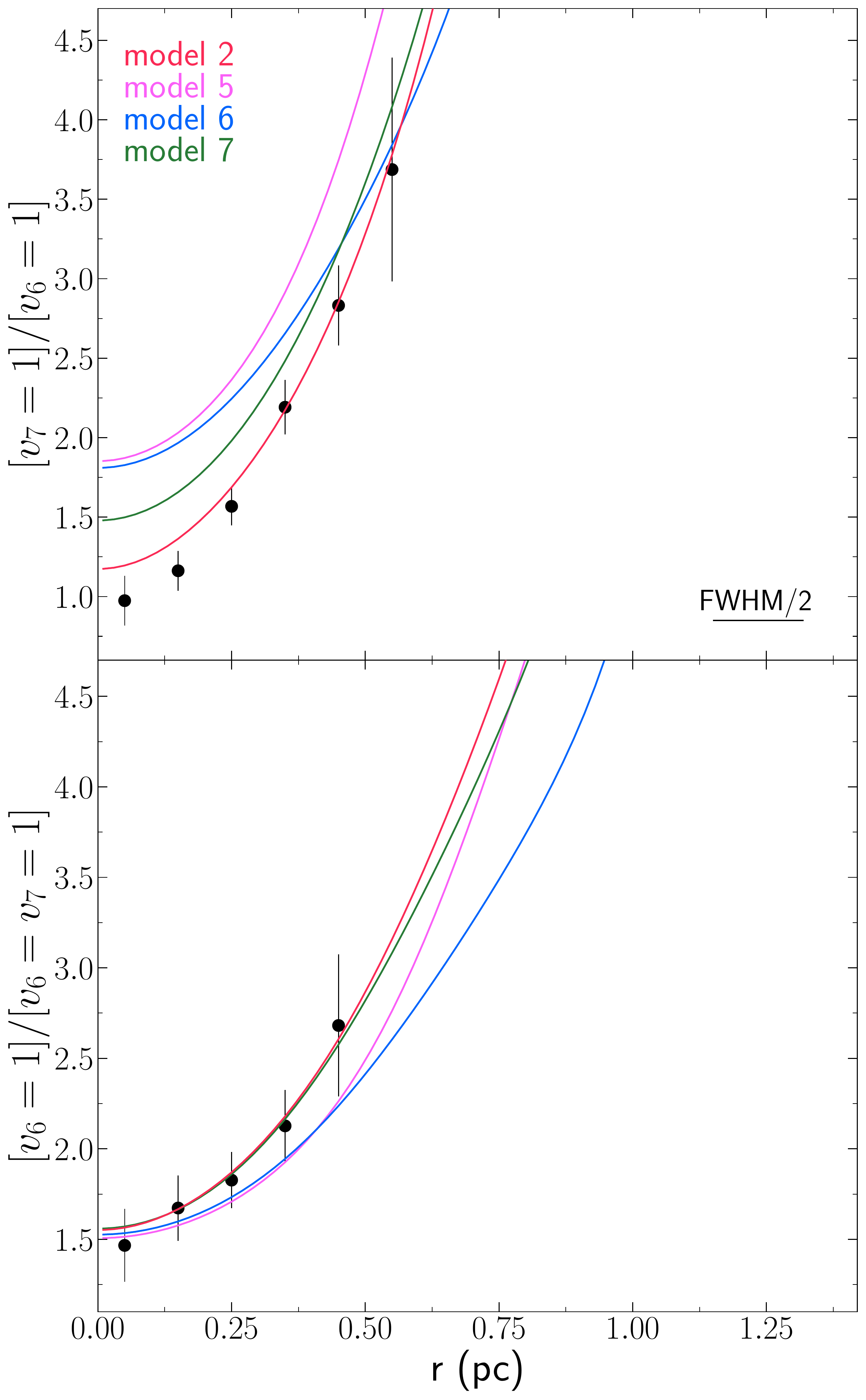}
  \caption[Proto-SSC\,$13$a observed and modelled HC$_3$N* line ratios profiles]{Same as Figure~\ref{fig:NGC253_HR_SLIM_SHC13_nLTE_ratios} but with Model 2 (in red), Model 5 (in magenta), Model 6 (in dark blue) and Model 7 (in dark green).
  }
  \label{fig:NGC253_HR_SLIM_SHC13_nLTE_ratios_comp}
\end{figure}

Central star formation models suffer similar problems as Models 3 and 4: the high temperature in the central region boosts too much the high-energy transitions, requiring lower abundances than Model 2. To compensate this, we lower the column density to $5.6\times10^{24}$\,cm$^{-2}$ in Model 7, which allows to maintain similar abundances as in Model 2. However, lowering the column densities reduces the absorption effect of Model 2, so that Models 5, 6 and 7 strongly overestimate the emission of the low-energy lines (mostly the $v=0$). Figure~\ref{fig:NGC253_HR_SLIM_SHC13_nLTE_ratios_comp} shows the comparison of the  low-energy $[v_7=1]/[v_6=1]$ line ratio and the  high-energy $[v_6=1]/[v_6=v_7=1]$  line ratio between these models. Model 7, despite reproducing well the $[v_6=1]/[v_6=v_7=1]$ ratio, fails to reproduce the low-energy $[v_7=1]/[v_6=1]$ as it underestimates the $v_6=1$\:$(24,-1-23,1)$ emission at $r>0.1$\,pc and with its lower column density is unable to reproduce the absorption observed in the $v_7=1$ and $v=0$ line profiles. 

All the temperature profiles from our modelling show that massive star formation is concentrated in the central region of proto-SSC\,$13$a. However, we have discarded Model 7 (the best model with only star formation at the center), since it does not reproduce simultaneously the low-energy line profiles and the $[v_6=1]/[v_6=v_7=1]$ line ratio. For models with distributed star formation, Model 2  best matches the observations and both line ratios. Therefore, our modelling indicates that the most massive star formation is taking place in the central region of proto-SSC\,$13$a (with $r_e\sim0.4$\,pc), but star formation is distributed over a more extended region ($\sim80\%$ of the star formation takes place inside $r=0.9$\,pc). 

\subsection{Star formation scenarios}

A centrally peaked but distributed star formation would be in accordance with the competitive accretion scenario, where most massive stars are being formed at the center of cluster potential well and less massive stars at the outskirts \citep[e.g.][]{Bonnell1997, Bonnell2001, Murray2012}. Also, this would mean that mass segregation (i.e. the most massive stars in the center of a star cluster) observed in more evolved SSCs \citep{Bonnell2001, Bonnell2007}  is already set in place at the early ages of proto-SSC\,$13$a \citep[$t_\text{age}\sim4\times10^4$\,yr, see][]{RicoVillas2020} as the competitive accretion naturally predicts. Furthermore, it has also been suggested that primordial mass segregation is present in the star clusters of the Milky Way \citep[less massive than SSCs; e.g.][]{McMillan2007, Pavlik2019} or in the much older (and comparable in mass to SSCs) globular clusters of the Milky Way with typical half-light radii of $\sim3$\,pc \citep[e.g.][]{Haghi2014}.

\subsection{The \texorpdfstring{proto-SSC\,13a}{proto-SSC13a} velocity structure: cloud-cloud induced star formation}
\label{subsec:vel_grad}

From the 2D analysis of the HC$_3$N* emission in Section~\ref{subsec:2d_analysis} we have seen that there is a velocity structure in proto-SSC\,$13$a that can provide information on the triggering process of the formation of SSCs. We find a separation of $\sim21$\,km\,s$^{-1}$ between the minimum ($242$\,km\,s$^{-1}$) and maximum ($263$\,km\,s$^{-1}$) velocities in proto-SSC\,$13$a. Here we briefly discuss three possible origins: rotation, outflow or cloud-cloud collision.

\begin{figure}
\centering
    \includegraphics[width=0.99\linewidth]{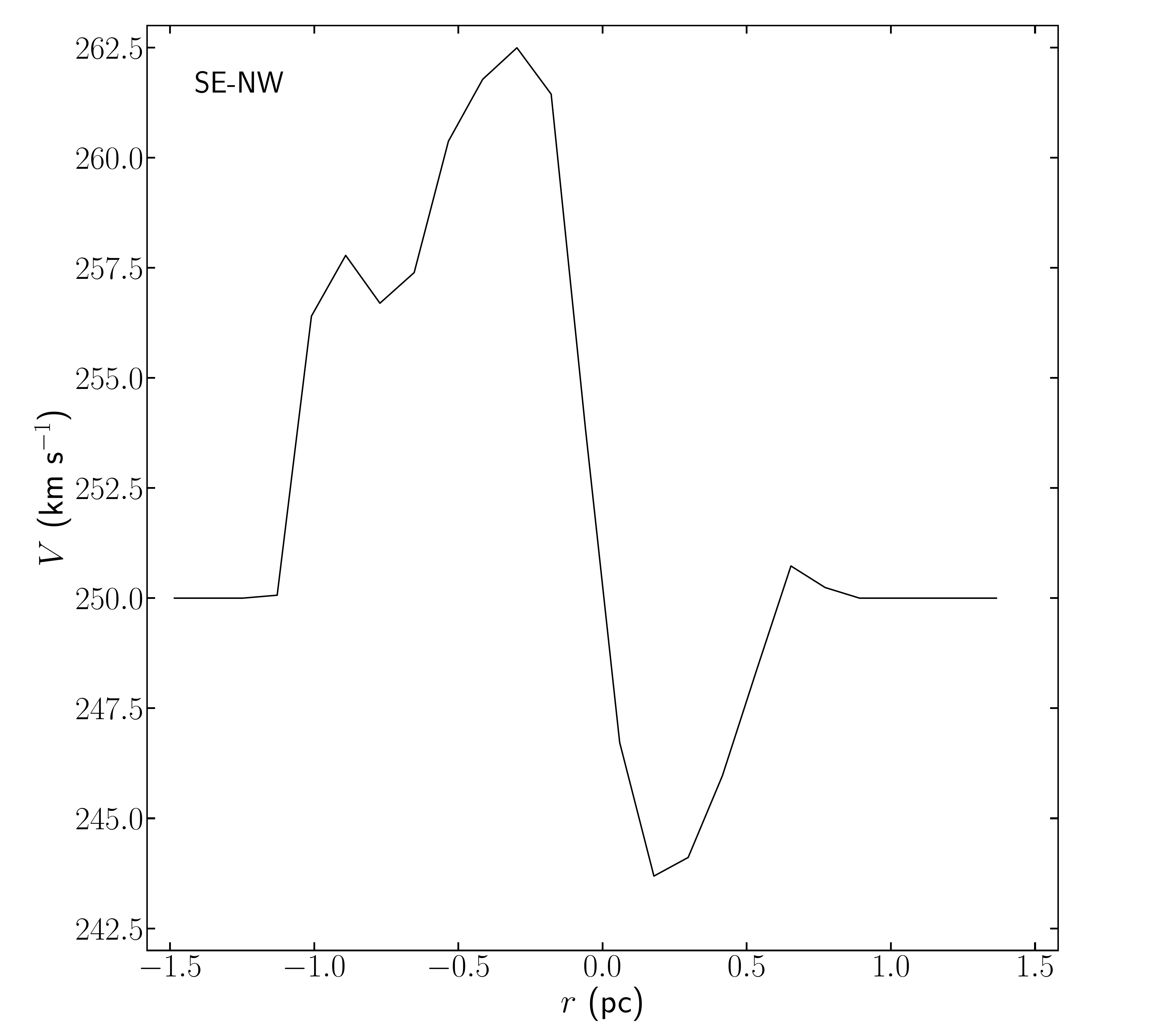}
  \caption[NGC\,253 proto-SSC\,13a velocity profiles]{Proto-SSC\,13a velocity profiles derived from the SLIM fitted values in the southeast-northwest direction (i.e. perpendicular to the galaxy rotation plane) in the left panel and in the northeast-southwest direction (i.e. parallel to the galaxy rotation plane) in the right panel.
  }
  \label{fig:NGC253_HR_SHC13_vel_profile}
\end{figure}

We can discard that the cloud is rotating. Figure~\ref{fig:NGC253_HR_SLIM_SHC13} shows that the local velocity change occurs in the southeast-northwest (SE-NW) direction, just perpendicular to the rotation of the galaxy disk in the northeast-southwest (NE-SW) direction. Such rotation pattern would be hard to explain in the context of the NGC\,$253$ disk kinematics and the conservation of angular momentum. Furthermore, in Figure~\ref{fig:NGC253_HR_SLIM_SHC13} we can see that the region with larger velocities ($250-263$\,km\,s$^{-1}$) is extended towards the northeast and also surrounds the region with lower velocities ($243-250$\,km\,s$^{-1}$) on the west, difficult to match with a rotation pattern. Figure~\ref{fig:NGC253_HR_SHC13_vel_profile} shows the velocity profile through the local velocity change (i.e. the SE-NW direction). It can be seen that there are two differentiated regions with different velocities: a larger cloud with velocities going from $\sim263-250$\,km\,s$^{-1}$ and a smaller cloud with $\sim250-242$\,km\,s$^{-1}$. 

If the local velocity change is caused by outflowing gas, we should see the characteristic P-Cygni profiles in some of the observed transitions, or at least line wings in the spectra of the lowest-energy lines. However, we do not observe such features despite the large column densities present in proto-SSC\,$13$a.
Also, as indicated by \citet{Levy2021}, some vibrationally excited emission lines are coincident with the blueshifted absorption (in our case the $v_5=1/v_7=3$ line is coincident with the $v=0$ line) could help to conceal the P-Cygni profiles. Because of this, \citet{Levy2021} do not observe clear P-Cygni profiles for SSC\,$13$a in their study of the CS $7-6$ and H$^{13}$CN $4-3$ transitions, as they do for SSCs $4$a, $5$a and $14$.

Since the outflow cannot be completely ruled out, we estimate some of its properties. 
We consider that the systemic velocity of the SSC is $250$\,km\,s$^{-1}$ and the gas is outflowing with a projected velocity of $10-15$\,km\,s$^{-1}$. Note that the outflow terminal velocity would be larger since the outflow might not be along the line of sight. Assuming a $45^\circ$ inclination, the terminal velocity would be about $20$\,km\,s$^{-1}$.
We can derive the mass of the outflowing small cloud with velocities $<250$\,km\,s$^{-1}$ from  the total gas mass ($M_{\text{H}_2}=9.5\times10^5$\,M$_\odot$) and density profile ($n_{\text{H}_2}\propto r^{-1.5}$) obtained in Section~\ref{subsec:model_disc} for Model 2. The outflowing mass for the blueshifted cloud with $<250$\,km\,s$^{-1}$ would be $M_{\text{out,H}_2}\sim7\times10^4$\,M$_\odot$.  
Unfortunately, the estimate of the mass of the redshifted cloud is more complicated, since it merges with the ambient gas which is not expected to be participating in the outflow. We thus guess that it has a similar mass that the blueshifted gas.
We can also estimate the dynamical age of the outflow by considering the distance of the outflowing cloud to its origin and the terminal velocity ($\delta v=21$\,km\,s$^{-1}$). For a distance of $\sim0.2$\,pc (i.e. the position of the blueshifted cloud from the expected origin) we obtain a dynamical age $t_\text{dyn}\sim10^4$\,yr. With these parameters we estimate its momentum flux by $\dot{P}_\text{out}=\dot{M}_\text{out}\delta v=9.3\times10^{32}$\,g\,cm\,s$^{-2}$, where $\dot{M}_\text{out}=M_{\text{out,H}_2}/t_\text{dyn}$. If we consider that the outflow is only driven by radiation, $\dot{P}_\text{out}$ should be compatible with $\sim L_\text{IR}/c=1.3\times10^{31}$\,g\,cm\,s$^{-2}$ \citep[see][]{Murray2005GalWinds}, which is not the case. The latter and the lack of observed P-Cygni profiles, are in agreement with the idea that proto-SSC\,$13$a is very young and where very likely the mechanical feedback from stars is not yet affecting significantly its gas content. 
%Although the value obtained for $L_\text{out}$ is compatible with an outflow powered by star formation with luminosity $L_\text{IR}\sim10^8$\,L$_\odot$, the lack of observed P-Cygni profiles agrees with the idea that proto-SSC\,$13$a is very young and where very likely the mechanical feedback from stars is not yet affecting significantly its gas content. 

We now consider the possibility of a cloud-cloud collision in which the gas with velocities $\sim242-250$\,km\,s$^{-1}$ would be in a cloud of $7\times10^4$\,M$_\odot$ which is driving into a larger cloud of $8.8\times10^5$\,M$_\odot$ with velocities $\sim250-263$\,km\,s$^{-1}$. The velocity of the small colliding cloud is also consistent with the non-disk component identified in the study  of NGC\,253 kinematics from CO emission by \citet{Krieger2019}. This scenario also fits within the observed velocity structure shown in Figures~\ref{fig:NGC253_HR_SLIM_SHC13}, with the smaller cloud being surrounded by the large cloud which still maintains its velocities, and is similar to what is expected in the early stages of a cloud-cloud collision \citep[e.g.][]{Habe1992, Takahira2014, Haworth2015}. At the collision front, a dense layer of material is developed as the small cloud advances through the large cloud, inducing the formation of gravitationally unstable cores that lead to massive star formation \citep{Inoue2013}. Such external triggering of massive star formation has been invoked for the formation of several young massive clusters (YMCs) of the Milky Way and of the Large Magallanic Cloud \citep[e.g.][]{Furukawa2009, Ohama2010, Fukui2016, Fukui2018, Fujita2021, Tsuge2021}.

\begin{figure}
\centering
    \includegraphics[width=\linewidth]{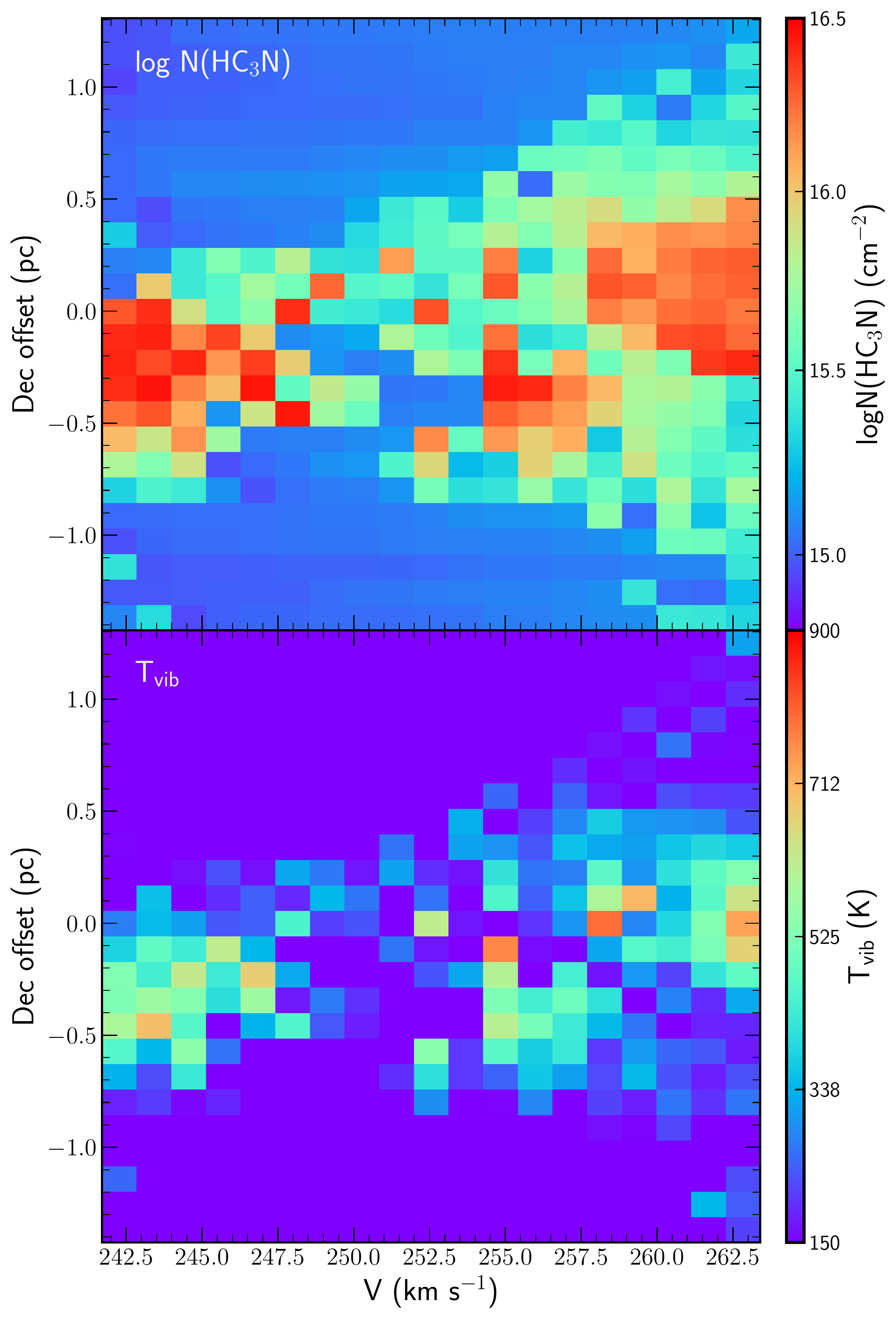}
  \caption[NGC\,253 proto-SSC\,13a position-velocity diagram]{Proto-SSC\,13a position velocity diagrams of the HC$_3$N column density (top panel) and the vibrational temperature (bottom panel) derived with \texttt{SLIM}.
  }
  \label{fig:NGC253_HR_SHC13_pvdiag}
\end{figure}

The compressed interface layer between the two clouds is usually translated into a region with intermediate velocities connecting the two clouds (i.e. a bridge feature) separated in velocity (but not spatially) in a position-velocity diagram (hereafter p-v diagram) like those shown in Figure~\ref{fig:NGC253_HR_SHC13_pvdiag}. In the HC$_3$N column density p-v diagram (Figure~\ref{fig:NGC253_HR_SHC13_pvdiag} left panel) we can see the bridge feature with intermediate velocities connecting the two colliding clouds separated in velocities. However, in the p-v diagram of the vibrational temperature (Figure~\ref{fig:NGC253_HR_SHC13_pvdiag} right panel), the bridge feature is less clear and no clear pattern of the highest temperatures, which one would expect to be located in the bridge feature, is observed. Still, the overall velocity structure seems to favour the cloud-cloud collision scenario rather than the outflow.

\subsection{Derived properties for \texorpdfstring{Proto-SSC\,13a}{Proto-SSC13a}}
\label{subsec:derived_properties}

The derived radius of $1.5$\,pc for proto-SSC\,$13$a is close to the smallest observed SSCs, which have typical radius between $1$ and $5$\,pc. %Together, with the lack of clear signatures of an outflow meaning that mechanical feedback has not yet started, reflects the very early stage of  proto-SSC\,$13$a.
%, and reflects its very early stage in which mechanical feedback has likely not started to expand the cluster. Indeed, as discussed previously, we do not observe clear signatures of an outflow.
From Model 2, we obtain for  proto-SSC\,$13$a  a luminosity of $L_\text{IR}  = 9.4\times10^7$\,L$_\odot$, a luminosity surface density of $1.3\times10^7$\,L$_\odot$\,pc$^{-2}$, and a molecular gas mass of $M_{\text{H}_2} =  9.5\times10^5$\,M$_\odot$ (see Table~\ref{tab:nlocal_modpars}).% Since we have varied $X_\text{dust}$ to fit the continuum emission, we note that the mass differ from the used to derive the $T_\text{dust}$ profile in \citep{GA19}, however we do not expect the differences to be more than $50\%$ in the derived luminosity. 

% For this radius, the luminosity ($L_\text{IR}$) and molecular gas mass ($M_{\text{H}_2}$) derived for proto-SSC\,$13$a (from Model 2) are $9.4\times10^7$\,L$_\odot$\ and $9.5\times10^5$\,M$_\odot$. Since we have varied $X_\text{dust}$ to fit the continuum emission, we note that the mass differ from the used to derive the $T_\text{dust}$ profile in \citep{GA19}, however we do not expect the differences to be more than $50\%$ in the obtained luminosity. 

\subsubsection{SSC formation through rapid collapse}

From the total gas mass obtained for proto-SSC\,$13$a we can derive its virial parameter $\alpha=5\sigma^2 R/GM_{\text{H}_2}$, where $\sigma$ is the velocity dispersion. From the mean of all the fitted FWHMs for each pixel in Section~\ref{subsec:2d_analysis} ($\sim25$\,km\,s$^{-1}$) we obtain a mean $\bar{\sigma}\sim11$\,km\,s$^{-1}$. Using this $\bar{\sigma}$, the virial parameter is $\alpha\approx0.21$, well below the value of $2$ to consider the cloud to be bounded and thus being supercritical. If we take into account the derived density profile in the innermost region ($n_{\text{H}_2}\propto r^{-1.5}$), $\alpha$ would be even smaller \citep{Kauffmann2013}. In this supercritical state ($\alpha \ll 2$), proto-SSC\,$13$a would be unstable and indicates a rapid collapse. Such a low virial parameter is typically  found in high-mass star forming clumps and cores \cite[see][for a review on regions with supercritical virial parameters]{Kauffmann2013}. 

We now investigate if the radiation pressure from the forming stars can disperse or at least halt the collapsing natal material of proto-SSC\,$13$a. From \citet{GA19}, the Eddington flux can be derived from $F_\text{Edd}=\Psi_\text{th}\Sigma_{\text{H}_2}/(1-\Psi_\text{th}/\Psi)$, where $\Sigma_{\text{H}_2}$ is the gas surface density, $\Psi$ the light-to-mass ratio of the current stellar population and $\Psi_\text{th}$ is the  light-to-mass ratio threshold value given by
\begin{align}
    \Psi_\text{th}=\frac{4\pi G c}{\kappa_\text{F}}=1.3\times10^3\left(\frac{\kappa_\text{F}}{10 \text{cm}^2\,\text{g}^{-1}}\right)^{-1}\:\text{L}_\odot\,\text{M}_\odot^{-1},
\end{align}
with $\kappa_\text{F}$ the Rosseland mean opacity. Assuming $\kappa_\text{F}$ to be $10$\,cm$^2$\,g$^{-1}$ \citep[typical for a hot starburst with $T>200$\,K, see][]{opacity_Andrews}, $\Psi_\text{th}=1.3\times10^3$\,L$_\odot$\,M$_\odot^{-1}$. If $\Psi<\Psi_\text{th}$, gravity overcomes the radiation pressure and the cloud can collapse further. For a model with $N_{\text{H}_2}=10^{25}$\,cm$^{-2}$, $\Sigma_\text{IR}=1.1\times10^8$\,L$_\odot$\,pc$^{-2}$ , \citet{GA19} shows that the outward force due to radiation pressure (red curve in panel b of their Figure 6) is close  to the inward gravity force (green lines in panel b of their Figure 6) only in the innermost regions, with gravity overcoming radiation pressure in the external regions,  even for a large $\Psi$ of $1700$\,L$_\odot$\,M$_\odot^{-1}$. Given that $\Sigma_\text{IR}$ for proto-SSC\,$13$a is lower, gravity would overcome radiation pressure more easily. Also, we can make a simpler estimation of the Eddington luminosity for proto-SSC\,$13$a from $L_\text{Edd}=4\pi c G M/\kappa_{F}$, which results in $10^9$\,L$_\odot$, an order of magnitude larger than the derived $L_\text{IR}$. From the above calculations, we conclude that the radiation pressure is not able to overcome gravity and prevent proto-SSC\,$13$a from collapse. 

\subsubsection{Short timescales and high SFE}

Since proto-SSC\,$13$a is collapsing, we can estimate its free-fall time from $t_\text{ff}=\sqrt{3\pi/\left(32\text{G}\bar{\rho}\right)}$, where $\bar{\rho}$ indicates the mean gas density. For proto-SSC\,$13$a, $\bar{\rho}=4.5\times10^{-18}$\,g\,cm$^{-3}$, which gives a free-fall time $t_\text{ff}=3\times10^{4}$\,yr, very similar to  its estimated age ($t_\text{age}\sim4\times10^4$\,yr) in \citet{RicoVillas2020}. The small $\alpha$ and a $t_\text{ff}\sim t_\text{age}\sim10^{4}$\,yr indicate that proto-SSC\,$13$a has just started to form stars and that it is doing so very rapidly. This would be in accordance with the lack of detection of outflow signatures, which would not have had enough time to develop and/or alter the gas content. 

Therefore, if we consider that no mechanical feedback is taking place and there has not been gas ejection, we can estimate the star formation efficiency (SFE) as in \citet{RicoVillas2020}
\begin{align}
    \text{SFE}\sim\frac{1}{1+M_{\text{H}_2}/(M_\text{p*}+M_*)},
\end{align}
where $M_*$ is the ZAMS stellar mass derived by \citet{Leroy2018}  from the $36$\,GHz continuum emission and assuming a light-to-mass ratio of $10^3$\,L$_\odot$\,M$_\odot^{-1}$ ($M_*=6.3\times10^4$\,M$_\odot$, see \citet{RicoVillas2020}), and $M_\text{p*}$ is the protostellar mass derived from $L_\text{IR}$ and assuming the same light-to-mass ratio ($M_\text{p*}=9.4\times10^4$\,M$_\odot$). The resulting SFE is $\sim0.14$ ($0.09$ if only the protostars were accounted for). The value is lower than that obtained in \citet{RicoVillas2020} since with the more detailed modelling of high-$v$ HC$_3$N* transitions we estimate a higher gas content. Despite the youth of proto-SSC\,$13$a, it has already been able to convert $\sim14\%$ of its gas content in stars with an \quotes{instantaneous} star formation rate (in a free-fall time) SFR$=5$\,M$_\odot$\,yr$^{-1}$ ($3$\,M$_\odot$\,yr$^{-1}$ taking into account only the protostars being formed). Given the derived SFR, the time to consume all the gas content ($t_\text{dep}$) would be  $2\times10^5$\,yr. This $t_\text{dep}$ is close to the timescale for cluster formation  of $5 t_\text{ff}$ found by \citet{Skinner2015} in their simulations. If we assume a more conservative final SFE of $0.5$, the time to convert  the $50\%$ of the remaining gas mass into stars is  $\sim7\times10^4$\,yr, on the order of $3t_\text{ff}$. 

The high SFR ($3-5$\,M$_\odot$\,yr$^{-1}$) derived for the individual proto-SSC\,$13$a (and also for the remaining SSCs seen in \citet{RicoVillas2020}) contrasts with the low SFR if derived from the IR luminosity following the relation given by \citet{Kennicutt1998}  (and other similar relations) of SFR$=4.5\times10^{-44}L_\text{IR}(\text{erg}\,\text{s}^{-1})\approx0.01$\,M$_\odot$\,yr$^{-1}$. This is because these relations are meant for entire galaxies and therefore, in order to calibrate their SFR-$L_\text{IR}$ relation, they assume continuous  star formation over $(10-100)\times10^6$\,yr. These large timescales are not comparable to the young ages of the SSCs forming stars in NGC\,253 much shorter ($(1-10)\times10^4$\,yr) timescales. Hence, the SFRs we derive in \citet{RicoVillas2020} and in this paper can be understood as an instantaneous SFR. In fact, the factor $\sim300$ between the SFR derived in this paper and that from the relation given by \citet{Kennicutt1998}, is similar to the factor between the different timescales considered of $3\times10^4$\,yr and $10^7$\,yr.

\subsubsection{Triggering the SSC formation}

It has been suggested that the formation of SSCs involve a rapid collapse and enhanced SFE, requiring extreme external pressures of $P_e/k\gtrsim10^8$\,K\,cm$^{-3}$ \citep{Elmegreen1997}. Following \citet{Johnson2015} and \citet{Finn2019}, the external pressure confining the cloud mass to the observed radius,  given its velocity dispersion, can be obtained from
\begin{align}
    P_e=\rho_e \sigma^2=\frac{3\Pi M \sigma^2}{4\pi r^3},
\end{align}
where $\rho_e=\Pi \bar{\rho}$ is the density at the edge of the cloud and where we use the same $0.5$ value for $\Pi$ as \citet{Johnson2015} to estimate the density at the cloud's edge from the average cloud density $\bar{\rho}$. For proto-SSC\,$13$a we obtain $P_e/k=1.8\times10^{10}$\,K\,cm$^{-2}$, two order of magnitudes greater than the required external pressure to avoid the dispersal of the forming SSC gas. But what is driving such a high external pressure? As discussed in Section~\ref{subsec:central_or_distr_SF}, the velocity structure  matches with a cloud-cloud collision. Such a mechanism could increase the pressure in proto-SSC\,$13$a to the observed values. A rough estimation of the ram pressure induced by the collision of the two colliding clouds separated by $21$\,km\,s$^{-1}$ is given by $P=\bar{\rho} v^2$ \citep{Finn2019}, which results in $P/k\sim10^{11}$\,K\,cm$^{-3}$ and shows that the cloud-cloud collision is able to provide the observed high external pressure to support SSCs formation.

\subsection{Comparison to other buried star forming regions}

Due to the deeply buried nature of proto-SSC\,$13$a in the SHC phase, we can compare the results obtained for proto-SSC\,$13$a with similar embedded sources. On the lower luminosity end, we can compare the SHC phase in proto-SSC\,$13$a to the Milky Way HCs. In fact, the SHC phase observed in proto-SSC\,$13$a seems to be a scaled-up version of Milky Way HCs. HCs, with luminosities $\lesssim10^7$\,L$_\odot$, are heated by a few massive protostars in a small cluster. In order to compare to proto-SSC\,$13$a, we selected the HCs sample from \citet{Rolffs2011a} (see Table~\ref{tab:HCs_BGNs}) since they derive their masses in a similar fashion as in our radiative transfer modelling (Section~\ref{ring_subsec_nlocal_mod}), taking into account the high opacity at the center of HCs and modelling the continuum emission accordingly (i.e. including the trapping of radiation or greenhouse effect) together with the  emission from vibrationally excited HCN. We use the IRAS luminosities and the FWHMs from the ATLASGAL survey at $870$\,$\upmu$m ($345$\,GHz) listed in \citet{Rolffs2011a} Table 1 (see references therein). Additionally,  for SgrB2(M) and SgrB2(N) HCs, we also use the values derived by \citet{Etxaluze2013} from Herschel observations.
On the higher luminosity end, we have the highly obscured nuclear regions of (U)LIRGs. These buried galactic nuclei \citep[hereafter BGNs; also known as CONs,][]{Falstad2021} have large column densities ($N_{\text{H}_2}>10^{24}$\,cm$^{-2}$) that span through the compact nuclear region  ($\lesssim100$\,pc) with very  high luminosities ($L_\text{IR}\gtrsim 10^{11}$\,L$_\odot$) powered by a starburst and/or an AGN. The BGNs parameters listed in Table~\ref{tab:HCs_BGNs} are taken from  \citet{GA19} and the masses from \citet{GA15} (see also references therein). For most sources, different values of $R$ and $L_\text{IR}$ are given, reflecting the uncertainty in their estimation.

\begin{table*}
\begin{center}
\caption[Parameters of Hot Cores and Buried Galactic Nuclei]{Parameters of HCs \citep[from][]{Rolffs2011a}, proto-SSC\,13 (this work) and BGNs \citep[from][]{GA15,GA19}}
\label{tab:HCs_BGNs}
\begin{threeparttable}
\begin{tabular}{lcccccc}
\hline \noalign {\smallskip}
HC	&	$D$	&	$R$		&	$L_\text{IR}$	&	$\Sigma_\text{IR}$		&	$M_{\text{H}_2}$ &	$L_\text{IR}/M_{\text{H}_2}$	 \\		
Source	    &	(kpc)	& (pc)	&	($10^5$\,L$_\odot$)		&	($10^5$\,L$_\odot$\,pc$^{-2}$)	&	($10^3$\,M$_\odot$)	&  (L$_\odot$/M$_\odot$) \\		
\hline \noalign {\smallskip}
IRAS12326-6245	& $	4.4	 $ & $	0.57	    $ & $	2.7	    $ & $	2.7	        $ & $	4.2		 $ & $64	 $ \\
G327.3-06   	& $	2.9	 $ & $	0.39	    $ & $	1.0	    $ & $	2.1	        $ & $	3.8		 $ & $26	 $ \\
IRAS16065-5158	& $	4.0	 $ & $	0.54	    $ & $	2.9	    $ & $	3.2	        $ & $	3.6		 $ & $81	 $ \\
NGC6334I	    & $	1.7	 $ & $	0.24	    $ & $	2.6	    $ & $	15	        $ & $	3.7		 $ & $70	 $ \\
IRAS17233-3606	& $	1.0	 $ & $	0.13	    $ & $	0.3     $ & $	5.0	        $ & $	6.2		 $ & $44   $ \\
SgrB2(N)$^{a}$	    & $	7.8	 $ & $	1.05 -	0.69$ & $	8.4-11	$ & $	2.4-7.4	$ & $	84-119   $ & $10 - 7  $ \\
SgrB2(M)$^{a}$	    & $	7.8  $ & $	1.23 -	0.69$ & $	63-50	$ & $	13-34   $ & $	220-167  $ & $29 - 42  $ \\
G10.47+0.03	    & $	10.6 $ & $	1.30	    $ & $	7.0     $ & $	1.3	        $ & $	23		 $ & $30	 $ \\
G31.41+0.31	    & $	7.9  $ & $	1.02	    $ & $	2.6	    $ & $	0.8	        $ & $	12		 $ & $22	 $ \\
G34.26+0.15	    & $	3.7  $ & $	0.52	    $ & $	4.7     $ & $	5.5	        $ & $	13		 $ & $36 $ \\
W51d	        & $	5.4  $ & $	0.79	    $ & $	24      $ & $	12	        $ & $	38		 $ & $63  $ \\
W51e	        & $	5.4  $ & $	0.76	    $ & $	12	    $ & $	6.6         $ & $	38		 $ & $32  $ \\
\hline \noalign {\smallskip}
\hline \noalign {\smallskip}

proto-SSC &	$D$	&	$R$		&	$L_\text{IR}$	&	$\Sigma_\text{IR}$		&	$M_{\text{H}_2}$	&	$L_\text{IR}/M_{\text{H}_2}$ \\		
Source	    &	(Mpc)	& (pc)	&	($10^{7}$\,L$_\odot$)		&	($10^7$\,L$_\odot$\,pc$^{-2}$)	&	($10^5$\,M$_\odot$)	&  (L$_\odot$/M$_\odot$) \\	
\hline \noalign {\smallskip}
proto-SSC\,13	        & $	3.5  $ & $	1.5 $ & $	9.4     $ & $	1.3	        $ & $	9.5	     $ & $ 99 $ \\	

\hline \noalign {\smallskip}
\hline \noalign {\smallskip}
BGN	&	$D$	&	$R$		&	$L_\text{IR}$	&	$\Sigma_\text{IR}$		&	$M_{\text{H}_2}$	&	$L_\text{IR}/M_{\text{H}_2}$ \\		
Source	    &	(Mpc)	& (pc)	&	($10^{11}$\,L$_\odot$)		&	($10^7$\,L$_\odot$\,pc$^{-2}$)	&	($10^9$\,M$_\odot$)	&  (L$_\odot$/M$_\odot$) \\	
\hline \noalign {\smallskip}
NGC4418	        & $	34  $ & $	11.7 - 13.5 $ & $	0.9-1.3     $ & $	22	        $ & $	0.3	     $ & $ 273-394 $ \\		
Arp220W	        & $	85	$ & $	47	 -	60	$ & $	7.6-12  	$ & $	11	        $ & $	5.5^{b}	 $ & $ 137-217 $ \\		
Arp220E	        & $	85	$ & $	87	 -	90	$ & $	1.3-2.4	    $ & $	0.5-1.0     $ & $	1.0^{b}	 $ & $ 127-234 $ \\		
Zw049.057	    & $	56	$ & $	15	 -	25	$ & $	1.0-1.4 	$ & $	5.0-20	    $ & $	1.1	     $ & $ 92-128  $ \\		
IC860	        & $	59	$ & $	14.5 -	20	$ & $	0.4-0.7 	$ & $	5.5         $ & $	0.2	     $ & $ 200-350 $ \\		
Mrk231	        & $	192 $ & $	55	 - 	73	$ & $	18-21       $ & $	11-22       $ & $	7.1	     $ & $ 254-296 $ \\
\hline \noalign {\smallskip}
\end{tabular}
\begin{tablenotes}
      \item $^a$ \small{In addition, the values from \citet{Etxaluze2013} are given for SgrB2(N) and SgrB2(M).}
      \item $^b$ \small{Masses for Arp\,220W and Arp\,220E have been roughly estimated from the Arp\,220 total mass ($6.6\times10^9$\,M$_\odot$) assuming each contributes in mass the same they do for the total} luminosity.  
\end{tablenotes}
\end{threeparttable}
\end{center}
\end{table*}

Due to the large difference, of several orders of magnitude, in radii, luminosities and masses between HCs and BGNs, we normalize these quantities by their size or by their gas mass to facilitate the comparison. Therefore, we estimate the luminosity surface density $\Sigma_\text{IR}=L_\text{IR}/(\pi r^2)$, mass surface density $\Sigma_{\text{H}_2}=M_{\text{H}_2}/(\pi r^2)$, and the luminosity per unit gas mass $L_\text{IR}/M_{\text{H}_2}$. The latter can be considered a proxy of the star formation efficiency as the derived luminosity arises from the embedded (proto)stars at the stage of SHC in which proto-SSC\,$13$a is at \citep[see][]{RicoVillas2020}. For proto-SSC\,$13$a these values, from Model 2, are: $\Sigma_\text{IR}=1.3\times10^7$\,L$_\odot$\,pc$^{-2}$, $\Sigma_{\text{H}_2}=1.3\times10^5$\,M$_\odot$\,pc$^{-2}$, and $L_\text{IR}/M_{\text{H}_2}=99$\,L$_\odot$\,M$_\odot^{-1}$.

\begin{figure*}
\centering
    \includegraphics[width=0.7\linewidth]{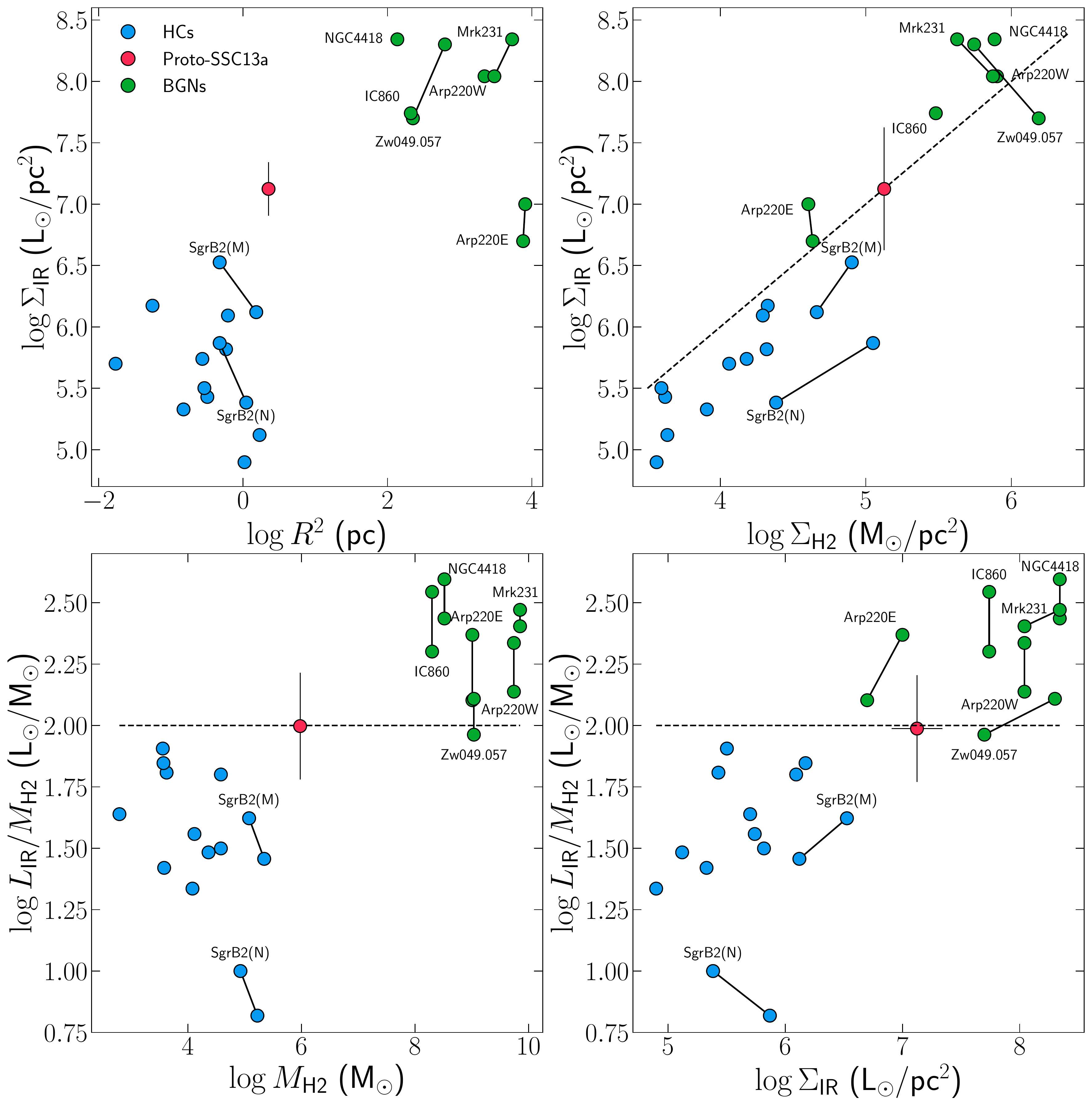}
  \caption[Proto-SSC\,$13$a comparison to HCs and BGNs]{Comparison between the derived properties of proto-SSC\,$13$a (red circle) to Milky Way HCs (blue circles) and BGNs from (U)LIRGs (green circles). Top left panel shows the luminosity surface density as a function of the squared source radius. Top right panel shows the luminosity surface density against the  surface density. The dashed line indicates the relationship between $\Sigma_\text{IR}$ and $\Sigma_{\text{H}_2}$ for $L_\text{IR}/M_{\text{H}_2}=100$\,L$_\odot$\,M$_\odot^{-1}$. Bottom left and bottom right panels show the luminosity per unit of gas mass as a function of the gas mass and the luminosity surface density, respectively. The error bars for proto-SSC\,$13$a show the assumed error of $50\%$ in luminosity due to varying $X_\text{d}$. Dashed lines indicate where $L_\text{IR}/M_{\text{H}_2}=100$\,L$_\odot$\,M$_\odot^{-1}$.
  }
  \label{fig:SHC13_HC_BGN}
\end{figure*}

In Figure~\ref{fig:SHC13_HC_BGN} we compare the properties of BGNs and Milky Way HCs with  proto-SSC\,$13$a in between. The upper left panel shows the luminosity surface density versus the squared radii of the sources. It can be seen that proto-SSC\,$13$a has an intermediate luminosity, at least one order of magnitude larger than HCs, but a size similar to the largest HCs. On the other hand, BGNs have much larger luminosities (three orders of magnitude larger than SHC\,$13$a) and also larger radii. On the right panel we plot the luminosity surface density now compared to the surface mass density. It can be seen that there is a trend with higher $\Sigma_\text{IR}$ for larger $\Sigma_{\text{H}_2}$, but for HCs and the proto-SSC\,$13$a the $L_\text{IR}/M_{\text{H}_2}$ ratio is $\lesssim 100$ (lower panels). This would point to a limit  on the amount of mass per surface area that embedded star formation can heat to produce the observed $\Sigma_\text{IR}$.

A similar trend is found in the lower left and right panels of Figure~\ref{fig:SHC13_HC_BGN}, where $L_\text{IR}/M_{\text{H}_2}$ increases as the mass and the luminosity surface increase, but staying $\lesssim100$ for HCs and proto-SSC\,$13$a. For the latter, this would indicate the  star formation proceeds much more efficiently than in most HCs (higher $L_\text{IR}/M_{\text{H}_2}$), close to the limit to what is observed in non-interacting or weakly-interacting star-forming galaxies \citep{Solomon1988, Gao2004, GraciaCarpio2011}. Instead, strong interacting or merging galaxies, like those hosting the BGNs, have $L_\text{IR}/M_{\text{H}_2}\gtrsim100$ (indeed they are located above the dashed lines in Figure~\ref{fig:SHC13_HC_BGN}). It has been proposed that this is due to a different mode of star formation, in which BGNs (i.e. (U)LIRGs) are able to convert an extraordinary amount of gas in extremely short times and in a very efficient manner (remember that $L_\text{IR}/M_{\text{H}_2}$ can be considered as a proxy to SFE) despite the probable presence of an AGN. 
We propose that, since massive star formation in HCs and proto-SSCs is already taking place very efficiently \citep{RicoVillas2020}, and that for the latter it does so in a very rapid and vigorous way, the excess of $L_\text{IR}/M_{\text{H}_2}$ above $\sim100$\,L$_\odot$\,M$_\odot^{-1}$ probably arises from the contribution of a buried AGN. However, the analysis of the remaining SSCs observed in NGC\,253 is required to confirm this hypothesis.

\section{Conclusions}

With very high resolution ALMA observations ($0.022\arcsec \times 0.020\arcsec\approx 0.37\,\text{pc}\times0.34\,\text{pc}$) at $219$\,GHz, we have resolved the young SSCs studied in \citet{RicoVillas2020}. Fragmentation is observed in most of them. %Using observations at $345$\,GHz with a resolution similar resolution to that at $219$\,GHz, we have fitted 2D Gaussians to the continuum emission of all sources, deriving the size (FWHMs) and integrated flux densities at $219$\,GHz and $345$\,GHz. The spectral index between these two frequencies reveal that most of these sources have large dust column densities. 
%With sparser resolution ($0.1\arcsec$) observations at $36$\,GHz and $111$\,GHz, we have also estimated the size of the free-free and non-thermal emission and its contribution to the $219$\,GHz flux density.
Due to its unique properties, we have focused on the study of the HC$_3$N* emission of proto-SSC\,$13$a because it is one of the brightest sources at $219$\,GHz and $345$\,GHz, it appears to have a circular symmetry around the center, and was already classified among the youngest proto-SSCs in \citet{RicoVillas2020}. Our main results for proto-SSC\,$13$a can be summarized as follows:

\begin{itemize}
    \item[(1)] The spectra shows the typically abundant molecular line emission observed towards HCs in the Milky Way. We have detected many rotational $J=24-23$ and $J=26-25$ HC$_3$N transitions from highly vibrationally excited states, covering a wide energy range from $131$\,K up to $1399$\,K. 
    
    \item[(2)] Using the \texttt{SLIM} module within \textsc{Madcuba}, we have carried out a 2D LTE analysis of the excitation of HC$_3$N*, obtaining maps of the HC$_3$N column density, vibrational temperature ($T_\text{vib}$), velocity and FWHMs. From the temperature map, we identify a small region with  $T_\text{vib}\gtrsim400$\,K towards the center of proto-SSC\,$13$a. From the velocity map we observe a velocity structure, with a large region with velocities of $250-263$\,km\,s$^{-1}$ and a smaller region with velocities of $243-250$\,km\,s$^{-1}$. 
    %Averaging the spectra in rings of $0.1$\,pc thickness up to $1.5$\,pc to increase the signal-to-noise ratio, specially in the high-$v$ lines, gives similar LTE results.
    The derived LTE dust temperatures overestimate the actual dust temperatures due to opacity effects and dust temperature gradients in the SHC.
    
    \item[(3)] We have also carried out, for the first time, a multi-transition non-local modelling of the spatial distribution of the  HC$_3$N* line emission profiles including the ground state and vibrationally excited states $v_7=1$, $v_7=2$, $v_6=1$, $v_5/v_7=3$, $v_6=v_7=1$, $v_4=1$, and $v_6=2$. The models are self-consistent with the continuum emission radial profile and take into account the greenhouse effect for different H$_2$ density profiles.
    Radial profiles of the observed HC$_3$N* line emission were produced by averaging the spectra in rings of $0.1$\,pc thickness up to a radius of $1.5$\,pc to increase the signal to noise ratio.
    Models only considering the continuum emission are very degenerated. The degeneration is broken when the radial profile of the HC$_3$N* emission is also considered.
    While the high-energy line emission can be explained either with high $T_\text{dust}$ or moderate $T_\text{dust}$ with high $N(\text{HC}_3\text{N})$, this degeneracy is solved when including the low-energy lines in the analysis, indicating that $T_\text{dust}\lesssim500$\,K and $N(\text{HC}_3\text{N})\sim3\times10^{17}$\,cm$^{-2}$ reproduce essentially all the data.
    %Despite the initial degeneracy of the models due to the large column densities present in proto-SSC\,$13$a, we were able to break it by taking into account simultaneously the continuum emission and the low-$v$ and high-$v$ transitions. The latter proves that thanks to the high-$v$ transitions we are able to observe deeper into the optically thick dust source and derive a temperature and density radial profile.
    From the model that best fits the data we obtain a luminosity of $L_\text{IR}=9.4\times10^7$\,L$_\odot$, similar to the value obtained in \citet{RicoVillas2020} and a mass of $M_{\text{H}_2}=9.5\times10^5$\,M$_\odot$, higher than that derived in \citet{RicoVillas2020}. 
    
    \item[(4)] The best-fit model assumes spatially distributed star formation proportional to the mass and density at each radii, with an H$_2$ density power law index of $q=1.5$ ($q=1$ in the outermost regions). In contrast, models with highly concentrated star formation only taking place at the center, despite reaching similar temperatures at the outer regions, they are unable to reproduce the high-$v$ and low-$v$ spatial profiles simultaneously. 
    %The distributed star formation model that best describes the observed spatial profiles has a density power-law index of $q=1.5$ in the innermost regions. For this $q$, the radius containing half of the star formation is $r_e\sim0.4$, i.e. more massive star formation is taking place at the center of proto-SSC\,$13$a, although star formation is still distributed (enclosing $\sim80\%$ of the total star formation at $r=0.9$\,pc).
    Therefore, the results from the best-fit  model would be in accordance with the competitive accretion scenario,  with the most massive stars located at the center and mass segregation is taking place at very early times.
    
    \item[(5)] We have discussed the possible nature of the velocity change observed from the 2D LTE analysis perpendicular to the rotation plane of NGC\,253.
   % We study then if its caused by an outflow. While the low mechanical luminosity derived for the possible outflow ($\sim10^5$\,L$_\odot$) could be explained in terms of stellar feedback, the lack of observed P-Cygni profiles \citep[also by][]{Levy2021} and other signs of mechanical feedback, make us lean towards a possible
   We conclude that cloud-cloud collision is the most likely explanation. Such a scenario would explain the velocity structure and the bridge feature observed in the $p-v$ diagrams. A rough analysis indicates that the collision between the two clouds provides the ram pressure  required to trigger the star formation in proto-SSC\,$13$a. This high pressure together with the shock resulting from the collision would be able to compress the gas and set the conditions for rapid massive star formation with high efficiencies. %The cloud-cloud collision mechanism is consistent with the possible external mechanism, briefly discussed in Chapter\ref{ch:paper_ngc253} for spatially unresolved SHCs, driving the formation of the SSCs in NGC\,253. Making a similar analysis he new high angular resolution data provides additional key constraints to the possible models aiming to explain the extreme mode of star formation in galaxies. 
    
    \item[(6)] From the derived $M_{\text{H}_2}$, we find that proto-SSC\,$13$a is in a supercriticial virial state with a virial parameter $\alpha\approx0.21$, suggesting that proto-SSC\,$13$a is going through a rapid collapse.
    %We investigate then if the radiation pressure from the forming stars can halt the collapse. From \citet{GA19}, we find that even for a high luminosity-to-mass ratio of $1700$\,L$_\odot$\,M$_\odot^{-1}$, the force exerted by radiation pressure is similar to the force of gravity only in the innermost regions.
    We also find that $L_\text{Edd}$ is an order of magnitude larger than the derived $L_\text{IR}$, meaning that radiation pressure is not able to halt the collapse. 
    
    \item[(7)] We infer a short free-fall time of $t_\text{ff}=3\times10^4$\,yr that, together with the derived small virial parameter, indicates that proto-SSC\,$13$ is indeed very young. Proto-SSC\,$13$a seems to be forming stars at SFR$=5$\,M$_\odot$\,yr$^{-1}$, which implies a short depletion time of $t_\text{dep}=2\times10^5\sim6t_\text{ff}$.
    
    \item[(8)] From the comparison of proto-SSC\,$13$a with other deeply embedded star forming regions in the Milky Way HCs and the BGNs of (U)LIRGs, we observe a trend with higher luminosity surface density ($\Sigma_\text{IR}$) for higher molecular gas mass surface density ($\Sigma_{\text{H}_2}$). However, for the pure star-forming HCs and proto-SSC\,$13$a, there seems to be an upper limit in $L_\text{IR}/M_{\text{H}_2}$ of $100$\,L$_\odot$\,M$_\odot^{-1}$, similar to that found for non-interacting star forming galaxies. Since SSCs are among the most condensed and extreme star formation modes known, we argue that the excess in $L_\text{IR}/M_{\text{H}_2}$ above $\sim 100$\,L$_\odot$\,M$_\odot^{-1}$ observed in BGNs most probably arises from the contribution of a buried AGN.% However, the latter needs to be confirmed with the analysis of the remaining SSCs in NGC\,253.
    
\end{itemize}

\section*{Acknowledgements}

This paper makes use of ALMA data in projects 2018.1.01395.S (PI: Rico-Villas, Fernando) and 2017.1.00433.S (PI:Bolatto, Alberto).
ALMA is a partnership of ESO (representing its member states), NSF (USA)
and NINS (Japan), together with NRC (Canada) and NSC and
ASIAA (Taiwan) and KASI (Republic of Korea), in cooperation
with the Republic of Chile. The Joint ALMA Observatory is operated by ESO, AUI/NRAO and NAOJ.

F.R.V acknowledges financial support to the Spanish Ministry of Science and Innovation (MCIN)
under grant number ESP2017-86582-C4-1-R, PID2019-105552RB-C41, and PhD fellowship BES-2016-078808, to the State Research
Agency (AEI) with DOI 10.13039/501100011033 under project number PID2019-105552RB-C41 and
to MDM-2017-0737 Unidad de Excelencia María de Maeztu.

V.M.R. is funded by the Agencia Estatal de Investigaci\'on (AEI) through the Ram\'on y Cajal programme (grant RYC2020-029387-I).

%%%%%%%%%%%%%%%%%%%%%%%%%%%%%%%%%%%%%%%%%%%%%%%%%%
\section*{Data Availability}

The data underlying this article were accessed from the ALMA Science Archive (\url{http://almascience.eso.org/asax/}), with the corresponding project identifiers listed on Table~\ref{tab:HRNGC253_continuums}. The derived data generated in this research will be shared on reasonable request to the corresponding author.



\bibliographystyle{mnras}
\bibliography{main_body} % if your bibtex file is called example.bib

%\begin{thebibliography}{99}
%\end{thebibliography}

%%%%%%%%%%%%%%%%%%%%%%%%%%%%%%%%%%%%%%%%%%%%%%%%%%

%%%%%%%%%%%%%%%%% APPENDICES %%%%%%%%%%%%%%%%%%%%%

\appendix

\section{\texorpdfstring{Proto-SSC\,13a}{Proto-SSC13a} averaged ring spectra and SLIM LTE model figures}
\label{ap:HR_spec}

To increase the signal-to-noise ratio of our data, we have averaged the spectra within rings of $0.1$\,pc thickness from the center of proto-SSC\,$13$a (i.e. $r=0$) up to a radius of $1.5$\,pc, increasing the signal-to-noise ratio for the higher energy HC$_3$N* transitions. We consider the center of the rings to be the position of the brightest continuum emission at $219$\,GHz, i.e. RA(J2000)$=00^\text{h}47^\text{m}33^\text{s}.1970$; Dec(J2000)$=-25^\circ17^\prime16\arcsec.733$. 

The following figures show the averaged spectra (black histogram lines) inside these rings towards proto-SSC\,$13$a. Overlaid is also the modelled \texttt{SLIM} LTE emission from the HC$_3$N* $J=24-23$ transitions (in green) and from the HC$_3$N* $J=26-25$ transitions. The sum of the \texttt{SLIM} fitted LTE profiles for all species is shown with  dashed red lines. With the fitted parameters,
\texttt{SLIM} models the emission of all the HC$_3$N* transitions even if not considered for the fit. Therefore, some undetected lines from high vibrational states show modelled emission within noise level. 
%We note that \texttt{SLIM} models the emission of the HC$_3$N* transition despite not being considered for the model fitting, and hence there are differences between the observed and modelled emission from the $v=0$ and $v_7=1$ transitions in the inner rings and some undetected transitions have modelled emission. 
The inner and outer radii of the rings, the number of pixels and the derived properties from the modelling are listed in Table~\ref{tab:NGC253_HR_SLIM_SHC13_TexLog_profiles}. The figures show the HC$_3$N* transitions vibrational state in black, with the quantum numbers of the upper level of the transition in parenthesis, i.e. ($l_\text{up}$) or ($l_\text{up}$, $k_\text{up}$). Labels for transitions from other species are indicated in grey. 

\begin{figure*}
\centering
    \includegraphics[width=0.97\linewidth]{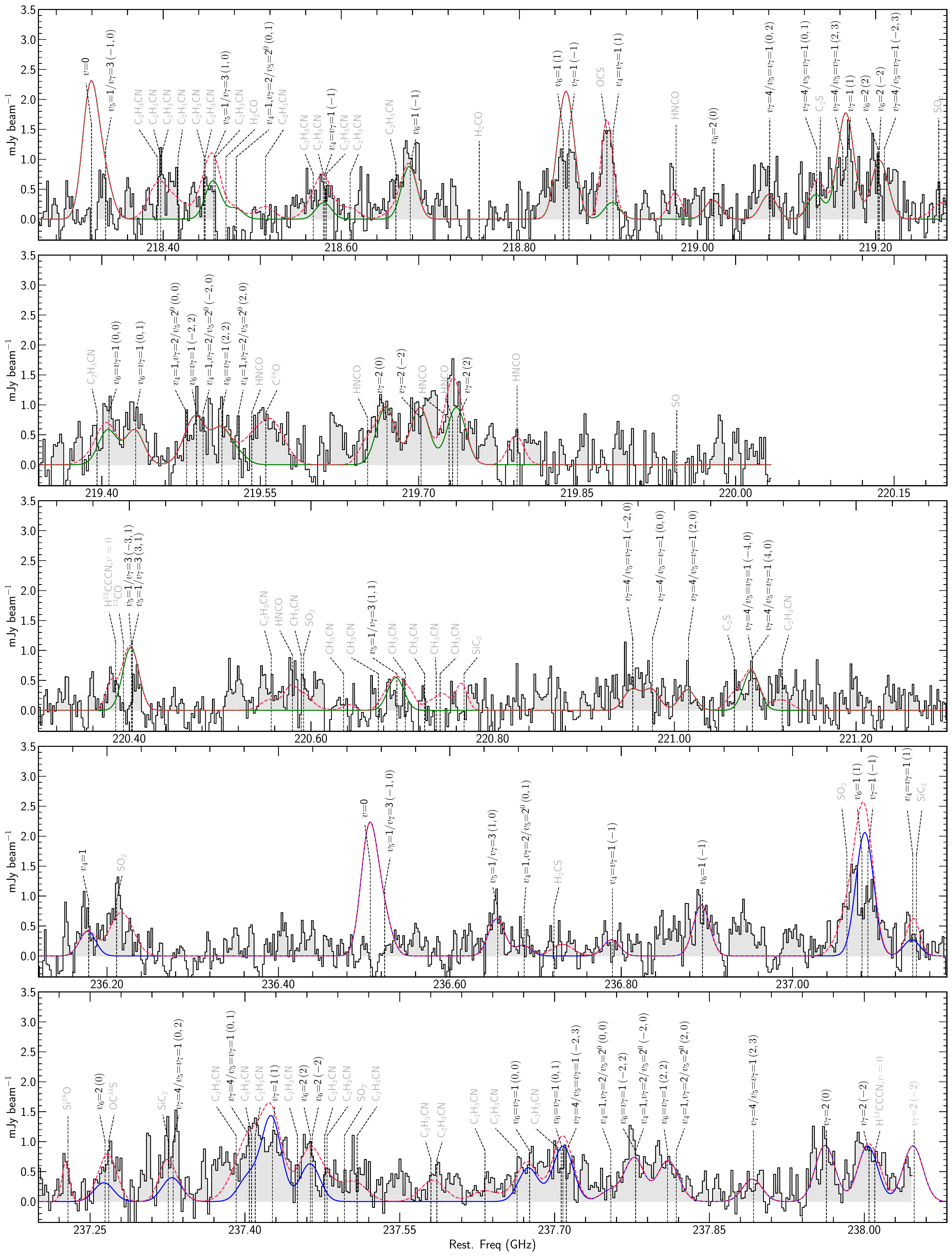}
  \caption[NGC\,253 proto-SSC\,$13$a HC$_3$N* ring $0.05$\,pc averaged spectra]{Proto-SSC\,$13$a HC$_3$N* averaged spectra and \texttt{SLIM} LTE emission (red line) for the different averaged ring emission from the ring enclosing the pixels with distances between $0.0$\,pc and $0.1$\,pc.
  }
  \label{ap:fig:NGC253_HR_SLIM_SHC13_ring0p05}
\end{figure*}

\begin{figure*}
\centering
    \includegraphics[width=0.97\linewidth]{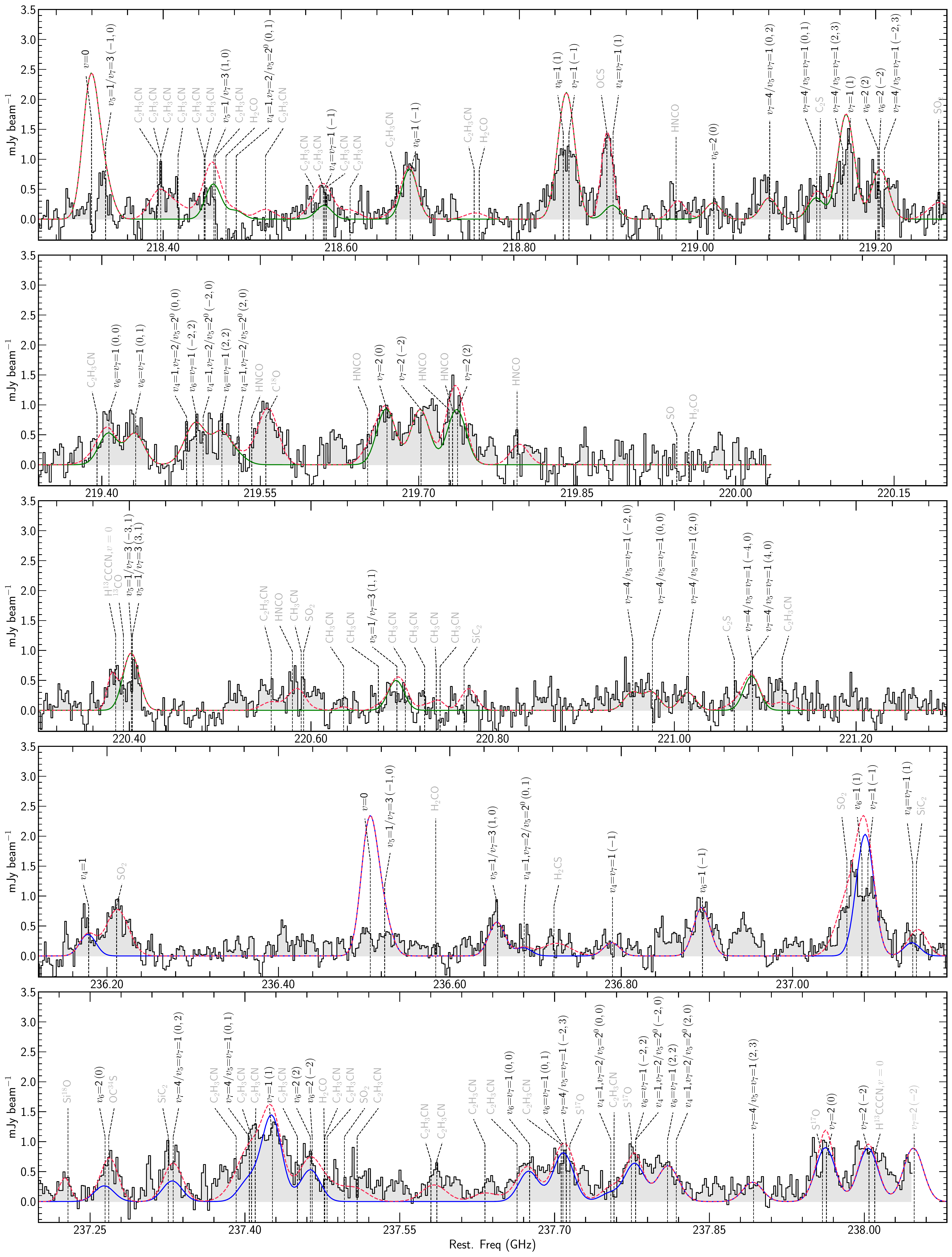}
  \caption[NGC\,253 proto-SSC\,$13$a HC$_3$N*  ring $0.15$\,pc  averaged spectra]{Proto-SSC\,$13$a HC$_3$N* averaged spectra and \texttt{SLIM} LTE emission (red line) for the different averaged ring emission from the ring enclosing the pixels with distances between $0.1$\,pc and $0.2$\,pc.
  }
  \label{ap:fig:NGC253_HR_SLIM_SHC13_ring0p15}
\end{figure*}

\begin{figure*}
\centering
    \includegraphics[width=0.97\linewidth]{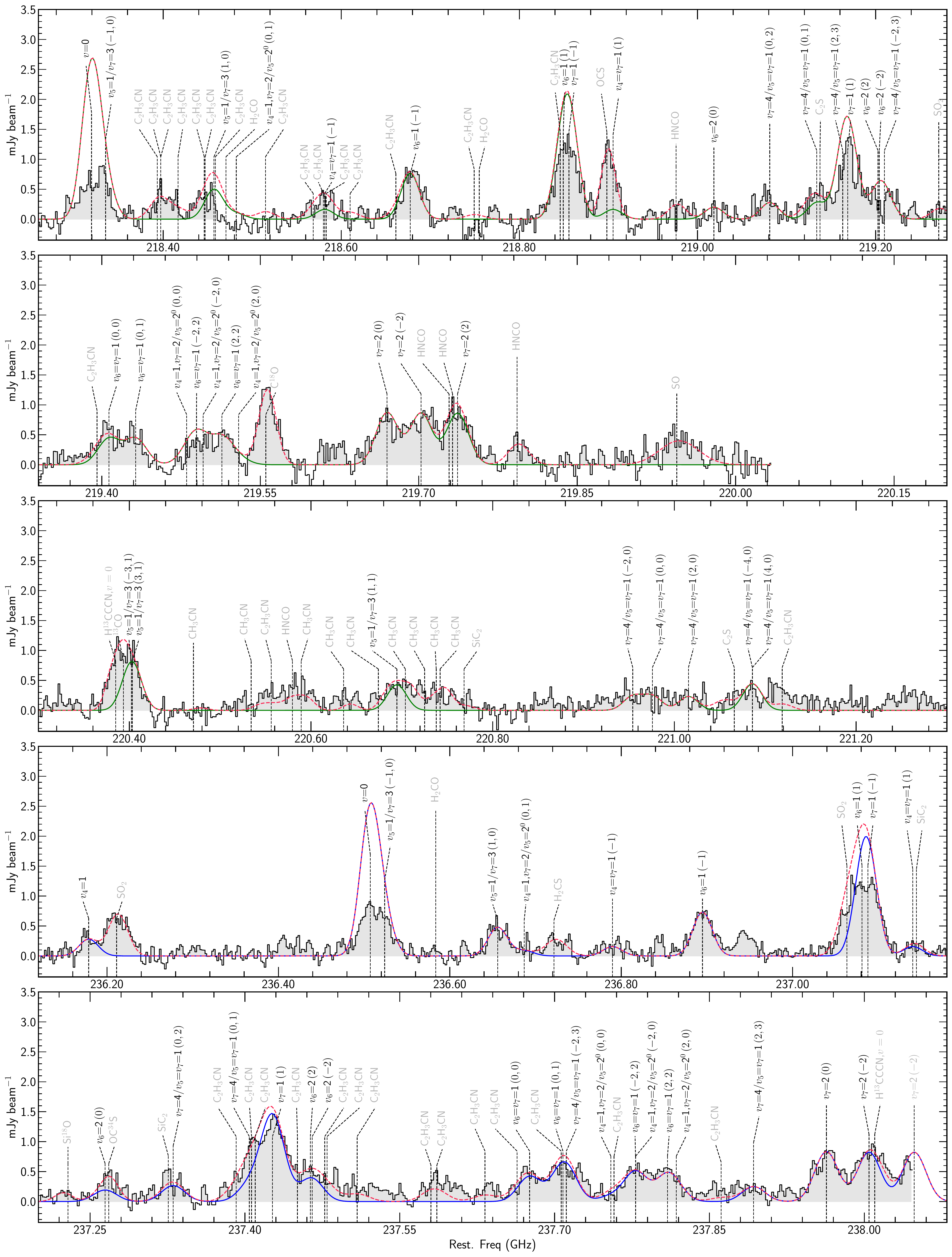}
  \caption[NGC\,253 proto-SSC\,$13$a HC$_3$N*  ring $0.25$\,pc averaged spectra]{Proto-SSC\,$13$a HC$_3$N* averaged spectra and \texttt{SLIM} LTE emission (red line) for the different averaged ring emission from the ring enclosing the pixels with distances between $0.2$\,pc and $0.3$\,pc. 
  }
  \label{ap:fig:NGC253_HR_SLIM_SHC13_ring0p25}
\end{figure*}

\begin{figure*}
\centering
    \includegraphics[width=0.97\linewidth]{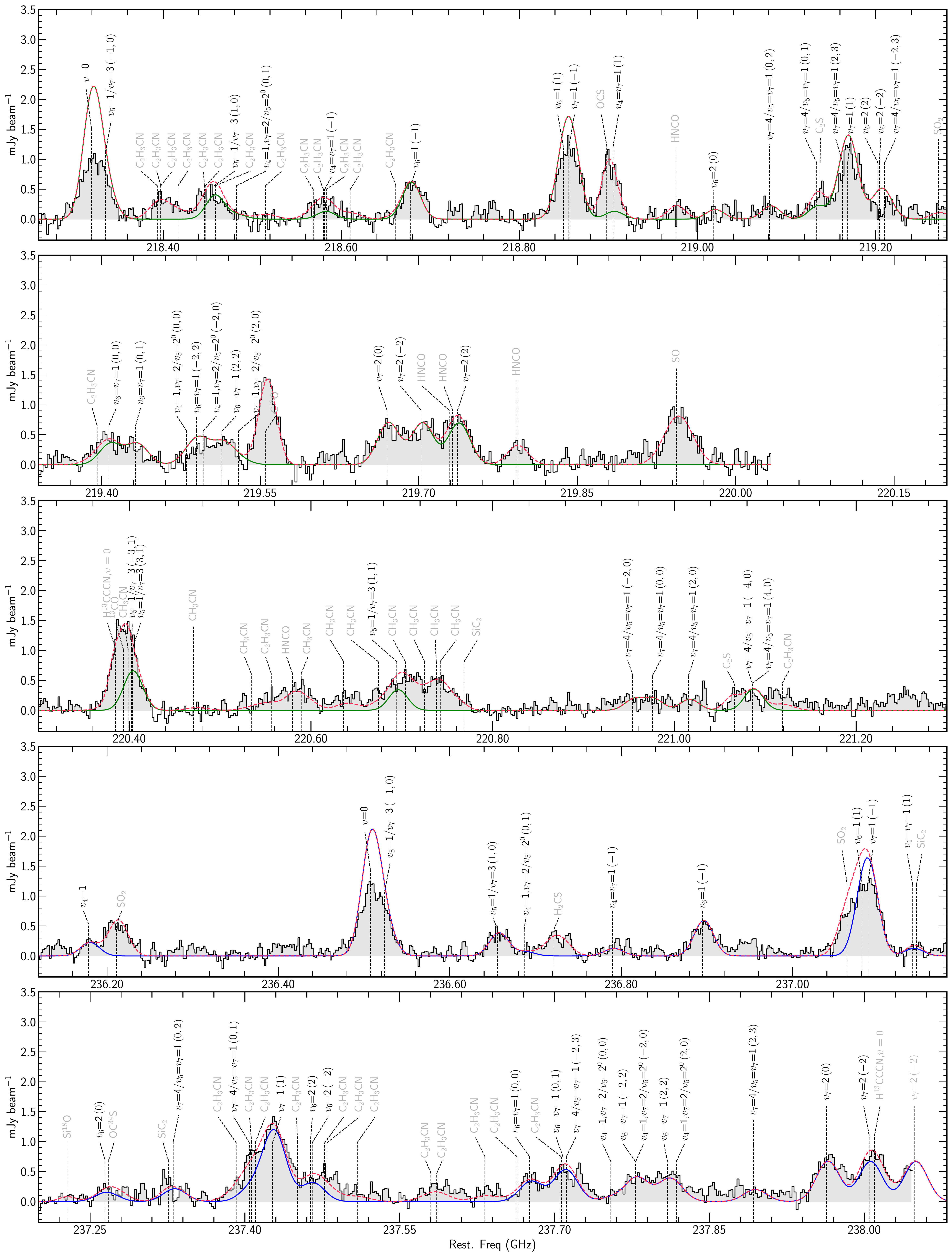}
  \caption[NGC\,253 proto-SSC\,$13$a HC$_3$N* ring $0.35$\,pc averaged spectra]{Proto-SSC\,$13$a HC$_3$N* averaged spectra and \texttt{SLIM} LTE emission (red line) for the different averaged ring emission from the ring enclosing the pixels with distances between $0.3$\,pc and $0.4$\,pc. 
  }
  \label{ap:fig:NGC253_HR_SLIM_SHC13_ring0p35}
\end{figure*}

\begin{figure*}
\centering
    \includegraphics[width=0.97\linewidth]{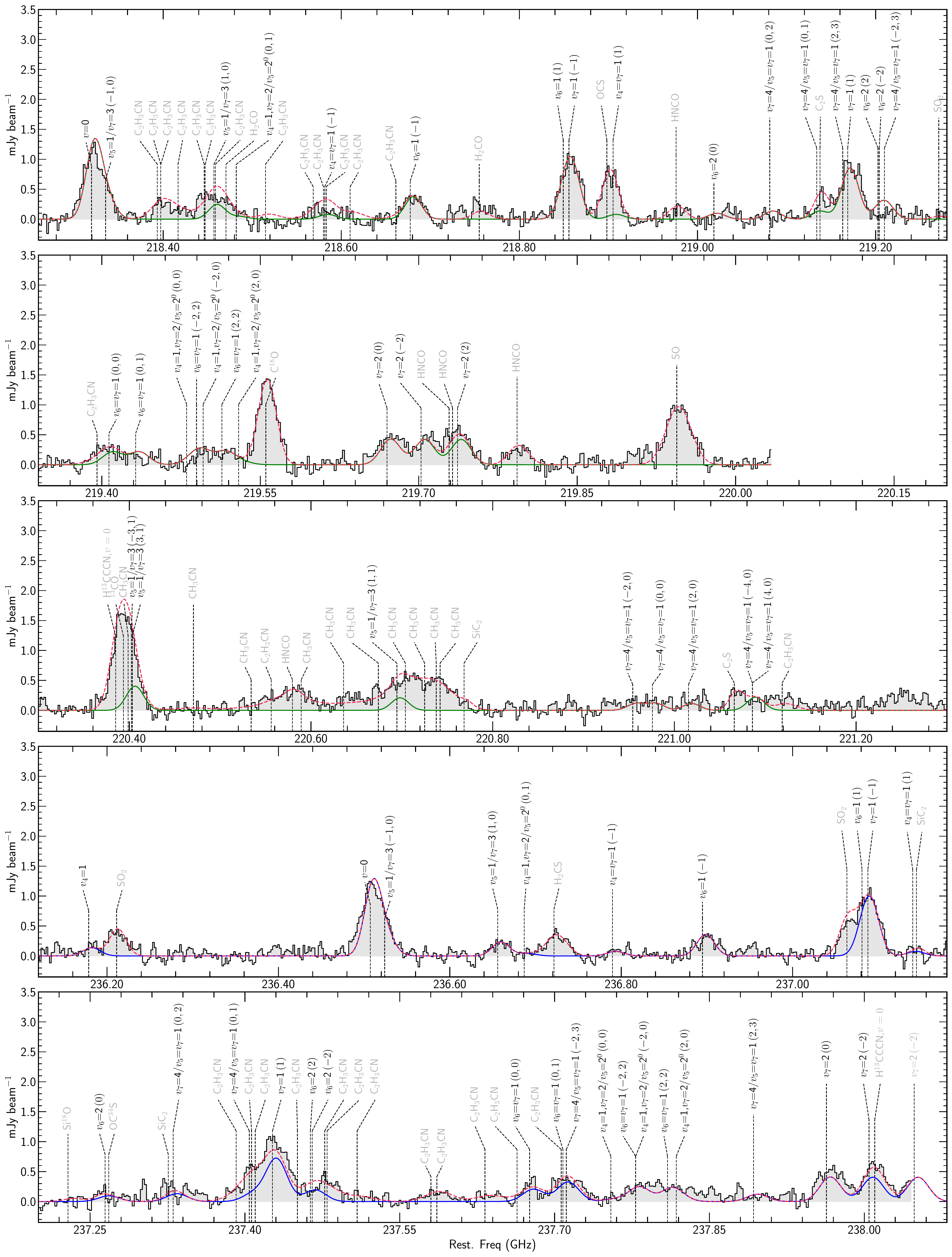}
  \caption[NGC\,253 proto-SSC\,$13$a HC$_3$N* ring $0.45$\,pc averaged spectra]{Proto-SSC\,$13$a HC$_3$N* averaged spectra and \texttt{SLIM} LTE emission (red line) for the different averaged ring emission from the ring enclosing the pixels with distances between $0.4$\,pc and $0.5$\,pc. 
  }
  \label{ap:fig:NGC253_HR_SLIM_SHC13_ring0p45}
\end{figure*}

\begin{figure*}
\centering
    \includegraphics[width=0.97\linewidth]{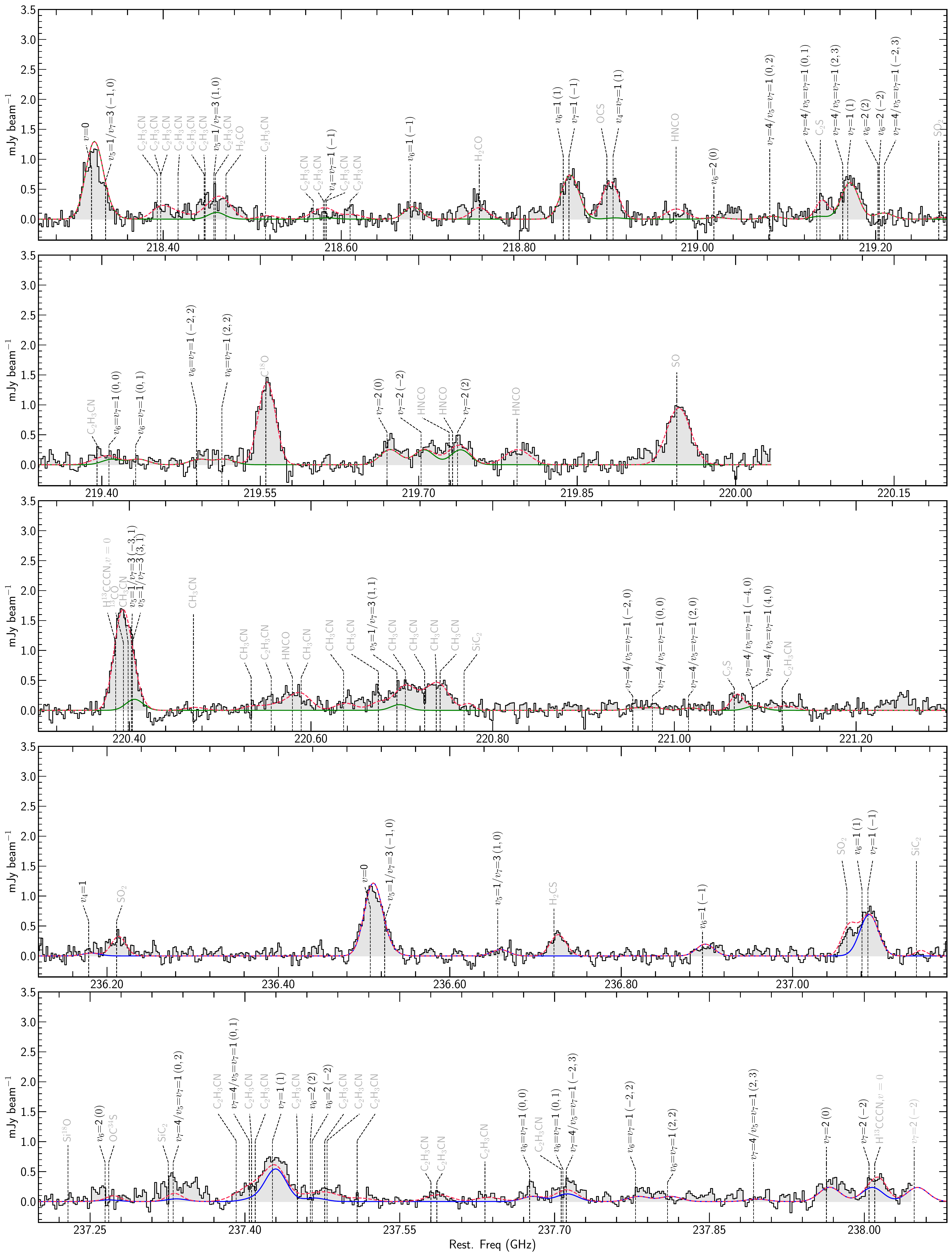}
  \caption[NGC\,253 proto-SSC\,$13$a HC$_3$N* ring $0.55$\,pc averaged spectra]{Proto-SSC\,$13$a HC$_3$N* averaged spectra and \texttt{SLIM} LTE emission (red line) for the different averaged ring emission from the ring enclosing the pixels with distances between $0.5$\,pc and $0.6$\,pc. 
  }
  \label{ap:fig:NGC253_HR_SLIM_SHC13_ring0p55}
\end{figure*}

\begin{figure*}
\centering
    \includegraphics[width=0.97\linewidth]{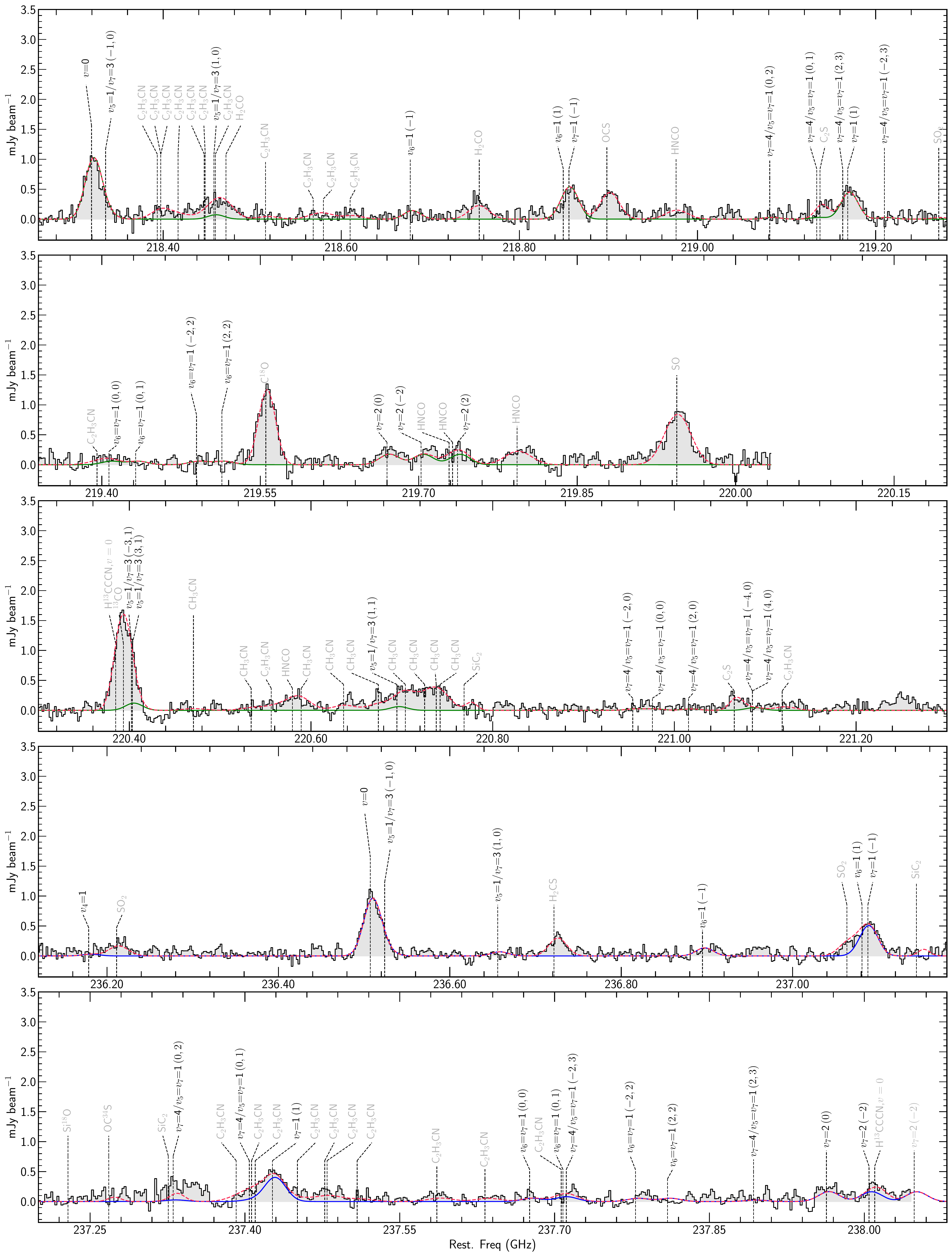}
  \caption[NGC\,253 proto-SSC\,$13$a HC$_3$N* ring $0.65$\,pc averaged spectra]{Proto-SSC\,$13$a HC$_3$N* averaged spectra and \texttt{SLIM} LTE emission (red line) for the different averaged ring emission from the ring enclosing the pixels with distances between $0.6$\,pc and $0.7$\,pc. 
  }
  \label{ap:fig:NGC253_HR_SLIM_SHC13_ring0p65}
\end{figure*}

\begin{figure*}
\centering
    \includegraphics[width=0.97\linewidth]{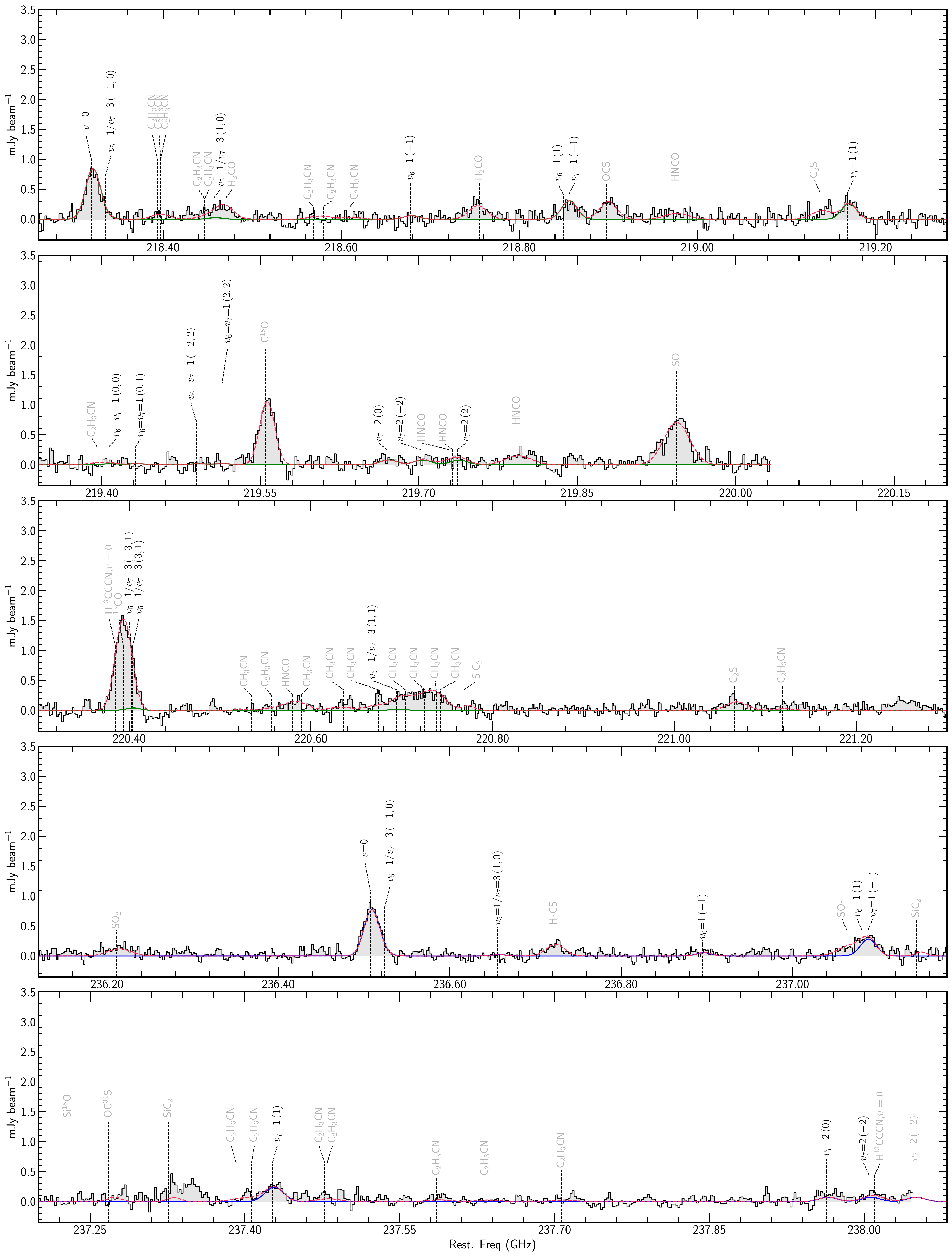}
  \caption[NGC\,253 proto-SSC\,$13$a HC$_3$N* ring $0.75$\,pc averaged spectra]{Proto-SSC\,$13$a HC$_3$N* averaged spectra and \texttt{SLIM} LTE emission (red line) for the different averaged ring emission from the ring enclosing the pixels with distances between $0.7$\,pc and $0.8$\,pc. 
  }
  \label{ap:fig:NGC253_HR_SLIM_SHC13_ring0p75}
\end{figure*}

\begin{figure*}
\centering
    \includegraphics[width=0.97\linewidth]{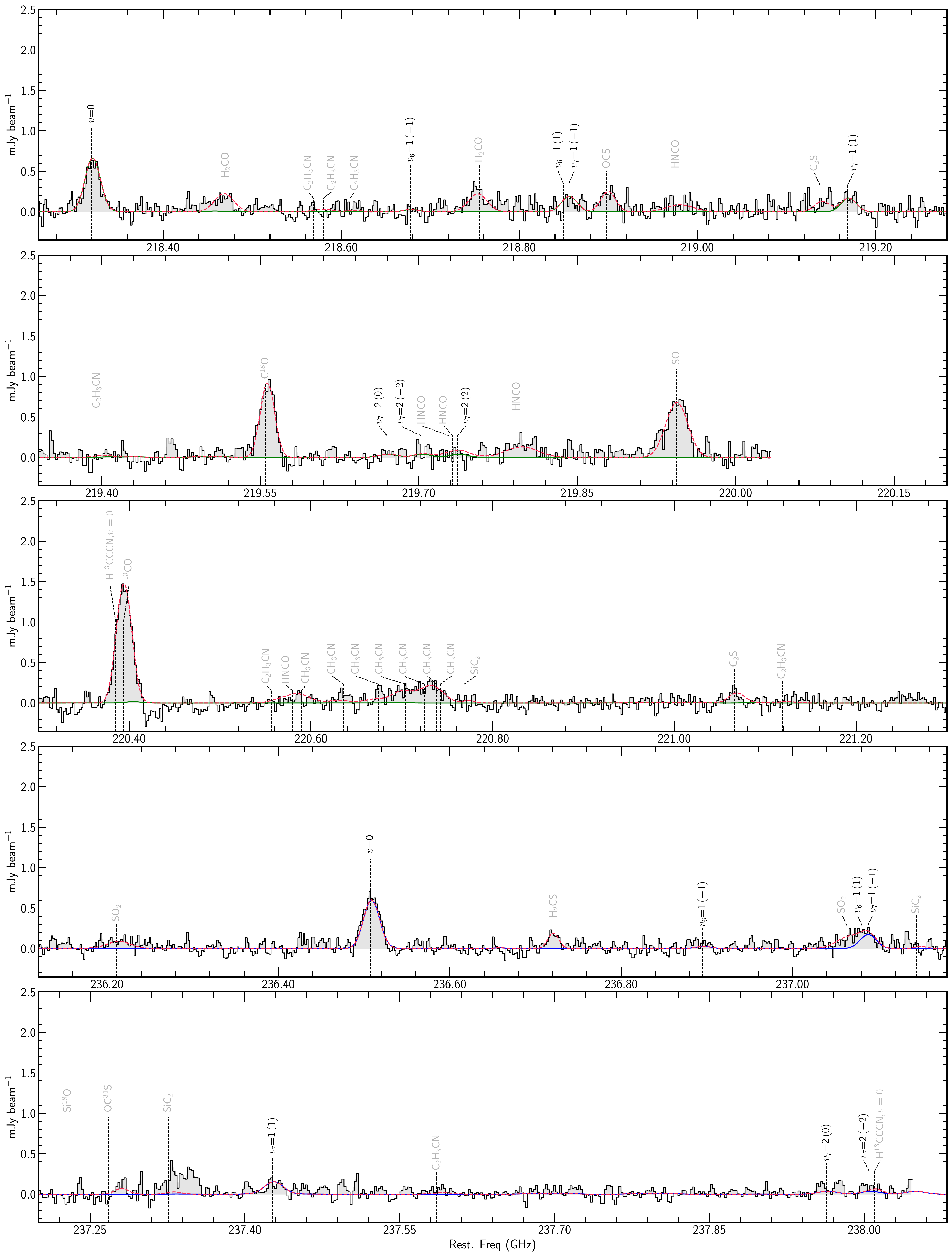}
  \caption[NGC\,253 proto-SSC\,$13$a HC$_3$N* ring $0.85$\,pc averaged spectra]{Proto-SSC\,$13$a HC$_3$N* averaged spectra and \texttt{SLIM} LTE emission (red line) for the different averaged ring emission from the ring enclosing the pixels with distances between $0.8$\,pc and $0.9$\,pc. 
  }
  \label{ap:fig:NGC253_HR_SLIM_SHC13_ring0p85}
\end{figure*}

\begin{figure*}
\centering
    \includegraphics[width=0.97\linewidth]{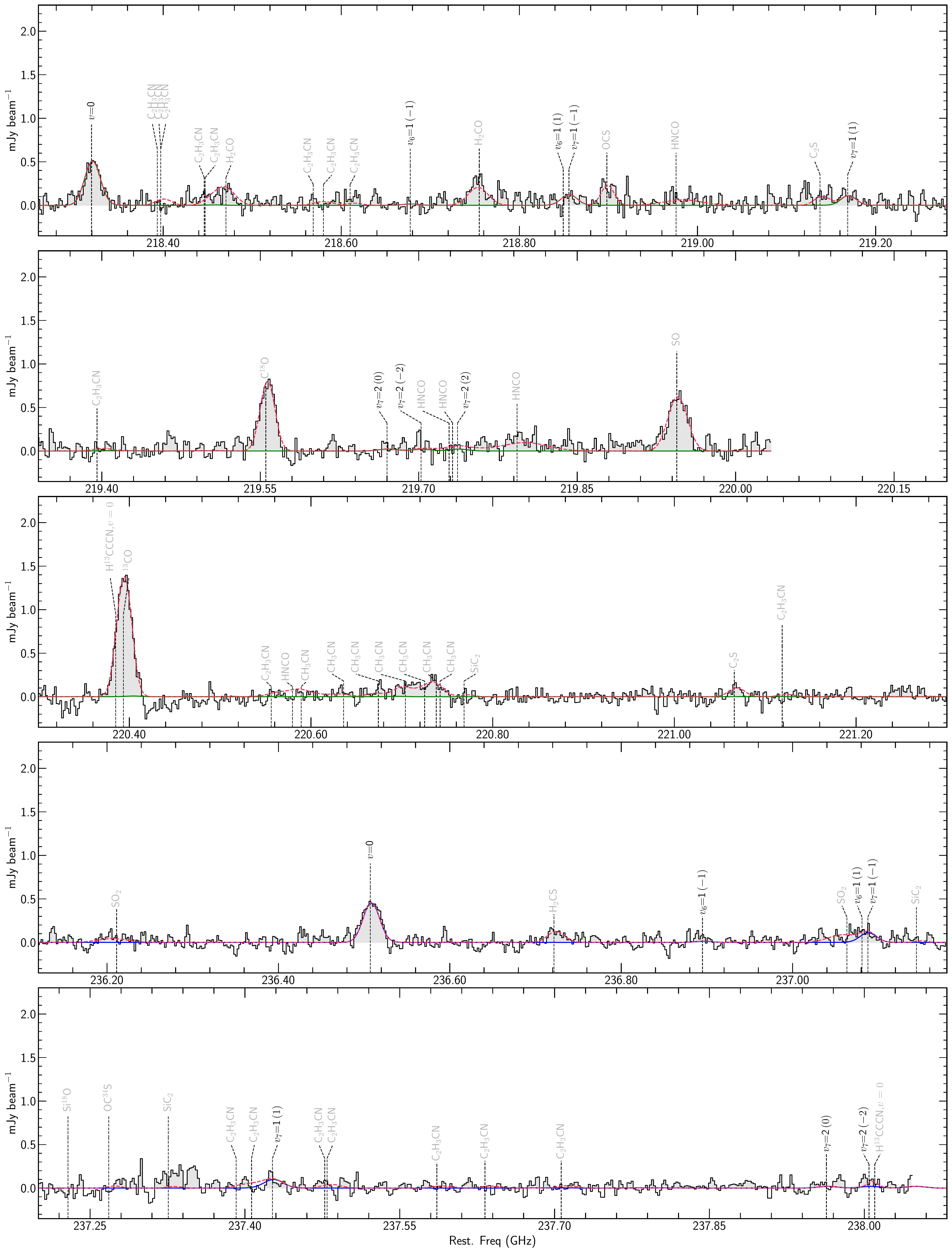}
  \caption[NGC\,253 proto-SSC\,$13$a HC$_3$N* ring $0.95$\,pc averaged spectra]{Proto-SSC\,$13$a HC$_3$N* averaged spectra and \texttt{SLIM} LTE emission (red line) for the different averaged ring emission from the ring enclosing the pixels with distances between $0.9$\,pc and $1.0$\,pc. 
  }
  \label{ap:fig:NGC253_HR_SLIM_SHC13_ring0p95}
\end{figure*}

\begin{figure*}
\centering
    \includegraphics[width=0.97\linewidth]{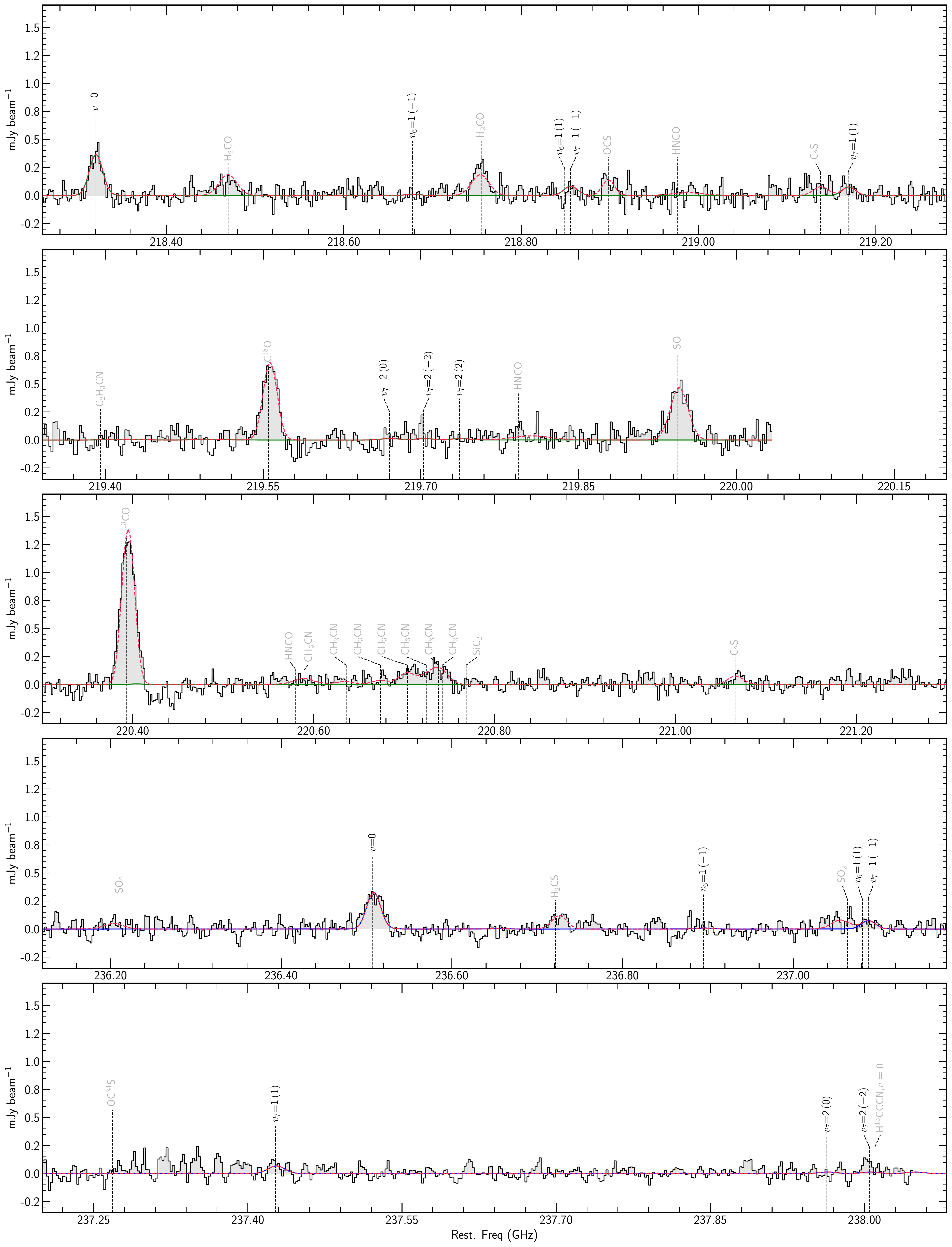}
  \caption[NGC\,253 proto-SSC\,$13$a HC$_3$N* ring $1.05$\,pc averaged spectra]{Proto-SSC\,$13$a HC$_3$N* averaged spectra and \texttt{SLIM} LTE emission (red line) for the different averaged ring emission from the ring enclosing the pixels with distances between $1.0$\,pc and $1.1$\,pc. 
  }
  \label{ap:fig:NGC253_HR_SLIM_SHC13_ring1p05}
\end{figure*}

\begin{figure*}
\centering
    \includegraphics[width=0.97\linewidth]{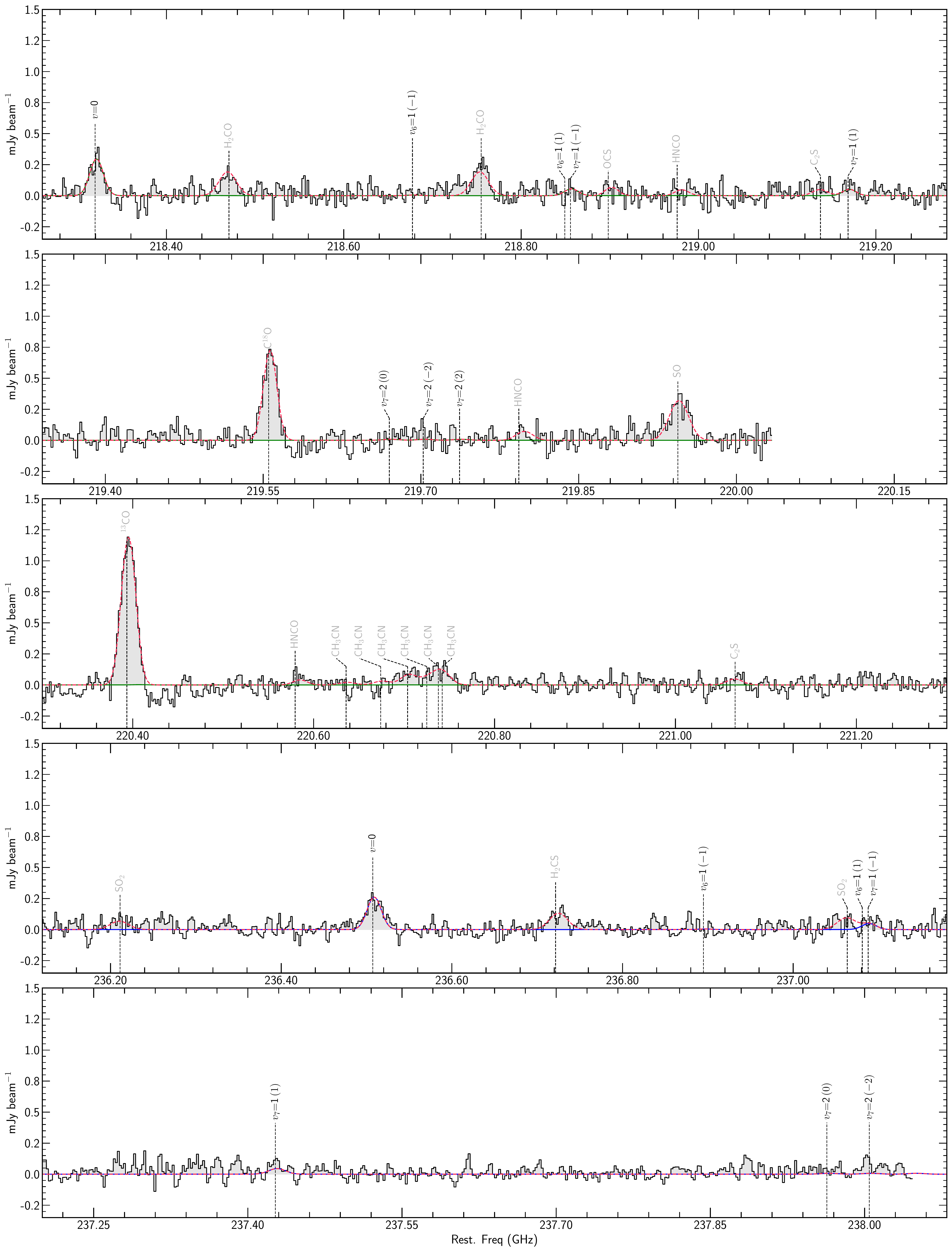}
  \caption[NGC\,253 proto-SSC\,$13$a HC$_3$N* ring $1.15$\,pc averaged spectra]{Proto-SSC\,$13$a HC$_3$N* averaged spectra and \texttt{SLIM} LTE emission (red line) for the different averaged ring emission from the ring enclosing the pixels with distances between $1.1$\,pc and $1.2$\,pc. 
  }
  \label{ap:fig:NGC253_HR_SLIM_SHC13_ring1p15}
\end{figure*}

\begin{figure*}
\centering
    \includegraphics[width=0.97\linewidth]{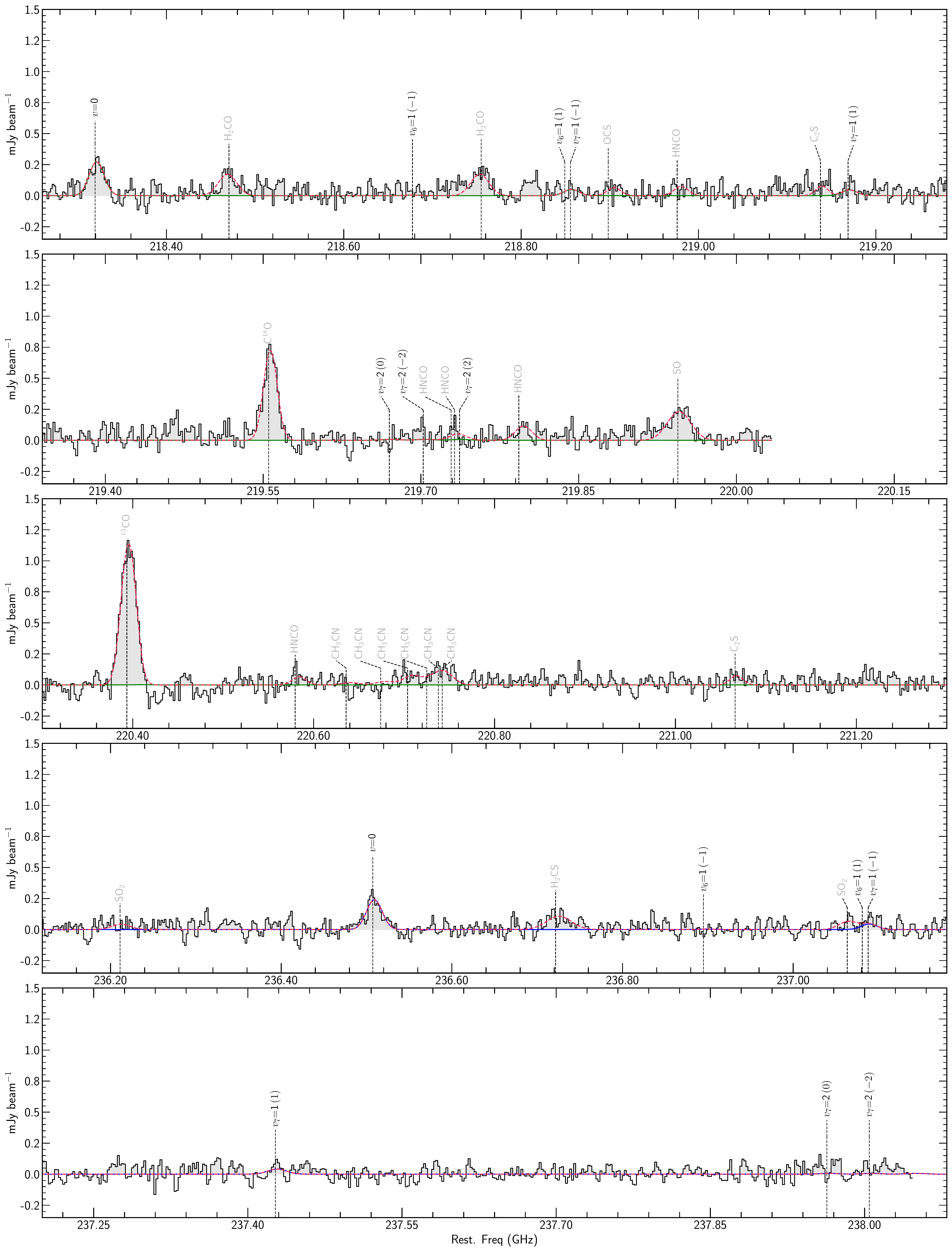}
  \caption[NGC\,253 proto-SSC\,$13$a HC$_3$N* ring $1.25$\,pc averaged spectra]{Proto-SSC\,$13$a HC$_3$N* averaged spectra and \texttt{SLIM} LTE emission (red line) for the different averaged ring emission from the ring enclosing the pixels with distances between $1.2$\,pc and $1.3$\,pc.
  }
  \label{ap:fig:NGC253_HR_SLIM_SHC13_ring1p25}
\end{figure*}

\begin{figure*}
\centering
    \includegraphics[width=0.97\linewidth]{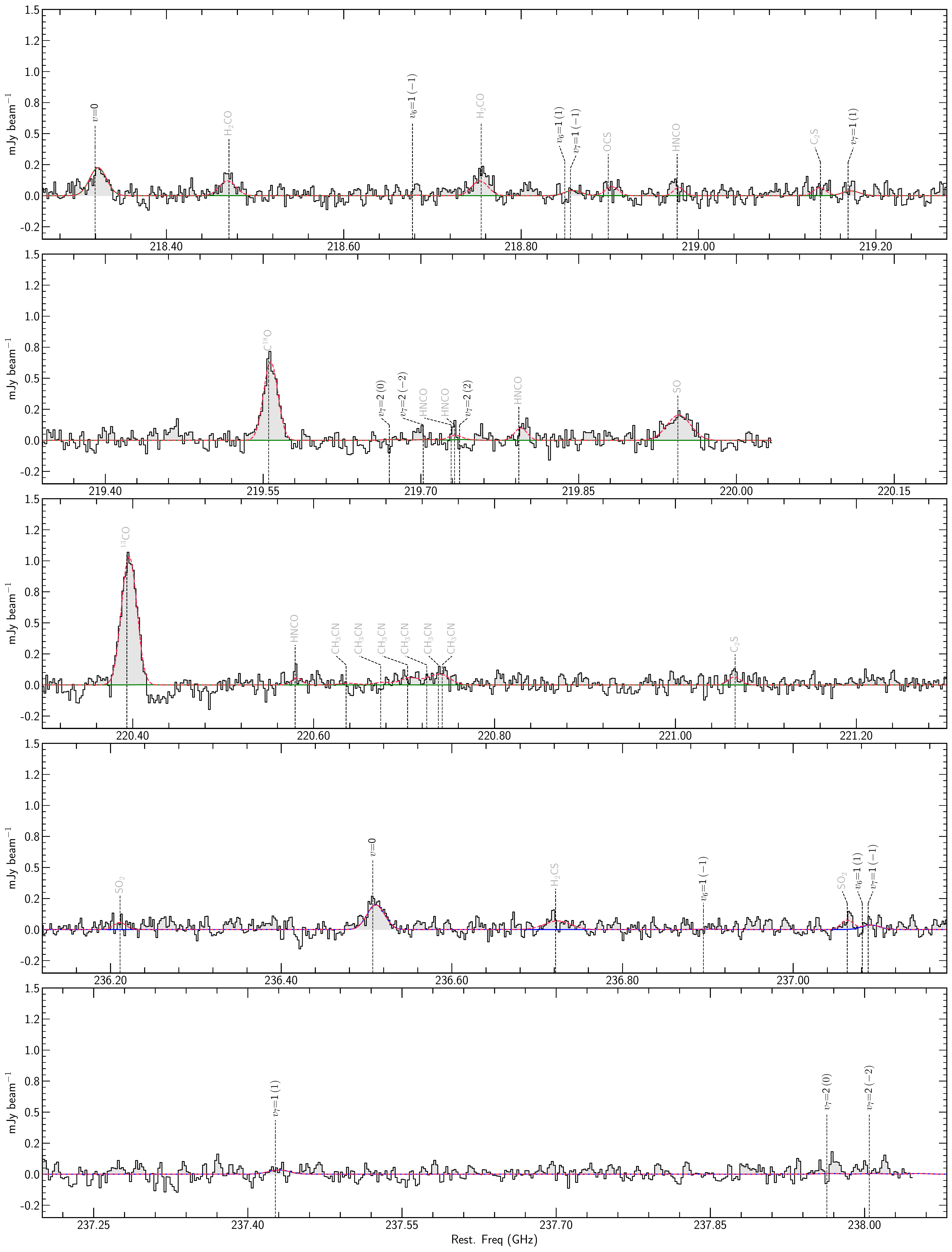}
  \caption[NGC\,253 proto-SSC\,$13$a HC$_3$N* ring $1.35$\,pc averaged spectra]{Proto-SSC\,$13$a HC$_3$N* averaged spectra and \texttt{SLIM} LTE emission (red line) for the different averaged ring emission from the ring enclosing the pixels with distances between $1.3$\,pc and $1.4$\,pc.
  }
  \label{ap:fig:NGC253_HR_SLIM_SHC13_ring1p35}
\end{figure*}

\begin{figure*}
\centering
    \includegraphics[width=0.97\linewidth]{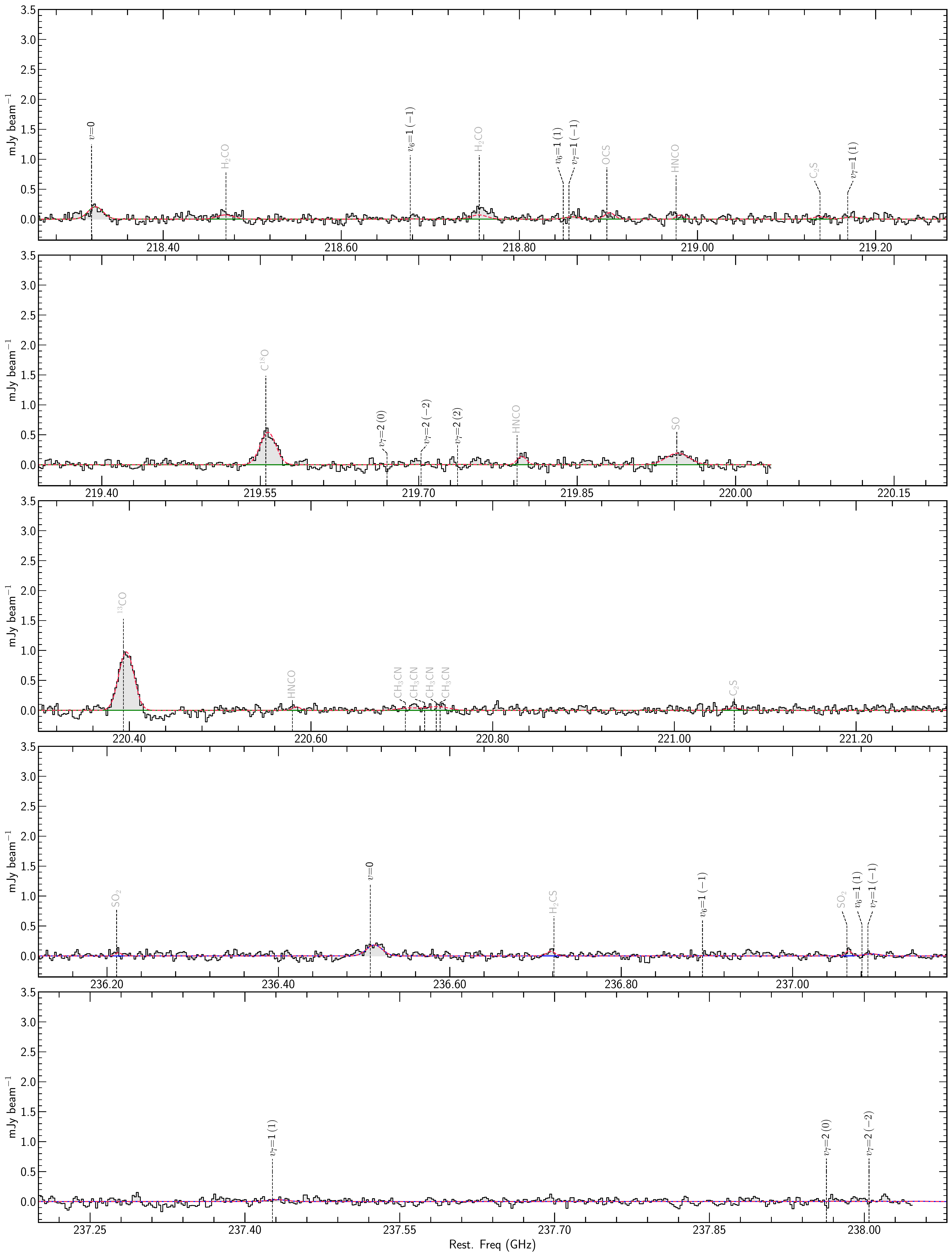}
  \caption[NGC\,253 proto-SSC\,$13$a HC$_3$N* ring $1.45$\,pc averaged spectra]{Proto-SSC\,$13$a HC$_3$N* averaged spectra and \texttt{SLIM} LTE emission (red line) for the different averaged ring emission from the ring enclosing the pixels with distances between $1.4$\,pc and $1.5$\,pc.
  }
  \label{ap:fig:NGC253_HR_SLIM_SHC13_ring1p45}
\end{figure*}

\section{Observed transitions}

Table~\ref{tab:HC3N_lines_info} lists the HC$_3$N* transitions vibrational state and quantum numbers in the observed frequency range.

\onecolumn
{\footnotesize
\setlength{\tabcolsep}{5pt}
\begin{longtable}{lcrccrcccccccc}
    \caption[HC$_3$N* transitions in the observed frequency range]{\normalsize
    HC$_3$N* transitions vibrational state and quantum numbers in the observed frequency range. Also indicated is the energy of the upper state ($E_\text{up}$), the frequency, the  logarithm of the Eintein-$A_\text{ul}$ coefficients, the line strength logarithm at $300$\,K ($\log \text{Int.}$) and if the line is detected (Det.) and contaminated (Cont.). When a line is both detected and contaminated it means that it is blended with another HC$_3$N transition or that it is slightly blended with another species.
    }\label{tab:HC3N_lines_info} \\
    \hline \noalign {\smallskip}
    Vib. State & $J_\text{up}$ & $l_\text{up}$ & $k_\text{up}$ & $J_\text{lo}$ & $l_\text{lo}$ & $k_\text{lo}$ & $g_\text{up}$ & $E_\text{up}$ &  $\nu$ &  $\log A_\text{ul}$ &  $\log \text{Int.}$ & Det. & Cont.  \\ 
    &  &  &  &  & &  &  & (K) &  (GHz) &  (s$^{-1}$)  & (nm$^2$\,MHz) & & \\ 
    \hline \noalign {\smallskip}
    \endfirsthead
    \caption[]{(continued from previous page)} \\
    \hline \noalign {\smallskip}
    Vib. State & $J_\text{up}$ & $l_\text{up}$ & $k_\text{up}$ & $J_\text{lo}$ & $l_\text{lo}$ & $k_\text{lo}$ & $g_\text{up}$ & $E_\text{up}$ &  $\nu$ &  $\log A_\text{ul}$ &  $\log \text{Int.}$ & Det. & Cont.  \\ 
    &  &  &  &  & &  &  & (K) &  (GHz) &  (s$^{-1}$)  & (nm$^2$\,MHz) & & \\ 
    \hline \noalign {\smallskip}
    \endhead
    \hline \noalign {\smallskip}
    \endfoot
    \hline \noalign {\smallskip}
    %\bottomrule
    \endlastfoot

$v=0$ 	&	24	&	  	&	  	&	23	&	  	&	  	&	49	&	131	&	218.32	&	-3.029	&	-1.295	&	 Y 	&	 N \\
$v=0$ 	&	26	&	  	&	  	&	25	&	  	&	  	&	53	&	153	&	236.51	&	-2.978	&	-1.222	&	 Y 	&	 N \\
$v_7=1$ 	&	24	&	-1	&	  	&	23	&	1	&	  	&	49	&	452	&	218.86	&	-3.083	&	-1.760	&	 Y 	&	 Y \\
$v_7=1$ 	&	24	&	1	&	  	&	23	&	-1	&	  	&	49	&	452	&	219.17	&	-3.081	&	-1.759	&	 Y 	&	 N \\
$v_7=1$ 	&	26	&	-1	&	  	&	25	&	1	&	  	&	53	&	474	&	237.09	&	-2.978	&	-1.688	&	 Y 	&	 Y \\
$v_7=1$ 	&	26	&	1	&	  	&	25	&	-1	&	  	&	53	&	475	&	237.43	&	-2.976	&	-1.687	&	 Y 	&	 N \\
$v_7=2$ 	&	24	&	0	&	  	&	23	&	0	&	  	&	49	&	774	&	219.68	&	-3.079	&	-2.224	&	 Y 	&	 N \\
$v_7=2$ 	&	24	&	-2	&	  	&	23	&	2	&	  	&	49	&	777	&	219.71	&	-3.082	&	-2.231	&	 Y 	&	 Y \\
$v_7=2$ 	&	24	&	2	&	  	&	23	&	-2	&	  	&	49	&	777	&	219.74	&	-3.082	&	-2.231	&	 Y 	&	 N \\
$v_7=2$ 	&	26	&	0	&	  	&	25	&	0	&	  	&	53	&	796	&	237.97	&	-2.974	&	-2.151	&	 Y 	&	 N \\
$v_6=1$ 	&	24	&	-1	&	  	&	23	&	1	&	  	&	49	&	849	&	218.68	&	-2.976	&	-2.335	&	 Y 	&	 N \\
$v_6=1$ 	&	24	&	1	&	  	&	23	&	-1	&	  	&	49	&	849	&	218.85	&	-3.082	&	-2.334	&	 Y 	&	 Y \\
$v_6=1$ 	&	26	&	-1	&	  	&	25	&	1	&	  	&	53	&	871	&	236.9	&	-2.978	&	-2.262	&	 Y 	&	 N \\
$v_6=1$ 	&	26	&	1	&	  	&	25	&	-1	&	  	&	53	&	871	&	237.09	&	-2.977	&	-2.261	&	 Y 	&	 Y \\
$v_5=1/v_7=3$ 	&	24	&	-1	&	0	&	23	&	1	&	0	&	49	&	1085	&	218.34	&	-3.090	&	-2.683	&	 Y 	&	 Y \\
$v_5=1/v_7=3$ 	&	24	&	1	&	0	&	23	&	-1	&	0	&	49	&	1085	&	218.46	&	-3.089	&	-2.682	&	 Y 	&	 Y \\
$v_5=1/v_7=3$ 	&	24	&	1	&	1	&	23	&	-1	&	1	&	49	&	1087	&	220.7	&	-3.076	&	-2.679	&	 Y 	&	 Y \\
$v_5=1/v_7=3$ 	&	24	&	-3	&	1	&	23	&	3	&	1	&	49	&	1094	&	220.41	&	-3.084	&	-2.677	&	 Y 	&	 Y \\
$v_5=1/v_7=3$ 	&	24	&	3	&	1	&	23	&	-3	&	1	&	49	&	1094	&	220.41	&	-3.084	&	-2.693	&	 Y 	&	 Y \\
$v_5=1/v_7=3$ 	&	26	&	-1	&	0	&	25	&	1	&	0	&	53	&	1107	&	236.53	&	-2.985	&	-2.693	&	 Y 	&	 Y \\
$v_5=1/v_7=3$ 	&	26	&	1	&	0	&	25	&	-1	&	0	&	53	&	1107	&	236.66	&	-2.984	&	-2.610	&	 Y 	&	 N \\
$v_6=v_7=1$ 	&	24	&	0	&	0	&	23	&	0	&	0	&	49	&	1168	&	219.41	&	-3.080	&	-2.610	&	 Y 	&	 Y  \\
$v_6=v_7=1$ 	&	24	&	0	&	1	&	23	&	0	&	1	&	49	&	1169	&	219.44	&	-3.080	&	-2.795	&	 Y 	&	 N  \\
$v_6=v_7=1$ 	&	24	&	-2	&	2	&	23	&	2	&	2	&	49	&	1169	&	219.49	&	-3.082	&	-2.797	&	 Y 	&	 N  \\
$v_6=v_7=1$ 	&	24	&	2	&	2	&	23	&	-2	&	2	&	49	&	1169	&	219.52	&	-3.082	&	-2.800	&	 Y 	&	 N  \\
$v_6=v_7=1$ 	&	26	&	0	&	0	&	25	&	0	&	0	&	53	&	1190	&	237.68	&	-2.975	&	-2.800	&	 Y 	&	 Y  \\
$v_6=v_7=1$ 	&	26	&	0	&	1	&	25	&	0	&	1	&	53	&	1192	&	237.71	&	-2.975	&	-2.723	&	 Y 	&	 Y  \\
$v_6=v_7=1$ 	&	26	&	-2	&	2	&	25	&	2	&	2	&	53	&	1192	&	237.78	&	-2.977	&	-2.724	&	 Y 	&	 N  \\
$v_6=v_7=1$ 	&	26	&	2	&	2	&	25	&	-2	&	2	&	53	&	1192	&	237.81	&	-2.977	&	-2.727	&	 Y 	&	 N  \\
$v_4=1$     	&	26	&	  	&	  	&	25	&	  	&	  	&	53	&	1399	&	236.18	&	-2.996	&	-3.042	&	 Y 	&	 N  \\
$v_7=4/v_5=v_7=1$ 	&	24	&	0	&	0	&	23	&	0	&	0	&	49	&	1404	&	220.98	&	-3.091	&	-3.150	&	 N 	&	 N \\
$v_7=4/v_5=v_7=1$ 	&	24	&	0	&	2	&	23	&	0	&	2	&	49	&	1406	&	219.09	&	-3.096	&	-3.155	&	 Y 	&	 N \\
$v_7=4/v_5=v_7=1$ 	&	24	&	-2	&	0	&	23	&	2	&	0	&	49	&	1407	&	220.96	&	-3.095	&	-3.159	&	 N 	&	 N \\
$v_7=4/v_5=v_7=1$ 	&	24	&	2	&	0	&	23	&	-2	&	0	&	49	&	1407	&	221.02	&	-3.095	&	-3.160	&	 N 	&	 N \\
$v_7=4/v_5=v_7=1$ 	&	24	&	2	&	3	&	23	&	-2	&	3	&	49	&	1407	&	219.17	&	-3.102	&	-3.162	&	 Y 	&	 Y \\
$v_7=4/v_5=v_7=1$ 	&	24	&	-2	&	3	&	23	&	2	&	3	&	49	&	1407	&	219.22	&	-3.100	&	-3.161	&	 Y 	&	 Y \\
$v_7=4/v_5=v_7=1$ 	&	24	&	0	&	1	&	23	&	0	&	1	&	49	&	1408	&	219.14	&	-3.096	&	-3.158	&	 Y 	&	 Y \\
$v_7=4/v_5=v_7=1$ 	&	24	&	-4	&	0	&	23	&	4	&	0	&	49	&	1416	&	221.09	&	-3.103	&	-3.180	&	 Y 	&	 Y \\
$v_7=4/v_5=v_7=1$ 	&	24	&	4	&	0	&	23	&	-4	&	0	&	49	&	1416	&	221.09	&	-3.103	&	-3.180	&	 Y 	&	 Y \\
$v_7=4/v_5=v_7=1$ 	&	26	&	0	&	2	&	25	&	0	&	2	&	53	&	1428	&	237.34	&	-2.992	&	-3.083	&	 N 	&	 Y \\
$v_7=4/v_5=v_7=1$ 	&	26	&	2	&	3	&	25	&	-2	&	3	&	53	&	1429	&	237.9	&	-3.017	&	-3.111	&	 N 	&	 Y \\
$v_7=4/v_5=v_7=1$ 	&	26	&	-2	&	3	&	25	&	2	&	3	&	53	&	1430	&	237.72	&	-3.002	&	-3.096	&	 N 	&	 Y \\
$v_7=4/v_5=v_7=1$ 	&	26	&	0	&	1	&	25	&	0	&	1	&	53	&	1430	&	237.41	&	-2.991	&	-3.085	&	 Y 	&	 Y \\
$v_6=2$ 	&	24	&	2	&	  	&	23	&	-2	&	  	&	49	&	1568	&	219.21	&	-3.098	&	-3.392	&	 Y 	&	 Y\\
$v_6=2$ 	&	24	&	-2	&	  	&	23	&	2	&	  	&	49	&	1568	&	219.21	&	-3.098	&	-3.392	&	 Y 	&	 Y\\
$v_6=2$ 	&	24	&	0	&	  	&	23	&	0	&	  	&	49	&	1589	&	219.02	&	-3.096	&	-3.420	&	 N 	&	 N\\
$v_6=2$ 	&	26	&	2	&	  	&	25	&	-2	&	  	&	53	&	1590	&	237.47	&	-2.993	&	-3.319	&	 N 	&	 Y\\
$v_6=2$ 	&	26	&	-2	&	  	&	25	&	2	&	  	&	53	&	1590	&	237.47	&	-2.993	&	-3.319	&	 N 	&	 Y\\
$v_6=2$ 	&	26	&	0	&	  	&	25	&	0	&	  	&	53	&	1612	&	237.27	&	-2.991	&	-3.348	&	 N 	&	 Y\\
$v_4=v_7=1$ 	&	24	&	-1	&	  	&	23	&	1	&	  	&	49	&	1715	&	218.59	&	-3.093	&	-3.598	&	 N 	&	 Y \\
$v_4=v_7=1$ 	&	24	&	1	&	  	&	23	&	-1	&	  	&	49	&	1715	&	218.91	&	-3.091	&	-3.597	&	 N 	&	 Y \\
$v_4=v_7=1$ 	&	26	&	-1	&	  	&	25	&	1	&	  	&	53	&	1737	&	236.8	&	-2.988	&	-3.525	&	 N 	&	 N \\
$v_4=v_7=1$ 	&	26	&	1	&	  	&	25	&	-1	&	  	&	53	&	1737	&	237.15	&	-2.985	&	-3.524	&	 N 	&	 Y \\
$v_4=1,v_7=2/v_5=2^{0}$ 	&	24	&	0	&	0	&	23	&	0	&	0	&	49	&	2032	&	219.49	&	-3.089	&	-4.055	&	 N 	&	 Y \\
$v_4=1,v_7=2/v_5=2^{0}$ 	&	24	&	-2	&	0	&	23	&	2	&	0	&	49	&	2036	&	219.5	&	-3.092	&	-4.063	&	 N 	&	 Y \\
$v_4=1,v_7=2/v_5=2^{0}$ 	&	24	&	2	&	0	&	23	&	-2	&	0	&	49	&	2036	&	219.53	&	-3.091	&	-4.063	&	 N 	&	 Y \\
$v_4=1,v_7=2/v_5=2^{0}$ 	&	24	&	0	&	1	&	23	&	0	&	1	&	49	&	2036	&	218.49	&	-3.095	&	-4.064	&	 N 	&	 Y \\
$v_4=1,v_7=2/v_5=2^{0}$ 	&	26	&	0	&	0	&	25	&	0	&	0	&	53	&	2055	&	237.76	&	-2.984	&	-3.982	&	 N 	&	 Y \\
$v_4=1,v_7=2/v_5=2^{0}$ 	&	26	&	-2	&	0	&	25	&	2	&	0	&	53	&	2058	&	237.78	&	-2.986	&	-3.990	&	 N 	&	 Y \\
$v_4=1,v_7=2/v_5=2^{0}$ 	&	26	&	2	&	0	&	25	&	-2	&	0	&	53	&	2058	&	237.82	&	-2.986	&	-3.990	&	 N 	&	 Y \\
$v_4=1,v_7=2/v_5=2^{0}$ 	&	26	&	0	&	1	&	25	&	0	&	1	&	53	&	2058	&	236.69	&	-2.990	&	-3.992	&	 N 	&	 Y \\
%$v_3=1$ & 26 &  &  & 25 &  &  & 53 & 3145 & 235.79 & -5.570 &-2.997 \\
%$v_2=1$ & 26 &  &  & 25 &  &  & 53 & 3424 & 235.39 & -5.977 &-3.000 \\
\end{longtable}
}
\twocolumn

%%%%%%%%%%%%%%%%%%%%%%%%%%%%%%%%%%%%%%%%%%%%%%%%%%

% Don't change these lines
\bsp	% typesetting comment
\label{lastpage}
\end{document}